\documentclass[a4paper,12pt]{article}
\usepackage{geometry}
\usepackage{setspace}

\usepackage{enumitem}
\usepackage{amsmath,amssymb,amsthm}
\usepackage{mathtools}

\newtheorem{theorem}{Theorem}
\newtheorem{lemma}{Lemma}
\newtheorem{proposition}{Proposition}
\newtheorem{corollary}{Corollary}
\theoremstyle{definition}
\newtheorem{assumption}{Assumption}
\newtheorem{subassumption}{Assumption}
\numberwithin{subassumption}{assumption}

\newtheorem{remark}{Remark}

\allowdisplaybreaks

\renewcommand{\d}{\mathrm{d}}
\newcommand\T{{ \mathrm{\scriptscriptstyle T} }}
\renewcommand{\Pr}{\mathrm{pr}}
\newcommand{\indep}{\mathrel{\reflectbox{\rotatebox[origin=c]{90}{$\models$}}}}
\makeatletter
\newcommand*\dott{\mathpalette\dott@{.75}}
\newcommand*\dott@[2]{\mathbin{\vcenter{\hbox{\scalebox{#2}{$\m@th#1\bullet$}}}}}
\makeatother

\usepackage{booktabs}
\usepackage{multirow}
\usepackage{array}

\usepackage{tikz}
\usetikzlibrary{arrows.meta,automata,calc,positioning,shapes.arrows,shapes.geometric,shapes.multipart,decorations.pathreplacing,decorations.pathmorphing,positioning,swigs}
\tikzset{vertex/.style={inner sep=0pt,minimum size=1em},
         sq/.style={draw,rectangle,inner sep=0pt,minimum size=1em},
         hidden/.style={draw,shape=circle,preaction={fill=gray!50,fill opacity=0.5},inner sep=1pt,minimum size=1em},
         swig vsplit={gap=3pt,line color right=red},
         ell/.style={draw,inner sep=0pt,shape=ellipse}}
\tikzset{snake it/.style={decorate, decoration={snake, amplitude=.4mm,segment length=1.5mm,post length=1mm,pre length=1mm}}}

\usepackage{natbib}

\usepackage{multibib}
\newcites{suppmat}{References}
\bibliographystylesuppmat{apalike}

\usepackage{times}
\usepackage{newtxmath}
\usepackage[cal=cm,bb=ams]{mathalpha}

\usepackage{authblk}

\usepackage{titlesec}
\titleformat*{\section}{\bfseries}
\titleformat*{\subsection}{\bfseries}
\titleformat*{\paragraph}{\bfseries}

\usepackage{minitoc}
\noptcrule

\usepackage{etoolbox}

\makeatletter
\patchcmd{\@maketitle}{\large}{\normalsize}{}{}
\patchcmd{\@maketitle}{\LARGE}{\large}{}{}
\makeatother

\title{Efficient estimation of the target population average treatment effect
  from multi-source data}
\date{}

\author[1]{Zehao Su\footnote{Corresponding author.}}
\author[1]{Helene Charlotte Rytgaard}
\author[2]{Henrik Ravn}
\author[1]{Frank Eriksson}
\affil[1]{Section of Biostatistics, University of Copenhagen, Copenhagen, Denmark}
\affil[2]{Novo Nordisk A/S, S\o{}borg, Denmark}

\begin{document}

\maketitle

\begin{abstract}
  We consider estimation of the target population average treatment effect (TATE) when outcome information is unavailable.
  Instead, we observe the outcome in multiple source populations and wish to combine the treatment effects therein to make inference on the TATE.
  In contrast to existing works that assume transportability on the conditional distribution of potential outcomes or conditional treatment-specific means, we work under a weaker form of effect transportability.
  Following the framework for causally interpretable meta-analysis, we assume transportability of conditional average treatment effects across multiple populations, which may hold with fewer standardization variables.
  Under this assumption, we derive the semiparametric efficiency bound of the TATE and characterize a class of doubly robust and asymptotically linear estimators.
  Within this class, an efficient estimator assigns optimal weights to observations from different data sources.
  Additionally, we suggest estimators of a low-dimensional summary of effect heterogeneity in the target population.
  We illustrate the use of the proposed estimators on a multicentre weight management clinical trial for semaglutide, a glucagon-like peptide-1 receptor agonist, on overweight or obese patients.
  Using outcome information from other regions, we estimate the weight loss effect of semaglutide in the United States subgroup.
  

\end{abstract}

\doparttoc[c]
\faketableofcontents

\section{Introduction}
Meta-analyses synthesize new evidence of intervention effects in a target population from a pool of existing data sources in the form of studies.
Usually there are non-negligible differences between the target population and the study population from which a data source is sampled.
Given an intervention effect of interest, meta-analysts then compile a set of baseline covariates that summarize the interpopulational differences.
These covariates are referred to as relevant variables for transportability.
Put simply, when individual participant data is available, the synthesized effect in the target population can often be obtained by standardizing the effect from some source population through adjustment of the relevant variables or by weighting the sample from the source population to account for possible covariate shifts of the relevant variables.
We follow up on the causally interpretable meta-analysis framework in the multi-source data setting \citep{dahabreh2023efficient}, where the parameter of interest is the target population average treatment effect (TATE).
Retaining the arguably weak identifiability assumptions in that work, we suggest a family of asymptotically linear estimators of the TATE that employs cross-fitting for estimation of the nuisance parameters.
We derive the efficient influence function (EIF) of the TATE and show that an efficient estimator emerges from this family of estimators.
Under the same assumptions, we also study a simple form of effect heterogeneity in the target population by projecting the conditional average treatment effect (CATE) onto a pre-specified basis expansion.

Two practical questions must be addressed before performing evidence synthesis.
The first concerns the scope of the relevant variables.
Evidence synthesis methods fundamentally rely on transportability of intervention effects across study populations.
For transportability on summary measures involving a single intervention, the relevant variables required are referred to as prognostic variables in the meta-analysis literature.
An instance of such summary measures is the conditional mean outcome of a treatment arm in a clinical trial.
In contrast, for transportability of effect measures, the required relevant variables are referred to as effect modifiers \citep{vanderweele2007four}.
These form only a subset of the prognostic variables, which makes them effective for evidence synthesis, as the variables simultaneously collected in all populations are often limited.
Moreover, transportability on a single-intervention summary measure is less likely to hold in the first place due to unmeasured prognostic factors that influence the outcome level, such as overall quality of healthcare.

The second question concerns the overlap between populations.
For example, if the source population is derived from a clinical trial, the target population may include individuals which do not meet the trial's inclusion criteria.
A lack of overlap results in biased synthesized effects, because the evidence synthesis involves extrapolating intervention effects to the target population.
Extrapolation may be avoided by trimming the original target population relative to the source population, thereby restoring overlap \citep{chen2023generalizability}.
However, the artificially trimmed target population may not correspond to any real-world cohort, which undermines interpretability of the synthesized effect.
When multiple source populations are available for the interventions of interest,  overlap may be achieved by treating these source populations as a single joint source population.
Although individual source populations may violate overlap with respect to the target population, pooling multiple source populations reduces the risk of unintended extrapolation of intervention effects.

Recently, there has been a large mass of research works on transportability in causal inference, as well as meta-analysis methods targeting causally interpretable estimands.
Most existing transportability and generalizability methods work with a single source population \citep{lee2022doubly,dahabreh2020extending,josey2021transporting,li2023note}.
While some methods can be extended to handle multiple source populations, they would typically rely on unnecessarily restrictive transportability assumptions for the intervention effect of interest.
In the context of using external controls in clinical trials, \citet{li2023improving} suggest efficient fusion estimator of the treatment effect assuming transportability of the conditional mean outcome under placebo.
\citet{li2023efficient} propose a general framework for data fusion with multi-source longitudinal data which hinges on overlap between every source population and the target population.
\citet{wang2024efficient} consider subgroup effect estimation under overlap between the target population and the joint source population in a multi-source setup and transportability of the conditional distribution of the potential outcome.
The aforementioned methods do not address the two questions that commonly arise in causal meta-analysis.
An exception is \citet{dahabreh2023efficient}, who establish identifiability of the TATE under the joint overlap condition and transportability of the CATE between the target population and every source population.
However, estimation procedures of the identified parameter have not yet been studied, a task we undertake in this work.

\section{Identifiability under CATE transportability}
\label{sec:identifiability}

The source data originates from a collection of \(m\) studies labelled by a discrete variable \(D\in[m]\), where \([m]\) is a shorthand for the index set \(\{1,\dots,m\}\).
Participants in every source study receive a binary treatment \(A\in\{0,1\}\), and their outcome \(Y\) is recorded at the end of the study.
We assume that the outcome is real-valued.
Additionally, all source studies measure a common set of baseline covariates \(X\) containing the relevant variables for transportability.
Within source study \(d\), the observed data is \(n_{0d}\) independent and identically distributed (i.i.d.) copies of the tuple \((Y,A,X)\).
From the target population, an i.i.d. sample of the covariates \(X\) of size \(n_1\) is collected.
The total sample size is thus \(n= \sum_{d\in[m]}n_{0d}+n_{1}\).
We further introduce a binary indicator \(G\) for whether an observation belongs to one of the source studies, in which case it is \(0\), or the target population, in which case it is \(1\).

The sample sizes of the source studies usually cannot be controlled by meta-analysts, as they reflect the practical data collection decisions in the respective studies.
A common example in randomized clinical trials is the minimum sample size determined from power calculations.
Although the actual sampling of data is specific to the study population, it is helpful to view the complete sample
as an i.i.d. sample from some joint distribution over the observed data \(O=\{(1-G)Y,(1-G)A,X,(1-G)D,G\}\).
We make the following assumption on the sample sizes throughout.
\begin{assumption}[Sampling proportions]
  \label{asn:sampling}
  There exist fixed values \(\alpha_{0d}\in(0,1)\) such that the proportions \(n_{0d}/n\to \alpha_{0d}\) for \(d\in[m]\) and \(n_{1}/n\to \alpha\) when \(n\to\infty\).
\end{assumption}

Then we have \(\Pr(D=d,G=0)=\alpha_{0d}\) and \(\Pr(G=1)=\alpha\), while the marginal probabilities of \(D\) and \(G\) do not reflect the relative sizes of the underlying study population nor the sampling mechanism adopted.
We can hypothesize the existence of an artificial super-population from which data is drawn.
Random sampling in the super-population is asymptotically equivalent to actual biased sampling.

To facilitate presentation, we introduce the following notations in the observed data distribution and compile them in table~\ref{tab:notation} for reference.
Define the selection score of being in the target population as \(\pi(x)= \Pr(G=1\mid X=x)\), the selection score of being in source study \(d\) when in the joint source population as \(\zeta(d\mid x)= \Pr(D=d\mid X=x,G=0)\), and the support of baseline covariates in the target population as \(\mathcal{X}=\{x:\pi(x)>0\}\).
When we condition on the event \(\{D=d\}\) that an individual belongs to source study \(d\), it is implied that we also condition on the event \(\{G=0\}\) that the individual belongs to the joint source population.
In source study \(d\), the propensity score of receiving intervention \(a\) is \(e(a\mid x,d)= \Pr(A=a\mid X=x,D=d)\), and the outcome has conditional mean \(\mu(a,x,d)=E(Y\mid A=a,X=x,D=d)\) and variance \(V(a,x,d)=\mathrm{var}(Y\mid A=a,X=x,D=d)\).

\begin{table}
  \caption{List of nuisance parameters in the observed data distribution.}
  \label{tab:notation}
  \centering
  \footnotesize
  \begin{tabular}{llp{.4\textwidth}}
    Notation & Definition & Meaning \\
    \(\alpha\) & \(\Pr(G=1)\) & Proportion of the target population \\
    \(\pi(x)\) & \(\Pr(G=1\mid X=x)\) & Selection score into the target population \\
    \(\zeta(d\mid x)\) & \(\Pr(D=d\mid X=x,G=0)\) & Selection scores within the joint source population\\
    \(\mathcal{X}\) & \(\{x:\pi(x)>0\}\) & Support of baseline covariates in the target population \\
    \(e(a\mid x,d)\) & \(\Pr(A=a\mid X=x,D=d)\) & Propensity score \\
    \(\mu(a,x,d)\) & \(E(Y\mid A=a,X=x,D=d)\) & Conditional outcome mean \\
    \(V(a,x,d)\) & \(\mathrm{var}(Y\mid A=a,X=x,D=d)\) & Conditional outcome variance\\
    \(w(x,d)\) & \(\{\mathrm{var}(U\mid X=x,D=d)\}^{-1}\) & Optimal weight function
  \end{tabular}
\end{table}

Let \(Y(a)\) denote the potential outcome of \(Y\) under intervention \(a=0,1\).
We take as the target parameter the TATE
\[
  \theta=E\{Y(1)-Y(0)\mid G=1\}.
\]
Identifiability of the TATE with the observed data requires the following assumptions.

\stepcounter{assumption}
\begin{subassumption}[External validity]
  \label{asn:transportability}
  \hfill
  \begin{enumerate}[nosep,label=(\roman*)]
  \item \label{asn:trial-positivity} (Overlap) \(\pi(x)<1\);
  \item \label{asn:trial-exchangeability} (CATE transportability)
    \(E\{Y(1)-Y(0)\mid X=x,G=1\}=E\{Y(1)-Y(0)\mid X=x,D=d\}\) for \(x\in\mathcal{X}\).
  \end{enumerate}
\end{subassumption}

Assumption~\ref{asn:transportability}\ref{asn:trial-positivity} states that the support of the baseline covariates in the target population must be a subset of that in the joint source population.
In other words, for any value \(x\) of the baseline covariates that may appear in the target population, it must be possible to observe this value in at least one source study.
This joint overlap condition is weaker than the pairwise overlap between the target population and every source study.
Hence, studies with cohorts less diverse than the target population may still contribute to identifiability of the TATE.
Transportability under a conditional effect measure, such as assumption~\ref{asn:transportability}\ref{asn:trial-exchangeability}, only requires the transportability assumption to hold conditioning on shifted effect modifiers instead of all prognostic variables \citep{colnet2024causal}.
It should be noted that the scope of effect modifiers is strongly tied to the chosen effect measure.
Therefore, assumption~\ref{asn:transportability}\ref{asn:trial-exchangeability} is conducive to identifiability of the TATE, but can be futile for identifiability of other marginal causal effects in the target population.
It is thus highly advisable to evaluate the plausibility of such transportability assumptions according to the concrete intervention effect under consideration.

\begin{subassumption}[Internal validity]
  \label{asn:causal}
  \hfill
  \begin{enumerate}[nosep,label=(\roman*)]
  \item \label{asn:consistency}
    (Consistency) \((1-G)Y(A)=(1-G)Y\);
  \item \label{asn:treatment-positivity} (Positivity)
    \(e(a\mid x,d)>0\) for \(a=0,1\);
  \item \label{asn:treatment-exchangeability} (Mean exchangeability)
    \(E\{Y(a)\mid X=x,D=d\}=E\{Y(a)\mid A=a',X=x,D=d\}\) for \(a=0,1\) and \(a'=0,1\).
  \end{enumerate}
\end{subassumption}

Though commonplace in the causal inference literature, assumption \ref{asn:causal}\ref{asn:consistency} should be carefully examined in the source studies for meta-analysis.
For instance, the presence of multiple versions of placebo in different clinical trials may invalidate consistency.
However, if the differences are negligible with regards to their acting mechanisms on the outcome, these placebos can be subsequently treated as the same intervention.
Assumption~\ref{asn:causal}\ref{asn:treatment-positivity} requires that the probability of an individual receiving every level of the interventions must be positive.
Assumption~\ref{asn:causal}\ref{asn:treatment-exchangeability} states that the conditional means of the potential outcome must be equal between the intervention groups.
In observational studies, mean exchangeability conditional on the shifted effect modifiers may not hold due to additional confounding.
In supplementary material \S\ref{sec:confounding-app}, we consider a potentially weaker version of assumption~\ref{asn:causal}\ref{asn:treatment-exchangeability} by allowing for the existence of study-specific confounders.
Proposition~\ref{ppn:identifiability-star} therein provides a parallel identifiability result to lemma~\ref{lem:identification} below under stronger positivity than assumption~\ref{asn:causal}\ref{asn:treatment-positivity}.

By assumptions~\ref{asn:transportability} and \ref{asn:causal}, when we fix the baseline characteristics, the difference between the conditional outcome means under the two interventions remains constant among the available source studies.
Define the difference function
\[
  \delta(x)=\sum_{d\in[m]}\zeta(d\mid x)\{\mu(1,x,d)-\mu(0,x,d)\}.
\]
For any \(x\in\mathcal{X}\), we have 
\begin{equation}
  \label{eqn:transportability}
  \mu(1,x,d)-\mu(0,x,d)=\delta(x).
\end{equation}
Note that the difference function \(\delta(x)\) is a function of the baseline covariates only and does not vary across the source studies for any \(x\in\mathcal{X}\).
Define the model \(\mathcal{P}\) as the collection of probability measures over \(O\) that respect the conditional mean difference restriction \eqref{eqn:transportability} for all \(x\).
Here for simplicity, we extend the restriction from holding on \(\mathcal{X}\) to holding on the support of the baseline covariates in the joint source population.
The extension does not impact the identifiability nor the estimation of the TATE.
\begin{lemma}[Identifiability]
  \label{lem:identification}
  Suppose assumptions~\ref{asn:transportability} and \ref{asn:causal} hold.
  The TATE is identifiable in the observed data model \(\mathcal{P}\) as
  \begin{equation}
    \label{eqn:g-formula}
    \theta = E\{\delta(X)\mid G=1\}.
  \end{equation}
  For any \(h(x,d)\) such that \(E\{h(X,D)\mid X=x,G=0\}=1\), the TATE is also identifiable as
  \begin{equation}
    \label{eqn:ipw}
    \theta = E\bigg\{\frac{1-G}{\alpha}\frac{\pi(X)}{1-\pi(X)}\frac{2A-1}{e(A\mid X,D)}h(X,D)Y\bigg\}.
\end{equation}
\end{lemma}

The g-formula representation \eqref{eqn:g-formula} of the target parameter with the observed data does not explicitly depend on the membership to any source study.
While the difference function \(\delta\) implicitly involves this information, it is defined strictly on the subset of the source studies where the supports of the baseline covariates contain \(x\).
The function \(h\) in the reweighting representation \eqref{eqn:ipw} assigns study-specific weights to subjects in the joint source population.
In the special case where \(h=1\), the outcomes from all source studies are weighted the same besides the inverse propensity score of their corresponding intervention.
This interpretation is helpful in the following section.

\section{Efficient estimation}
\label{sec:estimation}

\subsection{Semiparametric efficiency bound}

For the well-defined observed data target parameter \(\theta\) viewed as a functional that maps from the model \(\mathcal{P}\), it is natural to study its semiparametric efficiency bound as well as estimators that can achieve this bound asymptotically.
In the sequel, we assume that the data follows some true distribution \(P_0\) in the model \(\mathcal{P}\), since the semiparametric efficiency bound is derived under local regularity conditions at \(P_0\).
Quantities defined on \(P_0\) receive the subscript \(0\).
We first motivate the EIF of the TATE in the model \(\mathcal{P}\) through a class of candidate oracle estimators.

Lemma~\ref{lem:identification} characterizes two representations of \(\theta_0\) with the g-formula \eqref{eqn:g-formula} and reweighting \eqref{eqn:ipw}.
Suppose we have knowledge of the true nuisance parameters.
These representations together suggest a familiar class of augmented reweighting estimators
\begin{equation}
  \label{eq:augmented}
  \frac{1}{n_1}\sum_{i:G_i=0}\frac{\pi_0(X_i)}{1-\pi_0(X_i)}h(X_i,D_i)\frac{2A_i-1}{e_0(A_i\mid X_i,D_i)}\{Y_i-\mu_0(A_i,X_i,D_i)\}+\frac{1}{n_1}\sum_{i:G_i=1}\delta_0(X_i)
\end{equation}
for any \(h\) in \eqref{eqn:ipw}.
It is trivial to verify that any estimator in this class is unbiased.
Intuitively, to minimize the variance of estimators within this class, we should downweight the outcomes from source studies with a higher variance.
Define the weighted residual within each source study as
\[
  U_0=\frac{2A-1}{e_0(A\mid X,D)}\{Y-\mu_0(A,X,D)\}
\]
and consider the weight function
\begin{equation}
  \label{eq:optimal-weight}
  w_0(x,d)=\{\mathrm{var}_0(U_0\mid X=x,D=d)\}^{-1},
\end{equation}
the inverse conditional variance of \(U_0\).
After normalization, the function
\[
  \frac{w_0(x,d)}{\sum_{d'\in[m]}w_0(x,d')\zeta_0(d'\mid x)}
\]
can take the place of \(h\).
Indeed, this weighting function is optimal in terms of variance among estimators like \eqref{eq:augmented}.
In fact, this intuition carries over to the characterization of efficient estimators, since this weight function is optimal in the sense of lemma~\ref{lem:eif} below.
First we state the regularity conditions on the distribution \(P_0\) needed to establish the local semiparametric efficiency bound of \(\theta_0\).

\begin{assumption}[Regularity conditions]
  \label{asn:bounded}
  There exists a universal constant \(C>1\) such that \(|U_0|\leq C\) and \(w_0(X,D)\leq C\).
\end{assumption}

\begin{lemma}[EIF]
  \label{lem:eif}
  Suppose assumption~\ref{asn:bounded} holds.
  The EIF of \(\theta_0\) at the distribution \(P_0\) under the model \(\mathcal{P}\) is
  \begin{equation}
    \label{eqn:eif}
    \varphi(o,w_0) = \frac{1-g}{\alpha_0}\frac{\pi_0(x)}{1-\pi_0(x)}\frac{w_0(x,d)}{\sum_{d'\in[m]}\zeta_0(d'\mid x)w_0(x,d')}u_0
    + \frac{g}{\alpha_0}\{\delta_0(x)-\theta_0\}.
  \end{equation}
\end{lemma}

To understand the lemma, first consider the familiar average treatment effect
\[
  E_0\{\mu_0(1,X,d)-\mu_0(0,X,d)\mid D=d\}
\]
in source study \(d\) defined on the unrestricted model of distributions over \((1-G)I(D=d)O\).
The semiparametric efficiency bound of this parameter is \citep{hirano2003efficient}
\[
  E_0\{w_0^{-1}(X,d)\mid D=d\} + \mathrm{var}_0\{\mu_0(1,X,d)-\mu_0(0,X,d)\mid D=d\},
\]
which consists of an expectation part and a variance part.
The semiparametric efficiency bound \(\psi_0\) of \(\theta_0\) has a similar structure.
By some algebra, we have
\begin{multline*}
  \psi_0=E_0\{\varphi^2(O,w_0)\}\\
  =\frac{1-\alpha_0}{\alpha_0^2}E_0\bigg[\bigg\{\frac{\pi_0(X)}{1-\pi_0(X)}\bigg\}^2[E_0\{w_0(X,D)\mid X,G=0\}]^{-1}\biggm\vert G=0\bigg]\\
  +\frac{1}{\alpha_0}\mathrm{var}_0\{\delta_0(X)\mid G=1\}.
\end{multline*}
Each source study contributes to \(\psi_0\) with the weight \(w_0\), which are first weighted by the selection score \(\zeta_0\) and then inverted to give the inside of the expectation part of the semiparametric efficiency bound, up to the odds of selection score \(\pi_0\) and constants.

The following result is a direct consequence of lemma~\ref{lem:eif} and its proof.

\begin{corollary}[Influence functions]
  \label{cor:tangent-space-complement}
  Suppose assumption~\ref{asn:bounded} holds.
  The linear subspace
  \[
    \Lambda_0=\big\{(1-g)h(x,d)u_0:E_0\{h(X,D)\mid X,G=0\}=0\big\}
  \]
  is the orthogonal complement of the tangent space of the model \(\mathcal{P}\) at the distribution \(P_0\).
  Consequently, for any weight function \(\tilde{w}\) such that \(\big|\sum_{d'\in[m]}\zeta_0(d'\mid x)\tilde{w}(x,d')\big|> 0\) for \(x\in\mathcal{X}\),
  \(\varphi(o,\tilde{w})\) is an influence function of \(\theta_0\).
\end{corollary}

Corollary~\ref{cor:tangent-space-complement} indicates that if we wish to construct estimators based on influence functions, the weights used for the recalibration of observations from different source studies need not be the optimal weight function \(w_0\) in lemma \ref{lem:eif}.

\subsection{Estimators and asymptotics}
\label{sec:efficient-estimation}
The EIF \eqref{eqn:eif} of the target parameter \(\theta_0\) motivates an estimating equation.
In the following, we describe an estimation procedure with cross-fitted nuisance parameters \citep{zheng2011crossvalidated,chernozhukov2018double} and analyse the asymptotic behaviour of the resulting cross-fitted estimator of \(\theta_0\).
Consider a random partition of the data into \(K\) splits with index sets \(\mathcal{I}_k\) such that \(\cup_{k\in[K]}\mathcal{I}_{k}=[n]\).
On each training split of observations whose indices do not fall into \(\mathcal{I}_k\), we estimate a set of cross-fitted nuisance parameters \(\hat\eta_k=\{\hat\alpha_k,\hat\pi_k,\hat\zeta_k,\hat{w}_k,\hat{e}_k,\hat\mu_k\}\).

The proportion of samples from the target population \(\hat{\alpha}_k=n_1/n\) is a natural estimator of \(\alpha_0\).
The selection score into the target population \(\hat{\pi}_k\) approximates the probability \(\pi_0\), and we have a selection score model \(\hat{\zeta}_k\) among the source studies for the probability \(\zeta_0\).
Within each source study, the outcome regression model \(\hat{\mu}_k\) models the conditional outcome mean \(\mu_0\), and the propensity score for treatment assignment \(\hat{e}_k\) models the probability \(e_0\).
Additionally, we can choose possibly random study-specific weights \(\hat{w}_k\) that depend on the baseline covariates.
We do not require the sample version of restriction \eqref{eqn:transportability} to hold for the estimate \(\hat\mu_k\).
That is, we allow the difference between the fitted conditional means \(\hat{\mu}_k(1,x,d) - \hat{\mu}_k(0,x,d)\) to vary across \(d\).
Rather, we estimate the difference function \(\delta_0\) by
\[
  \hat{\delta}_k(x)=\sum_{d\in[m]}\frac{\hat{w}_k(x,d)\hat\zeta_k(d\mid x)\{\hat{\mu}_k(1,x,d) - \hat{\mu}_k(0,x,d)\}}{\sum_{d'\in[m]}\hat{w}_k(x,d')\hat\zeta_k(d'\mid x)}.
\]
In light of the structure of influence functions described in corollary~\ref{cor:tangent-space-complement}, we propose the estimator
\[
  \hat\theta=\frac{1}{n}\sum_{k\in[K]}\sum_{i\in\mathcal{I}_k}\ell(O_i,\hat\eta_k),
\]
where
\begin{align*}
  \ell(o,{\hat\eta_k})&=\frac{1-g}{\hat\alpha_k}\frac{\hat\pi_k(x)}{1-\hat\pi_k(x)}\frac{\hat{w}_k(x,d)}{\sum_{d'\in[m]}\hat\zeta_k(d'\mid x)\hat{w}_k(x,d')}\hat{u}_k+\frac{g}{\hat\alpha_k}\hat\delta_k(x),\\
  \hat{u}_k&=\frac{2a-1}{\hat{e}_k(a\mid x,d)}\{y-\hat\mu_k(a,x,d)\}.
\end{align*}

The asymptotic analysis of the proposed estimator relies on the following regularity conditions on the nuisance parameters and their estimators.
For a possibly random function \(f(o)\) of the observed data, let \(\|f(O)\|_{P_0}=\{\int f^2(o)\d P_0(o)\}^{1/2}\) denote its \(L_2(P_0)\)-norm.

\begin{assumption}[Regularity conditions]
  \label{asn:regularity}
  \hfill
  \begin{enumerate}[label=(\alph*),nosep]
  \item With probability approaching \(1\), \(\{x:\hat\zeta_k(d\mid x)>0\}\subset\{x:\zeta_0(d\mid x)>0\}\) for every \(d\in[m]\);
  \item There exist \(L_2(P_0)\)-functions \(\bar\pi\), \(\bar\zeta\), \(\bar{w}\), \(\bar{e}\), and \(\bar\mu\) such that
    \begin{align*}
      &\|(\hat{\pi}_k-\bar{\pi})(X)\|_{P_0}=o_{P_0}(1),\\
      &\|(\hat{\zeta}_k-\bar{\zeta})(d\mid X)\|_{P_0}=o_{P_0}(1),\\
      &\|I(D=d)(\hat{w}_k-\bar{w})(X,d)\|_{P_0}=o_{P_0}(1),\\
      &\|I(D=d)(\hat{e}_k-\bar{e})(a\mid X,d)\|_{P_0}=o_{P_0}(1),\\
      &\|I(D=d)(\hat{\mu}_k-\bar{\mu})(a,X,d)\|_{P_0}=o_{P_0}(1);
    \end{align*}
  \item There exists a universal constant \(C>1\) such that
    \begin{align*}
      &\alpha_0\geq C^{-1}, \hat{\alpha}\geq C^{-1}, \hat{\pi}_k(x)\leq 1-C^{-1}, e_0(a\mid x,d)\geq C^{-1}, \hat{e}_k(a\mid x,d)\geq C^{-1}, \\
      &|\hat{\mu}_k(a,x,d)|\leq C, V_0(a,x,d)\leq C, |\hat{w}_k(x,d)|\leq C, |\bar{w}(x,d)|\leq C,\\
      &\big|\textstyle\sum_{d'\in[m]}\hat{\zeta}_k(d'\mid x)\hat{w}_k(x,d')\big|\geq C^{-1}, \big|\textstyle\sum_{d'\in[m]}\zeta_0(d'\mid x)\bar{w}(x,d')\big|\geq C^{-1}.
    \end{align*}
  \end{enumerate}
\end{assumption}

\stepcounter{assumption}
\begin{subassumption}[Correct specifications]
  \label{asn:model-consistency}
  \hfill
  \begin{enumerate}[label=(\roman*)]
  \item \(\bar{\mu}=\mu_0\); or
  \item \(\bar{e}=e_0\),  \(\bar{\zeta}=\zeta_0\), and \(\bar{\pi}=\pi_0\).
  \end{enumerate}
\end{subassumption}

\begin{subassumption}[Rate conditions]
  \label{asn:model-linearity}
  \hfill
  \begin{enumerate}[label=(\roman*)]
  \item\label{asn:plim} \(\bar{\mu}=\mu_0\), \(\bar{e}=e_0\), \(\bar{\zeta}=\zeta_0\), and \(\bar{\pi}=\pi_0\);
  \item \(\|I(D=d)(\hat{\mu}_k-\bar{\mu})(a,X,d)\|_{P_0}\big\{\|I(D=d)(\hat{e}_k-\bar{e})(a\mid X,d)\|_{P_0}+\|(\hat{\zeta}_k-\bar{\zeta})(d\mid X)\|_{P_0}+\|(\hat{\pi}_k-\bar{\pi})(X)\|_{P_0}\big\}=o_{P_0}(n^{-1/2})\).
  \end{enumerate}
\end{subassumption}

\begin{theorem}[Asymptotic behaviour]
  \label{thm:asymptotic}
  Suppose assumption~\ref{asn:regularity} holds.
  Then:
  \begin{enumerate}[label=(\alph*)]
  \item \(\hat{\theta}-\theta_0=o_{P_0}(1)\) under assumption~\ref{asn:model-consistency};
  \item \(\hat{\theta}-\theta_0=n^{-1}\sum_{i=1}^{n}\varphi(O_i,\bar{w})+o_{P_0}(n^{-1/2})\) under assumption~\ref{asn:model-linearity}.
    Furthermore, \(\hat\theta\) achieves the local semiparametric efficiency bound \(\psi_0\) if there exists a function \(c(x)\neq 0\) such that
    \[
      \bar{w}(x,d)=c(x)w_0(x,d)
    \]
    and assumption~\ref{asn:bounded} holds.
  \end{enumerate}
\end{theorem}

The cross-fitted estimator \(\hat{\theta}\) is doubly robust in the sense that it is consistent when either the outcome regression model \(\hat\mu_k\) is correctly specified or when the selection score model for the target population \(\hat\pi_k\), the selection score model among source studies \(\hat\zeta_k\), and the propensity score \(\hat{e}_k\) are all correctly specified.
For example, in randomized clinical trials, the propensity score \(e_0\) is usually known or estimable with parametric rates.
However, the nuisance parameters \(\pi_0\) and \(\zeta_0\) are generally more challenging to estimate because they inherit the complexity of the sampling procedure, as well as the pragmatic choice of the source studies to include in the meta-analysis.
Therefore, it can be important that the conditional mean \(\mu_0\) is correctly modelled for consistency of \(\hat\theta\).
The asymptotic linearity of \(\hat\theta\) further requires that all nuisance parameter estimators except possibly \(\hat{w}_k\) converge to the truth at a reasonably fast rate, such as \(o_{P_0}(n^{-1/4})\).
This rate is achieved by flexible curve-fitting algorithms, such as the highly adaptive lasso estimator \citep{benkeser2016highly} and the sieve neural network \citep{chen1999improved}.
Again in clinical trials, prior knowledge about the propensity score often renders the product term \(\|\hat{e}_k-e_0\|_{P_0}\|\hat{\mu}_k-\mu_0\|_{P_0}\) asymptotically negligible.

An estimator of the asymptotic variance of \(\hat\theta\) is given by the cross-fitted squared empirical \(L_2\)-norm of the influence function
\[
  \hat{\psi}=\frac{1}{n}\sum_{k\in[K]}\sum_{i\in\mathcal{I}_k}\bigg\{\ell(O_i,{\hat\eta_k})-\frac{G_i}{\hat\alpha_k}\hat\theta\bigg\}^2.
\]
We show its consistency for \(\bar\psi=E_0\{\varphi^2(O,\bar{w})\}\) in supplementary material \S\ref{sec:proof-app}.
An asymptotic \((1-2\kappa)\)-confidence interval (CI) of \(\theta_0\) is given by
\[
  \hat\theta\pm \Phi^{-1}(1-\kappa)\bigg(\frac{\hat\psi}{n}\bigg)^{1/2},
\]
where \(\Phi\) is the distribution function of a standard normal random variable.

In supplementary material \S\ref{sec:confounding-app}, we propose a class of estimators of the TATE in the presence of study-specific confounders.
Compared with the estimators proposed in this section, these estimators require an additional regression step to handle disparate covariates from the source studies.

\subsection{Efficiency and choice of weight functions}
The choice of the weight function \(\hat{w}_k\) does not affect asymptotic linearity of the resulting estimator beyond assumption~\ref{asn:regularity}.
However, theorem~\ref{thm:asymptotic} dictates that only the particular specification of the weight function \(w_0\) yields an efficient estimator, up to some scaling function.
In practice, we can pick non-study-specific weights for convenience.
However, precision can be gained from estimation of the optimal weight function.
Suppose there exist two regular asymptotically linear estimators for \(\theta_0\), one with influence function \(\varphi(o,w_0)\), and the other with influence function \(\varphi(o,\tilde{w})\) using a weight function \(\tilde{w}\) that is constant in \(d\).
The difference between their asymptotic variances, shown in supplementary material \S\ref{sec:diff-asymp-var-app}, is
\begin{multline*}
  E_0\{\varphi^{2}(O,w_0)\}-E_0\{\varphi^2(O,{\tilde{w}})\} \\
  = E_0\bigg(\frac{\pi_0^2(X)}{\alpha_0^2\{1-\pi_0(X)\}}[E_0\{w_0(X,D)\mid X,G=0\}]^{-1}-E_0\{w_0^{-1}(X,D)\mid X,G=0\}\bigg)\leq 0,
\end{multline*}
where the bound follows from the Cauchy-Schwarz inequality.

The optimal weight function can be expanded as
\[
  w_0(x,d)= \bigg\{\frac{V_0(1,x,d)}{e_0(1\mid x,d)}+\frac{V_0(0,x,d)}{e_0(0\mid x,d)}\bigg\}^{-1}.
\]
A straightforward estimator of \(w_0\) can be constructed by plugging in the nuisance parameter estimates \(\hat{e}_k\) and \(\hat{V}_k\) for \(e_0\) and \(V_0\), respectively.
However, inverting the estimated conditional variances may introduce numerical instability to the subsequent estimator \(\hat{\theta}\) when \(\hat{V}_k\) is small.
Nonparametric conditional variance estimators, such as the local kernel linear regression \citep{fan1998efficient}, may deteriorate finite sample performance of the estimator.
To circumvent the inconvenience, we follow \citet{hines2024optimally} and notice that \(w_0\) as given in \eqref{eq:optimal-weight} is the minimizer of the weighted regression loss
\[
  E_0[U_0^{2}\{U_0^{-2}-w(X,D)\}^{2}\mid G=0].
\]
Given a generic function class \(\mathcal{F}\), we propose to use a two-stage estimator that minimizes the empirical loss such that
\begin{equation}
  \label{eq:weight-erm}
  \hat{w}_{k}(x,d) = \arg\min_{f\in \mathcal{F}} \sum_{i:i\in\mathcal{I}_k,G_i=0}\{-2f(X_i,D_i)+\hat{U}_{ki}^2f^{2}(X_i,D_i)\},
\end{equation}
where
\[
  \hat{U}_{ki}=\frac{2A_i-1}{\hat{e}_k(A_i\mid X_i,D_i)}\{Y_i-\hat\mu_k(A_i,X_i,D_i)\}.
\]
Under the conditions of theorem~\ref{thm:asymptotic}, if the estimated weight function is consistent for the optimal weight function, the resulting TATE estimator will be efficient.
Notably, efficiency does not invoke any convergence rate of the weight function beyond its consistency with respect to the \(L_2(P_0)\)-norm.
Nonetheless, if either \(\hat\mu_k\) or \(\hat{e}_k\) is misspecified or if the hypothesis class \(\mathcal{F}\) does not approximate \(w_0\) arbitrarily well as the sample size grows, there is no guarantee that the asymptotic variance of the resulting TATE estimator will be lower than the estimator using constant weights. 
When both \(\hat\mu_k\) or \(\hat{e}_k\) are consistent for \(\mu_0\) and \(e_0\), the asymptotically variance will typically not be worse, as long as \(\mathcal{F}\) includes the null model.

\subsection{Connections to existing works}
\label{sec:other-transportability}

In the remainder of the section, we draw some links between this work and selected literature on data fusion.

We first present the semiparametric efficiency bounds of the TATE under observed data models induced by stronger transportability assumptions than the CATE transportability we assume here.
These assumptions typically induce models smaller than \(\mathcal{P}\).
Let the model \(\mathcal{P}^{\dagger}\) contain all probability measures over \(O\) satisfying the conditional mean restriction
\begin{equation}
  \label{eqn:transportability-dagger}
  \mu(a,x,d)=\mu^\dagger(a,x),
\end{equation}
for all \(x\) and \(a=0,1\), where \(\mu^\dagger(a,x)=E(Y\mid A=a,X=x,G=0)\) is the conditional outcome mean in the joint source population.
Restriction \eqref{eqn:transportability-dagger} on \(\mathcal{X}\) is implied by assumptions~\ref{asn:transportability}\ref{asn:trial-positivity} and \ref{asn:causal}, as well as transportability of the conditional treatment-specific means,
\[
  E\{Y(a)\mid X=x,G=1\}=E\{Y(a)\mid X=x,D=d\}.
\]
We state the regularity conditions on \(P_0\) needed to establish the semiparametric efficiency bound of \(\theta_0\) under the model \(\mathcal{P}^\dagger\).
For comparability of notations, we denote the relevant nuisance parameters by \(e_0^\dagger(a\mid x)=P_0(A=a\mid X=x,G=0)\), \(\zeta_0^\dagger(d\mid a,x)=P_0(D=d\mid A=a,X=x)\), \(w^{\dagger}_0(a,x,d)=V_0^{-1}(a,x,d)\), and \(\delta^\dagger_0(x)=\mu^{\dagger}_0(1,x)-\mu^{\dagger}_0(0,x)\) and the residual by \(U^\dagger_0=Y-\mu^\dagger_0(A,X)\).
\begin{assumption}[Regularity conditions]
  \label{asn:bounded-li}
  There exists a universal constant \(C>1\) such that \(|U^\dagger_0|\leq C\) and \(w_0^\dagger(A,X,D)\leq C\).
\end{assumption}

\begin{proposition}[EIF]
  \label{ppn:eif-li}
  Suppose assumption~\ref{asn:bounded-li} holds.
  The EIF of \(\theta_0\) at the distribution \(P_0\) under the model \(\mathcal{P}^{\dagger}\) is
  \begin{multline*}
    \varphi^{\dagger}(o,w_{0}^\dagger) = \frac{1-g}{\alpha_0}\frac{\pi_0(x)}{1-\pi_0(x)}\frac{2a-1}{e^\dagger_0(a\mid x)}\frac{w^{\dagger}_0(a,x,d)}{\sum_{d'\in[m]}w^{\dagger}_0(a,x,d')\zeta^\dagger_0(d'\mid a,x)}u^\dagger_0  \\
    + \frac{g}{\alpha_0}\{\delta^\dagger_0(x)-\theta_0\}.
  \end{multline*}
  The orthocomplement of the tangent space of \(\mathcal{P}^\dagger\) at the distribution \(P_0\) is
  \[
    \Lambda^\dagger_0=\big\{(1-g)h(a,x,d)u^\dagger_0:E_0\{h(A,X,D)\mid A,X,G=0\}=0\big\}.
  \]
  For any \(\tilde{w}^\dagger\) such that \(|\sum_{d'\in[m]}\tilde{w}^\dagger(a,x,d')\zeta_0(d'\mid a,x)|> 0\),
  \(\varphi^\dagger(o,\tilde{w}^\dagger)\) is an influence function of \(\theta_0\).
\end{proposition}
The optimal weights \(w^{\dagger}_0\) can differ for individuals with the same level of baseline covariates receiving different interventions in the same source study, in contrast to the optimal weights \(w_0\) seen previously in lemma~\ref{lem:eif}, which are only study-specific.
Proposition~\ref{ppn:eif-li} is a generalization of the efficiency result from \citet{li2023improving} to accommodate multiple source studies.

Now consider the model \(\mathcal{P}^{\ddagger}\) consisting of probability measures over \(O\) satisfying the conditional independence
\[
  Y\indep D\mid \{A,X,G=0\}.
\]
This corresponds to the distribution-level transportability
\[
  Y(a)\indep \{G,(1-G)D\}\mid X
\]
in \citet{wang2024efficient}.
They show that the EIF of \(\theta_0\) at the distribution \(P_0\) under the model \(\mathcal{P}^{\ddagger}\) is
\[
  \varphi^{\ddagger}(o) = \frac{1-g}{\alpha_0}\frac{\pi_0(x)}{1-\pi_0(x)}\frac{2a-1}{e^\dagger_0(a\mid x)}u^\dagger_0 + \frac{g}{\alpha_0}\{\delta^\dagger_0(x)-\theta_0\}.
\]
For completeness, we provide the orthocomplement of the tangent space of the model \(\mathcal{P}^\ddagger\) at \(P_0\), which is
\begin{multline*}
  \Lambda^{\ddagger}_0 = \big\{(1-g)h(y,a,x,d):E_0\{h(Y,A,X,D)\mid A,X,D\}=0,\\
  E_0\{h(Y,A,X,D)\mid Y,A,X,G=0\}=0\big\}.
\end{multline*}

The three models discussed so far are nested such that \(\mathcal{P}^{\ddagger}\subset \mathcal{P}^{\dagger}\subset \mathcal{P}\).
Modulo local regularity conditions, \(\varphi(o,w_0)\) from lemma~\ref{lem:eif} and \(\varphi(o,\tilde{w})\) from corollary~\ref{cor:tangent-space-complement} are valid influence functions of \(\theta_0\) if the true distribution \(P_0\) belongs to the smaller models \(\mathcal{P}^\dagger\) or \(\mathcal{P}^\ddagger\).
However, if in fact \(P_0\in\mathcal{P}\setminus\mathcal{P}^{\dagger}\), estimators constructed from estimating equations based on \(\varphi^\dagger(o,w_0^\dagger)\) \(\varphi^\dagger(o,\tilde{w}^\dagger)\), or \(\varphi^\ddagger(o)\) will be inconsistent in general.
The true difference function \(\delta_0\) in this case is not equal to the difference function \(\delta^\dagger_0\), and the influence functions under the smaller models do not have zero mean.

Most akin to our work is \citet{li2024efficient}, which studies the semiparametric efficiency bound of the TATE combining observational and experimental data under models with conditional mean restrictions.
In that work, the difference between conditional outcome means is constant in the two data sources through a possibly non-identity link function.
The intersection of the current setting and the one in that work sits where the number of source studies is two and the restriction is in the form of \eqref{eqn:transportability}.
In supplementary material \S\ref{sec:li-app}, we verify that the orthocomplement of the tangent space stated in that work is algebraically equivalent to \(\Lambda_0\) in corollary~\ref{cor:tangent-space-complement}.
In contrast to the proposal in that work, we do not enforce restriction \eqref{eqn:transportability} in estimation of the TATE and thereby avoid choosing a single source study to obtain an estimate of \(\delta_0\).
Furthermore, for construction of an efficient estimator, we suggest direct estimation of the optimal weight function rather than treating the conditional outcome variance as a nuisance parameter.

\section{Effect heterogeneity}

We have assumed CATE transportability to make inference on the TATE.
In meta-analyses, the CATE function has been referred to as the effect-size surface and may be of interest for various good reasons \citep{rubin1992metaanalysis}.
With the combined sample size of multi-source data, we may have enough power to ascertain a larger body of evidence than a single-parameter summary of the intervention effect in the target population such as the TATE.
For example, practitioners and policymakers might seek to identify specific subgroups within the target population who could either benefit from or be harmed by the intervention, while the outcome and intervention information is only available from existing clinical trials.

Parametric formulations of causal effect heterogeneity are useful in many clinical settings thanks to their interpretability.
In some applications, it may be reasonable to describe effect heterogeneity in a semiparametric model where \(\delta(x)\) is known up to a Euclidean parameter.
In supplementary material \S\ref{sec:parametric-app}, we derive the efficient score of this parameter.
In this section, we describe an alternative finite-dimensional parameterization of effect heterogeneity related to the CATE and present an efficient estimation strategy.

For a subset of the baseline covariates \(Z\subset X\), consider the basis function
\[
  b(z)=\{b_1(z),\dots,b_q(z)\}^\T
\]
with a fixed dimension \(q\geq 1\).
A natural description of effect heterogeneity is the projected CATE \citep{semenova2021debiased,cui2023estimating}, which is the best approximation of the CATE \(\delta_0\) in the linear span of the basis \(b\).
The function \(\gamma_0^\T b(z)\), where
\[
  \gamma_0 \in \arg\min_{\gamma\in\mathbb{R}^q}E_0[\{\delta_0(X)-\gamma^\T b(Z)\}^2\mid G=1],
\]
is a low-dimensional characterization of the CATE in the target population in the nonparametric model without structural assumption on \(\delta_0\).
If \(Z\) comprises only categorical variables and the basis \(b\) dummy variables, the coefficients in \(\gamma_0\) reduce to the subgroup-specific TATEs studied in \citet{wang2024efficient}.
Let \(\|v\|\) denote the Euclidean norm of \(v\in\mathbb{R}^q\).
Under the following assumption, \(\gamma_0\) is uniquely identifiable.

\begin{assumption}[Uniqueness]
  \label{asn:invertible}
  \(\|b\|\in L_2(P_0)\) and \(E_0\{b(Z)b^\T(Z)\mid G=1\}\) is invertible.
\end{assumption}

\begin{proposition}[EIF]
  \label{ppn:eif-pcate}
  Suppose assumptions~\ref{asn:bounded} and \ref{asn:invertible} hold.
  The EIF of \(\gamma_0\) at the distribution \(P_0\) under the model \(\mathcal{P}\) is
  \begin{multline*}
    \phi(o,w_0)=\big[E_0\{b(Z)b^\T(Z)\mid G=1\}\big]^{-1}b(z)\\
    \bigg[\frac{1-g}{\alpha_0}\frac{\pi_0(x)}{1-\pi_0(x)}\frac{w_0(x,d)}{\sum_{d'\in[m]}\zeta_0(d'\mid x)w_0(x,d')}u_0+\frac{g}{\alpha_0}\{\delta_0(x)-\gamma_0^\T b(z)\}\bigg].
  \end{multline*}
\end{proposition}

The EIF \(\phi(o,w_0)\) motivates the cross-fitted least-squares estimator
\[
  \hat\gamma = \bigg\{\frac{1}{n_1}\sum_{i:G_i=1}b(Z_i)b^\T(Z_i)\bigg\}^{-1}\bigg\{\frac{1}{n}\sum_{k\in[K]}\sum_{i\in\mathcal{I}_k}b(Z_i)\ell(O_i,{\hat\eta_k})\bigg\}.
\]
Denote the support of \(Z\) in the target population by \(\mathcal{Z}\).
We have the following result for \(\hat\gamma\).

\begin{theorem}[Asymptotic behaviour]
  \label{thm:asymptotic-tpcate}
  Suppose assumptions~\ref{asn:regularity} and \ref{asn:invertible} hold and that\linebreak \(\sup_{z\in\mathcal{Z}}\|b(z)\|\leq C\) for a universal constant \(C>0\).
  Then:
  \begin{enumerate}[label=(\roman*)]
  \item \(\hat{\gamma}-\gamma_0=o_{P_0}(1)\) under assumption~\ref{asn:model-consistency};
  \item \(\hat{\gamma}-\gamma_0=n^{-1}\sum_{i=1}^{n}\phi(O_i,\bar{w})+o_{P_0}(n^{-1/2})\) under assumption~\ref{asn:model-linearity}, where \(\bar{w}\) is the probability limit of \(\hat{w}_k\).
    Furthermore, \(\hat\gamma\) achieves the local semiparametric efficiency bound \(\Psi_0=E_0\{\phi(O,w_0)\phi^\T(O,w_0)\}\) if there exists a function \(c(x)\neq 0\) such that
    \[
      \bar{w}(x,d)=c(x)w_0(x,d)
    \]
    and assumption~\ref{asn:bounded} holds.
  \end{enumerate}
\end{theorem}
A pointwise asymptotic \((1-2\kappa)\)-CI of \(\gamma_0^\T b(z)\) for any \(z\in\mathcal{Z}\) can be constructed as
\[
  \hat\gamma^\T b(z)\pm \Phi^{-1}(1-\kappa)\bigg\{\frac{b^{\T}(z)\hat{\Psi}b(z)}{n}\bigg\}^{1/2},
\]
where we use the sandwich estimator
\begin{multline*}
  \hat{\Psi} = \bigg\{\frac{1}{n_1}\sum_{i:G_i=1}b(Z_i)b^\T(Z_i)\bigg\}^{-1}\bigg[\frac{1}{n}\sum_{k\in[K]}\sum_{i\in\mathcal{I}_k}b(Z_i)b^\T(Z_i)\bigg\{\ell(O_i,{\hat\eta_k})-\frac{G_i}{\hat\alpha}\hat\gamma^\T b(Z_i)\bigg\}^2\bigg]\\
  \bigg\{\frac{1}{n_1}\sum_{i:G_i=1}b(Z_i)b^\T(Z_i)\bigg\}^{-1}
\end{multline*}
of the asymptotic variance \(\bar\Psi=E_0\{\phi(O,\bar{w})\phi^\T(O,\bar{w})\}\) of \(\hat\gamma\).
We show its consistency in supplementary material \S\ref{sec:proof-app}.

Under stronger regularity assumptions, we establish uniform inference of the projected CATE over \(\mathcal{Z}\).
Let \(\lambda_{\mathrm{min}}(M)\) and \(\lambda_{\mathrm{max}}(M)\) denote the minimum and maximum eigenvalues of a matrix \(M\).
\begin{assumption}[Bounded basis]
  \label{asn:bounded-psi}
  There exists a universal constant \(C \geq 1\) such that \(C^{-1}\leq \inf_{z\in\mathcal{Z}}\|b(z)\|\leq \sup_{z\in\mathcal{Z}}\|b(z)\|\leq C\) and \(C^{-1}\leq \lambda_{\mathrm{min}}(\bar\Psi)\leq \lambda_{\mathrm{max}}(\bar\Psi)\leq C\).
\end{assumption}

\begin{corollary}[Weak convergence]
  \label{cor:asymptotic-tpcate-uniform}
  Suppose assumptions~\ref{asn:regularity}, \ref{asn:model-linearity}, \ref{asn:invertible}, and \ref{asn:bounded-psi} hold.
  Then
  \[
    \frac{n^{1/2}b^\T(z)(\hat\gamma-\gamma_0)}{\{b^{\T}(z)\hat{\Psi}b(z)\}^{1/2}}\leadsto \mathbb{T}(z)\qquad \text{in } \ell^\infty(\mathcal{Z}),
  \]
  where \(\ell^{\infty}(\mathcal{Z})\) is the space of bounded functions over \(\mathcal{Z}\), and \(\mathbb{T}(z)\) is a mean-zero Gaussian process over \(\mathcal{Z}\) with covariance function
  \[
    \mathrm{cov}_0\{\mathbb{T}(z),\mathbb{T}(z')\}=\frac{b^\T(z)\bar\Psi b(z')}{\{b^{\T}(z)\bar\Psi b(z)\}^{1/2}\{b^{\T}(z')\bar\Psi b(z')\}^{1/2}}.
  \]
\end{corollary}
A uniform asymptotic \((1-\kappa)\)-confidence interval of \(\gamma_0^\T b(z)\) is given by
\[
  \hat\gamma^\T b(z)\pm n^{-1/2}c_{T}(1-\kappa)\{b^{\T}(z)\hat{\Psi}b(z)\}^{1/2},
\]
where \(c_T\) is the quantile function of \(T=\sup_{z\in \mathcal{Z}}|\mathbb{T}(z)|\), the supremum of the Gaussian process over \(\mathcal{Z}\).
The distribution of \(T\) depends on unknown nuisance parameters.
In practice, it can be approximated via Monte-Carlo methods, such as \(t\)-bootstrap \citep{belloni2015new} and multiplier bootstrap \citep{belloni2018uniformly}.

\section{Simulation}
\label{sec:simulation}

We consider \(m=3\) source studies in the simulation study.
We generate the observed data \(O=\{(1-G)Y,(1-G)A,X,(1-G)D,G\}\) as follows.
The baseline covariates \(X=(X_{1},X_{2},X_{3})^{\T}\) include three correlated and uniformly distributed measurements between \(-1\) and \(1\).
The nuisance parameters \(\pi_0\) and \(\eta_0\) correspond to a binary and a multinomial logistic model, respectively.
Exact specifications are listed in supplementary material \S\ref{sec:details}.
The binary treatment is randomized within each source study with \(e_{0}(1\mid x,d)= I(d=1)0.5+I(d=2)0.4+I(d=3)0.6\).
Finally, the outcome \(Y\) is drawn from the normal distribution
\[
  Y\mid \{A,X,D\} \sim \mathrm{Normal}\{A \delta_0(X) +\mu_0(0,X,D), V_0(A,X,D)\}.
\]
where
\begin{align*}
  \delta_0(x)&=1+0.5x_{1}-0.2x_{2}+0.4x_{3}+\exp(0.3x_1)+\sin(0.25x_2)+\cos(0.5x_3), \\
  \mu_0(0,x,d)&=0.25d+0.7x_1-0.1x_2-0.3x_3+(d-2)(-0.2x_1+0.2x_2-0.1x_3).
\end{align*}
The conditional variance \(V_0(a,x,d)\) does not depend on the treatment and is to be specified later.

For simplicity, we assume the propensity score \(e_0\) is known, which has no impact on the local semiparametric efficiency bound of the parameter \(\theta_0\).
We considered four estimators with different weight functions that share the form
\[
  \hat\theta=\frac{1}{n}\sum_{k=1}^{5}\sum_{i\in\mathcal{I}_k}\ell(O_i,\hat{\eta}_{k}),
\]
where the set of nuisance parameter estimates \(\hat{\eta}_{k}\) includes different specifications for the weight function.
We refer to \(\hat{w}_{k}(x,d)=V_0^{-1}(a,x,d)e_0(1\mid x,d)e_0(0\mid x,d)\) as the oracle weight function and use it as the benchmark.
We call \(\hat{w}_{k}(x,d)=e_0(1\mid x,d)e_0(0\mid x,d)\) the overlap weight function, assuming homoscedasticity of the outcome across source studies.
The constant weight function \(\hat{w}_{k}(x,d)=1\) further ignores the difference between propensity scores.
Finally, the learned optimal weight function corresponds to the strategy where it is estimated through empirical risk minimization \eqref{eq:weight-erm}.

To evaluate the performance of the estimators under different weight functions, we simulated data under three specifications of conditional variances of the outcome listed in table~\ref{tab:variance-sim}, corresponding to homoscedastic error within each study (setting I), study-specific covariate-dependent variance (setting II), and completely homoscedastic error (setting III), respectively.

\begin{table}
  \caption{Specifications of the conditional variance in three simulation settings.}
  \label{tab:variance-sim}
  \centering
  \footnotesize
  \begin{tabular}{lccc}
     & \multicolumn{3}{c}{Setting} \\
    Variance & I & II & III \\
    \(V_0(a,x,1)\) & \(2\) & \(4\) & \(1\) \\
    \(V_0(a,x,2)\) & \(1\) & \(2\{1+0.1(x^\T1+3)\}^{-1}\) & \(1\) \\
    \(V_0(a,x,3)\) & \(2\) & \(4\{1+0.1(x^\T1+3)\}^{-1}\) & \(1\)
  \end{tabular}
\end{table}

To investigate robustness against model misspecifications, we computed the estimators under four scenarios: all nuisance models are correctly specified (experiment 1), only \(\hat\mu_k\) is misspecified (experiment 2), only \(\hat\zeta_k\) and \(\hat\pi_k\) are misspecified (experiment 3), and \(\hat\mu_k\), \(\hat\zeta_k\), and \(\hat\pi_k\) are misspecified (experiment 4).
All nuisance models except the propensity score were fitted with the super learner ensemble learning algorithm \citep{vanderlaan2007super}, using random forest, a generalized additive model, lasso, and the null model as base learners.
Model misspecifications were performed by replacing the original \(X\) with element-wise absolute values \(|X|\).
We simulated datasets of sizes \(n=1250,2500\).
For each sample size, we generated \(1000\) datasets.
All estimators were obtained with \(5\)-fold cross-fitting.
The standard errors (SEs) were computed as \((\hat\psi/n)^{1/2}\) by plugging in the nuisance parameter estimates including the weight function.

Summary statistics calculated from the simulation are displayed in table~\ref{tab:sim-1} and tables~\ref{tab:sim-2} and \ref{tab:sim-3} of the supplementary material.
Under setting I, the estimators have no Monte-Carlo bias except in experiment 4 when all nuisance parameter models are misspecified.
Under experiment 1 of no model misspecification, the plug-in SE estimates lead to the desired \(95\%\)-CI coverage.
Results from settings II and III show similar conclusions.
As expected in experiment 1, the estimators using oracle and optimal weights have lower Monte-Carlo mean squared errors compared to the estimators using overlap and constant weights in settings I and II.
No meaningful improvement of the standard error was observed in setting III by using the oracle or the learned optimal weight.
Similar evidence was found in table~\ref{tab:prop-smaller-se}, where we calculated the proportion of plug-in SE estimates of the estimators using the oracle and optimal weights being smaller than those of estimators using the overlap and the constant weights.
We highlight the importance of cross-fitting for nominal coverage of the confidence interval.
In tables~\ref{tab:sim-1-nocf}--\ref{tab:sim-3-nocf} of the supplementary material, we show that in all settings, the plug-in SE estimator in experiment 1 underestimated the uncertainty of the TATE estimator without cross-fitting.

\begin{table}
  \caption{Summary of simulation results for \(\hat\theta\) in setting I with cross-fitting.}
  \label{tab:sim-1}
  \footnotesize
  \centering
  
\begin{tabular}{rllrrrrr}
{\(n\)} & {Experiment} & {Weight} & {Mean} & {Bias} & {RMSE} & {SE} & {Coverage}\\
\(1250\) & 1 & Constant & \(3.12\) & \(7.30\) & \(10.06\) & \(9.85\) & \(94.9\)\\
 &  & Overlap & \(3.12\) & \(7.26\) & \(10.08\) & \(9.87\) & \(94.9\)\\
 &  & Learned & \(3.11\) & \(6.54\) & \(10.04\) & \(9.60\) & \(94.6\)\\
 &  & Oracle & \(3.11\) & \(6.68\) & \(9.81\) & \(9.42\) & \(94.8\)\\
 & 2 & Constant & \(3.09\) & \(-20.89\) & \(12.67\) & \(10.51\) & \(93.6\)\\
 &  & Overlap & \(3.09\) & \(-21.12\) & \(12.79\) & \(10.54\) & \(93.7\)\\
 &  & Learned & \(3.09\) & \(-21.35\) & \(12.80\) & \(10.51\) & \(93.3\)\\
 &  & Oracle & \(3.09\) & \(-19.68\) & \(11.82\) & \(10.25\) & \(93.5\)\\
 & 3 & Constant & \(3.11\) & \(1.80\) & \(10.66\) & \(9.62\) & \(92.8\)\\
 &  & Overlap & \(3.11\) & \(1.64\) & \(10.72\) & \(9.65\) & \(93.2\)\\
 &  & Learned & \(3.11\) & \(1.59\) & \(10.57\) & \(9.37\) & \(93.6\)\\
 &  & Oracle & \(3.11\) & \(1.83\) & \(10.12\) & \(9.19\) & \(93.6\)\\
 & 4 & Constant & \(2.94\) & \(-170.43\) & \(21.75\) & \(10.12\) & \(57.0\)\\
 &  & Overlap & \(2.93\) & \(-172.82\) & \(22.06\) & \(10.16\) & \(56.1\)\\
 &  & Learned & \(2.95\) & \(-161.72\) & \(21.18\) & \(10.19\) & \(61.4\)\\
 &  & Oracle & \(2.96\) & \(-151.97\) & \(19.45\) & \(9.87\) & \(62.7\)\\
\(2500\) & 1 & Constant & \(3.12\) & \(8.21\) & \(6.83\) & \(6.80\) & \(94.9\)\\
 &  & Overlap & \(3.12\) & \(8.18\) & \(6.84\) & \(6.81\) & \(95.0\)\\
 &  & Learned & \(3.11\) & \(7.14\) & \(6.63\) & \(6.54\) & \(95.4\)\\
 &  & Oracle & \(3.12\) & \(7.30\) & \(6.60\) & \(6.49\) & \(95.3\)\\
 & 2 & Constant & \(3.10\) & \(-5.96\) & \(7.60\) & \(7.21\) & \(93.3\)\\
 &  & Overlap & \(3.10\) & \(-6.18\) & \(7.61\) & \(7.23\) & \(93.3\)\\
 &  & Learned & \(3.10\) & \(-5.70\) & \(7.55\) & \(7.10\) & \(92.8\)\\
 &  & Oracle & \(3.10\) & \(-5.62\) & \(7.53\) & \(7.07\) & \(93.4\)\\
 & 3 & Constant & \(3.11\) & \(3.44\) & \(6.97\) & \(6.58\) & \(93.7\)\\
 &  & Overlap & \(3.11\) & \(3.31\) & \(7.00\) & \(6.60\) & \(93.7\)\\
 &  & Learned & \(3.11\) & \(2.95\) & \(6.66\) & \(6.34\) & \(94.8\)\\
 &  & Oracle & \(3.11\) & \(3.05\) & \(6.62\) & \(6.29\) & \(94.9\)\\
 & 4 & Constant & \(2.94\) & \(-172.32\) & \(18.61\) & \(6.83\) & \(29.9\)\\
 &  & Overlap & \(2.93\) & \(-174.87\) & \(18.85\) & \(6.84\) & \(29.1\)\\
 &  & Learned & \(2.95\) & \(-158.23\) & \(17.32\) & \(6.82\) & \(38.2\)\\
 &  & Oracle & \(2.95\) & \(-153.11\) & \(16.81\) & \(6.73\) & \(38.7\)\\
\end{tabular}

  \medskip
  {Mean: average of estimates; Bias: Monte-Carlo bias, \(10^{-3}\); RMSE: root mean squared error, \(10^{-2}\); \\SE: average of standard error estimates, \(10^{-2}\); Coverage: \(95\%\) confidence interval coverage, \(\%\);
  Experiment 1: all nuisance models correctly specified; Experiment 2: \(\hat\mu_k\) misspecified; Experiment 3: \(\hat\zeta_k\) and \(\hat\pi_k\) are misspecified; Experiment 4: \(\hat\mu_k\), \(\hat\zeta_k\), and \(\hat\pi_k\) misspecified.}
\end{table}

For comparison, we also computed the \(5\)-fold cross-fitted transported TATE estimators that are asymptotically efficient under alternative probability models  \(\mathcal{P}^\dagger\) and \(\mathcal{P}^\ddagger\) discussed in \S\ref{sec:other-transportability}.
Summary statistics for these estimators can be found in table~\ref{tab:sim-alt} of the supplementary material.
The data generating mechanism here is compatible with the testable implication \eqref{eqn:transportability} of CATE transportability in the observed data distribution.
However, the stronger transportability assumptions discussed in \S\ref{sec:other-transportability} are violated, so that these estimators suffer from severe bias in this setting.

For the target population projected CATE, we chose \(Z=\{X_1\}\) and the cubic polynomial basis \(b(x_1)=(1,x_1,x_1^2,x_1^3)^\T\).
We computed the cross-fitted least-squares estimates \(\hat\gamma=(\hat\gamma_{1},\hat\gamma_{2},\hat\gamma_{3},\hat\gamma_{4})^\T\) with different weight functions.
Summary statistics of individual entries \(\hat\gamma_{j}\) can be found in tables~\ref{tab:sim-beta-1}--\ref{tab:sim-beta-4} of the supplementary material, demonstrating the expected behaviour of these estimators.
We used \(t\)-bootstrap to approximate the \(95\%\)-quantile of the Gaussian process supremum.
In table~\ref{tab:sim-pcate-uniform-coverage} of the supplementary material, we show that the uniform \(95\%\)-CI for the projected CATE \(\gamma^\T b(x_1)\) over \(x_1\in[-1, 1]\) achieves the nominal coverage under experiment 1 with no model misspecification across all settings.

\section{Data example}

STEP-1 (ClinicalTrials.gov ID NCT03548935, \citealp{wilding2021onceweekly}) is a multicentre randomized controlled trial comparing the weight loss effect of once-weekly semaglutide, \(2.4\) mg to placebo in addition to a lifestyle intervention.
The main inclusion criteria of STEP-1 are body-mass index (BMI) of at least \(30\) or at least \(27\) with weight-related co-morbidities, and no diabetes.
As an illustration of our method, we regroup the trial data into four regions and take the study population recruited in the United States (US) as the target population.
The three source populations are defined by the subjects recruited from the European Union area (EU, containing Belgium, Denmark, Finland, France, Germany, and Poland), the United Kingdom (UK), and East Asia (EA, containing Japan and Taiwan).
We choose the target population based on the assumption that the demographic composition of the United States is the most varied in these four regions.

The outcome \(Y\) is the percentage change in body weight from baseline (week \(0\)) to week \(68\).
The target parameter is the TATE \(\theta=E\{Y(1)-Y(0)\mid G=1\}\) in the study population of STEP-1 recruited in the US (\(G=1\)) of semaglutide (\(A=1\)) versus placebo (\(A=0\)).
The baseline covariates \(X\) used are body weight, age, sex, BMI, waist circumference, smoking status, diabetic history, and haemoglobin A1c.
The analysis assumes cross-regional CATE transportability conditioning on \(X\).
Note that the interpretation of the treatment effect defined on the original scale of body weight and that defined on the percentage change differ.
However, the transportability assumption is the same on both scales, since the baseline body weight is included in \(X\).

We perform a complete-case analysis where subjects with missing body weight measurements at week \(68\) are removed.
The percentage of missing outcomes and the number of complete cases per region are reported in table~\ref{tab:missing}.
All nuisance parameters except the propensity score are fitted with super learning, using gradient-boosted trees, random forest, a generalized additive model, lasso, and the null model as base learners.
The propensity score within each source region is computed as the proportion of subjects receiving the active treatment.
In particular, the outcome model is fitted on each combination of treatment and source region, such that the conditional means of the body weight change under the same treatment are allowed to vary freely across regions.
We computed three transported estimators of the TATE using the constant weights, the overlap weights, and the learned optimal weights.
All estimators were computed with \(10\)-fold cross-fitting, and plug-in SE estimates were used for the construction of \(95\%\)-CIs.

\begin{table}
  \caption{Percentage of missing outcomes and complete cases by region.}
  \label{tab:missing}
  \centering
  \footnotesize
  \begin{tabular}{lrrrr}
    & US & EU & UK & EA \\
    Missing (\%) & \(8.3\) & \(4.7\) & \(16.5\) & \(1.5\) \\
    Complete (\(N\)) & \(700\) & \(367\) & \(182\) & \(133\)
  \end{tabular}
\end{table}

The results are displayed in table~\ref{tab:tate}.
Since outcome and treatment information is directly available in the target population, we report the estimate given by the augmented inverse probability weighting (AIPW) estimator based only on the target population in table~\ref{tab:ate} for reference.
In addition, the estimates within the source regions based on AIPW estimators are also displayed.
The TATE estimator with learned optimal weights gave a weight reduction of \(12.57\) percentage points from baseline.
The estimates using the overlap weights and the constant weights both indicated a reduction of \(12.75\) percentage points after rounding.
These two estimators are practically identical, since the overlap weights are nearly constant in a well-implemented multi-site clinical trial.
The plug-in SE of the estimator using the learned optimal weights is slightly higher than that of the estimators using the overlap weights and the constant weights.
However, this may be observed even in situations where variance reduction should be expected; see table~\ref{tab:prop-smaller-se} of the supplementary material.
The point estimates given by the transported TATE estimators roughly agree and are about \(0.5\) percentage points higher than the estimate given by the non-transported TATE estimator in the US study population.
Besides possible failure to account for all shifted effect modifiers, the discrepancy between the transported and non-transported TATE estimates may result from the deletion of missing observations, since patients' dropout rates clearly differ among the regions.

\begin{table}
  \caption{Results for the transported TATE with \(10\)-fold cross-fitting.}
  \label{tab:tate}
  \centering
  \footnotesize
  
\begin{tabular}{lrrr}
{Weight} & {Estimate} & {SE} & {\(95\%\)-CI}\\
Overlap & \(-12.75\) & \(0.73\) & \((-14.18, -11.32)\)\\
Constant & \(-12.75\) & \(0.73\) & \((-14.18, -11.31)\)\\
Learned & \(-12.57\) & \(0.75\) & \((-14.03, -11.10)\)\\
\end{tabular}

\end{table}

\begin{table}
  \caption{Results for the region-specific average treatment effects based on AIPW with \(10\)-fold cross-fitting.}
  \label{tab:ate}
  \centering
  \footnotesize
  
\begin{tabular}{lrrr}
{Region} & {Estimate} & {SE} & {\(95\%\)-CI}\\
EA & \(-11.17\) & \(1.37\) & \((-13.86, -8.47)\)\\
EU & \(-12.43\) & \(0.93\) & \((-14.25, -10.62)\)\\
UK & \(-14.32\) & \(1.25\) & \((-16.77, -11.88)\)\\
US & \(-13.21\) & \(0.66\) & \((-14.50, -11.92)\)\\
\end{tabular}

\end{table}

The effect estimates given by the cross-fitted estimators depend on the number of splits.
In the sensitivity analysis, we computed the same estimators for the TATE with \(5\)-fold cross-fitting, \(2\)-fold cross-fitting, and no cross-fitting.
The results are displayed in table~\ref{tab:tate-extra} of the supplementary material.
The point estimates are mostly similar to the preceding results.
The plug-in SEs of the transported TATE estimators remain comparable under \(5\)-fold cross-fitting, but become higher under \(2\)-fold cross-fitting.
Notably, the plug-in SEs of the transported TATE estimators without cross-fitting appear much lower than those obtained with cross-fitting.
However, the SE of the non-transported TATE estimator barely changes.
See supplementary material \S\ref{sec:details} for further discussions.
We recommend the results obtained by cross-fitting with at least \(5\) folds.

To investigate treatment effect heterogeneity in the US study population, we calculated three target population projected CATEs onto the basis formed by baseline body weight, age, and BMI, respectively.
In all cases, a polynomial basis of order \(3\) was applied.
The nuisance parameters were estimated identically as in the estimation of the transported TATE, and only the learned optimal weights were applied.
The results are plotted in figure~\ref{fig:tpcate}.
The percentage reduction in body weight tends to be smaller with a higher body weight at baseline, while there is no clear trend with respect to BMI.
The treatment effect is less pronounced for elder individuals up to around \(60\) years old.
We also observe a plateau of weight loss effect for those above \(60\) in age, but the estimated projected CATE shows much statistical uncertainty.
The discrepancy in sex is most striking, with females on average losing nearly \(6.5\) percentage points more body weight from semaglutide than males.

\begin{figure}
  \centering
  {
    \scalebox{.8}{\input{./art/tpcate-weight.tex}}
    \scalebox{.8}{\input{./art/tpcate-age.tex}}
    \scalebox{.8}{
\begin{tikzpicture}[x=1pt,y=1pt]
\definecolor{fillColor}{RGB}{255,255,255}
\path[use as bounding box,fill=fillColor,fill opacity=0.00] (0,0) rectangle (216.81,252.94);
\begin{scope}
\path[clip] ( 49.20, 61.20) rectangle (191.61,203.75);
\definecolor{fillColor}{RGB}{0,0,0}

\path[fill=fillColor] ( 98.43,125.06) circle (  2.25);

\path[fill=fillColor] (142.38,159.18) circle (  2.25);
\end{scope}
\begin{scope}
\path[clip] (  0.00,  0.00) rectangle (216.81,252.94);
\definecolor{drawColor}{RGB}{0,0,0}

\path[draw=drawColor,line width= 0.4pt,line join=round,line cap=round] ( 49.20, 61.20) --
	(191.61, 61.20) --
	(191.61,203.75) --
	( 49.20,203.75) --
	cycle;
\end{scope}
\begin{scope}
\path[clip] (  0.00,  0.00) rectangle (216.81,252.94);
\definecolor{drawColor}{RGB}{0,0,0}

\node[text=drawColor,anchor=base,inner sep=0pt, outer sep=0pt, scale=  1.00] at (120.41, 15.60) {Sex};

\node[text=drawColor,rotate= 90.00,anchor=base,inner sep=0pt, outer sep=0pt, scale=  1.00] at ( 10.80,132.47) {Weight change (p.p.)};
\end{scope}
\begin{scope}
\path[clip] (  0.00,  0.00) rectangle (216.81,252.94);
\definecolor{drawColor}{RGB}{0,0,0}

\path[draw=drawColor,line width= 0.4pt,line join=round,line cap=round] ( 98.43, 61.20) -- (142.38, 61.20);

\path[draw=drawColor,line width= 0.4pt,line join=round,line cap=round] ( 98.43, 61.20) -- ( 98.43, 55.20);

\path[draw=drawColor,line width= 0.4pt,line join=round,line cap=round] (142.38, 61.20) -- (142.38, 55.20);

\node[text=drawColor,anchor=base,inner sep=0pt, outer sep=0pt, scale=  1.00] at ( 98.43, 39.60) {F};

\node[text=drawColor,anchor=base,inner sep=0pt, outer sep=0pt, scale=  1.00] at (142.38, 39.60) {M};

\path[draw=drawColor,line width= 0.4pt,line join=round,line cap=round] ( 49.20, 66.48) -- ( 49.20,198.47);

\path[draw=drawColor,line width= 0.4pt,line join=round,line cap=round] ( 49.20, 66.48) -- ( 43.20, 66.48);

\path[draw=drawColor,line width= 0.4pt,line join=round,line cap=round] ( 49.20, 92.88) -- ( 43.20, 92.88);

\path[draw=drawColor,line width= 0.4pt,line join=round,line cap=round] ( 49.20,119.27) -- ( 43.20,119.27);

\path[draw=drawColor,line width= 0.4pt,line join=round,line cap=round] ( 49.20,145.67) -- ( 43.20,145.67);

\path[draw=drawColor,line width= 0.4pt,line join=round,line cap=round] ( 49.20,172.07) -- ( 43.20,172.07);

\path[draw=drawColor,line width= 0.4pt,line join=round,line cap=round] ( 49.20,198.47) -- ( 43.20,198.47);

\node[text=drawColor,rotate= 90.00,anchor=base,inner sep=0pt, outer sep=0pt, scale=  1.00] at ( 34.80, 66.48) {\(-25\)};

\node[text=drawColor,rotate= 90.00,anchor=base,inner sep=0pt, outer sep=0pt, scale=  1.00] at ( 34.80, 92.88) {\(-20\)};

\node[text=drawColor,rotate= 90.00,anchor=base,inner sep=0pt, outer sep=0pt, scale=  1.00] at ( 34.80,119.27) {\(-15\)};

\node[text=drawColor,rotate= 90.00,anchor=base,inner sep=0pt, outer sep=0pt, scale=  1.00] at ( 34.80,145.67) {\(-10\)};

\node[text=drawColor,rotate= 90.00,anchor=base,inner sep=0pt, outer sep=0pt, scale=  1.00] at ( 34.80,172.07) {\(-5\)};

\node[text=drawColor,rotate= 90.00,anchor=base,inner sep=0pt, outer sep=0pt, scale=  1.00] at ( 34.80,198.47) {\(0\)};
\end{scope}
\begin{scope}
\path[clip] ( 49.20, 61.20) rectangle (191.61,203.75);
\definecolor{drawColor}{RGB}{0,0,0}

\path[draw=drawColor,line width= 0.4pt,line join=round,line cap=round] ( 98.43,116.03) -- ( 98.43,134.09);

\path[draw=drawColor,line width= 0.4pt,line join=round,line cap=round] (142.38,145.27) -- (142.38,173.09);
\end{scope}
\end{tikzpicture}}
    \scalebox{.8}{\input{./art/tpcate-bmi.tex}}
  }
  \caption{Transported weight loss effect of semaglutide measured in percentage body weight change in the US study population, conditional on baseline body weight, age, sex, and BMI, respectively.
    For body weight, age, and BMI, the solid lines are point estimates, while the pointwise and uniform \(95\%\)-CIs are drawn with dashed and dotted lines.
    For sex, the point estimates and \(95\%\)-CIs are displayed as solid dots and error bars.
    F: female; M: male; p.p.: percentage point.
  }
  \label{fig:tpcate}
\end{figure}

The outcome scale is paramount to the interpretation of effect heterogeneity.
In the present example, the definition of the outcome, percentage change in body weight, explicitly involves the body weight at baseline.
Figure~\ref{fig:tpcate-alt} of the supplementary material displays the target population projected CATE estimates using body weight at week \(68\) as the outcome.
In contrast, the treatment effect under this outcome appears to increase with both baseline body weight and BMI.
An informed treatment decision should preferably be based on multiple outcome scales of clinical value.

\section{Discussion}

In this work, we study efficient estimation of the TATE using multi-source data.
CATE transportability allows for identifiability of this parameter but does not constrain the conditional counterfactual distributions nor the conditional treatment-specific means across the source studies.
However, if the outcome is bounded, for instance, binary or positive, the CATE transportability we assume induces potential variational dependence in the counterfactual distributions.
In the real data example, the outcome is clearly bounded from below.
This is mostly harmless when the conditional effect size is relatively small, but it may introduce undesired implicit assumptions when the effect is large.

Certain effect measures do not suffer from this problem.
However, all effect measures do not naturally lend themselves to transportability.
As is pointed out by \citet{colnet2023risk}, the causal odds ratio fails to disentangle the risk under placebo, even under a complete lack of treatment effect heterogeneity.
Moreover, non-collapsibility of the odds ratio leads to a mismatch between conditional and marginal effect measures.
For positive outcomes, transportability on the ratio between conditional treatment-specific means presents a possibility.
This assumption has been adopted for the estimation of target population causal mean ratio from a single source study \citep{wang2024causal}, but may be extended to the multi-source setup in light of \citet{li2024efficient} and the current work.

\citet{shyr2024multistudy} propose a multi-study R-learner precisely to leverage the overlap between the study populations.
The samples are given the same weight in the multi-study R-learner, and it is not robust against the misspecification of nuisance parameter models.
However, by modifying the influence functions of the TATE, we may obtain pseudo-outcomes that are orthogonal to the nuisance parameters \citep{foster2023orthogonal} and hence exihibit the desired robustness.
A multi-study DR learner regresses the pseudo-outcomes onto the baseline covariates to produce estimates of the CATE.
We leave the investigation of finite sample performance of this proposal for future work.
An alternative line of work is higher-order meta-learners for CATE combining ideas from the minimax rate results from \citet{kennedy2024minimax}.

\section*{Conflict of interest}
Zehao Su is funded by a research gift from Novo Nordisk A/S to the Section of Biostatistics, University of Copenhagen.
Henrik Ravn is employed by Novo Nordisk A/S.

\section*{Supplementary material}
The supplementary material contains an extension of the method, justifications for some claims made in the main text, details in the simulation and the real data example, and proofs of theoretical results.

\bibliography{./multi-trial-transport.bib}

\newpage

\part{}
\begin{center}
  {\large Supplementary material for ``Efficient estimation of the target population average treatment effect
  from multi-source data''}
\end{center}

\bigskip

\renewcommand{\theequation}{S\arabic{equation}}%
\renewcommand{\thesection}{S\arabic{section}}%
\renewcommand{\thetable}{S\arabic{table}}%
\renewcommand{\thefigure}{S\arabic{figure}}%
\renewcommand{\thetheorem}{S\arabic{theorem}}%
\renewcommand{\theproposition}{S\arabic{proposition}}%
\renewcommand{\thelemma}{S\arabic{lemma}}%
\renewcommand{\theassumption}{S\arabic{assumption}}%
\renewcommand{\theremark}{S\arabic{remark}}%
\renewcommand{\thecorollary}{S\arabic{corollary}}%
\renewcommand{\thepage}{S\arabic{page}}

\setcounter{equation}{0}
\setcounter{section}{0}
\setcounter{table}{0}
\setcounter{figure}{0}
\setcounter{theorem}{0}
\setcounter{proposition}{0}
\setcounter{lemma}{0}
\setcounter{assumption}{0}
\setcounter{remark}{0}
\setcounter{corollary}{0}
\setcounter{page}{1}

\parttoc

To simplify presentation, we define some notations.
For nonnegative sequences \((a_n)\) and \((b_n)\), we write \(a_n\lesssim b_n\) if there is a universal constant \(C>0\) such that \(a_n\leq Cb_n\).
If the constant \(C\) further depends on some constant \(M\), we write \(a_n\lesssim_M b_n\).
Let \(L_2^0(P_0)\) denote the space of mean-zero \(L_2(P_0)\)-functions.
Without loss of generality, we assume that every cross-fitting fold \(k\in[K]\) has sample size \(n/K\).
For any possibly random function \(f(o)\) of the data and \(k\in[K]\), let
\[
  \mathbb{P}_{n,k}f=\frac{K}{n}\sum_{i\in\mathcal{I}_k}f(O_i).
\]

\section{Additional confounding within the source studies}
\label{sec:confounding-app}

\subsection{Setup and identifiability}

In this section, we consider the setup where residual confounding may be present within each source population after controlling for the baseline covariates \(X\), such that assumption~\ref{asn:causal}\ref{asn:treatment-exchangeability} may no longer hold.
In this case, we may need to control for additional baseline covariates \(X_d\) observed in source study \(d\) to justify mean exchangeability of the intervention.

\begin{assumption}[Internal validity]
  \label{asn:causal-confounding}
  \hfill
  \begin{enumerate}[nosep,label=(\roman*)]
  \item (Positivity) \(\Pr(A=a\mid X_D=x_d,X=x,D=d)>0\) for \(a=0,1\);
  \item (Mean exchangeability) \(E\{Y(a)\mid A=a',X_D=x_d,X=x,D=d\}=E\{Y(a)\mid X_D=x_d,X=x,D=d\}\) for \(a=0,1\) and \(a'=0,1\).
  \end{enumerate}
\end{assumption}

Conceivably, these potential confounders highly depend on the source studies.
In register-based studies, the covariates \(X_d\) may be of high dimension.
Whereas in randomized controlled trials, these covariates may simply be taken to be undefined.
Hence, the observed data may appear rather heterogeneous.
To be specific, the sample consists of \(n_{0d}\) i.i.d. observations of \((Y,A,X_d,X)\) from source population \(d\) and \(n_1\) i.i.d. observations of \(X\) from the target population.
Under assumption~\ref{asn:sampling}, the observed data can be treated as an i.i.d. sample of size \(n\) of
\begin{multline*}
  O^*=\{(1-G)Y,(1-G)I(D=1)X_1,\dots,(1-G)I(D=m)X_m,\\
  (1-G)A,(1-G)D,X,G\}.
\end{multline*}
Let \(\mu^*(a,x_d,x,d)=E(Y\mid A=a,X_D=x_d,X=x,D=d)\) denote the mean outcome under exposure \(a\) in source population \(d\) conditional on baseline covariates \(x\) and confounders \(x_d\).
Under assumptions~\ref{asn:causal}\ref{asn:consistency} and \ref{asn:causal-confounding}, CATE transportability and the associated overlap condition (assumption~\ref{asn:transportability}) imply the restriction in the observed data distribution that for \(x\in\mathcal{X}\),
\begin{equation}
  \label{eq:cate-restriction}
  E\{\mu^*(1,X_D,X,D)-\mu^*(0,X_D,X,D)\mid X=x,D=d\}=\delta^*(x),
\end{equation}
where
\[
  \delta^{*}(x)=E\{\mu^*(1,X_D,X,D)-\mu^*(0,X_D,X,D)\mid X=x,G=0\}.
\]
Denote by \(\mathcal{P}^*\) the set of probability distributions over \(O^*\) that obey restriction~\eqref{eq:cate-restriction}.
We have the following identifiability result.

\begin{proposition}[Identifiability]
  \label{ppn:identifiability-star}
  Suppose assumptions~\ref{asn:transportability}, \ref{asn:causal}\ref{asn:consistency}, and \ref{asn:causal-confounding} hold.
  The TATE is identifiable in the observed data model \(\mathcal{P}^*\) as
  \[
    \theta=E\{\delta^*(X)\mid G=1\}.
  \]
\end{proposition}

\subsection{Estimation}

We do not undertake the study of the semiparametric efficiency bound of the parameter \(\theta\) in the model \(\mathcal{P}^*\).
Compared with the restriction on the mean differences \eqref{eqn:transportability} in the main text, restriction~\eqref{eq:cate-restriction} constrains the means of the mean differences.
Therefore, the strategy for constructing the maximal tangent space in the model \(\mathcal{P}\) in the proof of lemma~\ref{lem:eif} cannot be applied.
Instead, we present a class of scores that can be used to construct estimating equations for the parameter.
For consistency in notations, we focus on the true observed data distribution \(P_0\in\mathcal{P}\).
Consider any function \(w^*(x,d)\) such that \(\sum_{d\in[m]}\zeta_0(d\mid x)w^*(x,d)\neq 0\).
Consider the class of functions of the observed data
\begin{align*}
  \MoveEqLeft\ell^*(o^*,\eta^*)\\
  &= \frac{1-g}{\alpha}\frac{\pi(x)}{1-\pi(x)}\frac{w^*(x,d)}{\sum_{d'\in[m]}\zeta(d'\mid x)w^*(x,d')}\frac{2a-1}{e^*(a\mid x_d,x,d)}\{y-\mu^*(a,x_d,x,d)\}\\
  &\hphantom{={}}+\begin{multlined}[t][.85\textwidth]
    \frac{1-g}{\alpha}\frac{\pi(x)}{1-\pi(x)}\frac{w^*(x,d)}{\sum_{d'\in[m]}\zeta(d'\mid x)w^*(x,d')}\\
    \bigg\{\mu^*(1,x_d,x,d)-\mu^*(0,x_d,x,d)-\sum_{d'\in[m]}\frac{w^*(x,d')\zeta(d'\mid x)}{\sum_{d''\in[m]}w^*(x,d'')\zeta(d''\mid x)}\tau(x,d')\bigg\}
  \end{multlined} \\
  &\hphantom{={}}+\frac{g}{\alpha}\sum_{d'\in[m]}\frac{w^*(x,d')\zeta(d'\mid x)}{\sum_{d''\in[m]}w^*(x,d'')\zeta(d''\mid x)}\tau(x,d')
\end{align*}
indexed by the set of nuisance parameters
\[
  \eta^*=\{\alpha,\pi(x),\zeta(d\mid x),w^*(x,d),e^*(a\mid x_d,x,d),\mu^*(a,x_d,x,d),\tau(x,d)\}.
\]
Here the nuisance parameter \(e^*(a\mid x_d,x,d)\) is a placeholder for estimators of the propensity score \(e_0^*(a\mid x_d,x,d)=P_0(A=a\mid X_D=x_d,X=x,D=d)\) in source study \(d\).
The nuisance parameter \(\tau(x,d)\) is a placeholder for estimators of the difference function \(\delta_0^*(x)\) but is allowed to vary across studies.

For estimation, we describe a \(K\)-fold cross-fitting procedure similar to that given in \S\ref{sec:efficient-estimation} of the main text.
In particular, for each fold \(k\in[K]\), the estimators of the nuisance parameters \(\hat\alpha_k\), \(\hat\pi_k\) and \(\hat\zeta_k\) are fitted as in \S\ref{sec:efficient-estimation} of the main text.
The estimators \(\hat{e}^*_k\) and \(\hat\mu^*_{k}\) are fitted with the additional variables \(X_d\) within every source study.
With estimators \(\hat\tau_k(x,d)\) obtained from regressing  \(\hat\mu^*(1,X_d,X,d)-\hat\mu^*(0,X_d,X,d)\) onto \(X\) within source study \(d\),
define
\[
  \hat\delta^*_k(x)=\sum_{d\in[m]}\frac{\hat{w}^*_k(x,d)\hat\zeta_k(d\mid x)}{\sum_{d'\in[m]}\hat{w}^*_k(x,d')\hat\zeta_k(d'\mid x)}\hat\tau_k(x,d)
\]
An estimator of \(\theta_0\) is given by
\begin{equation}
  \hat\theta^*=\frac{1}{n}\sum_{k\in[K]}\sum_{i\in\mathcal{I}_k}\ell^*(O^*_i,\hat\eta^*_k).\label{eq:theta-star}
\end{equation}
Let the conditional variance of the outcome in source study \(d\) be \(V_0^*(a,x_d,x,d)=\mathrm{var}_0(Y\mid A=a,X_D=x_d,X=x,D=d)\).

\begin{assumption}[Regularity conditions]
  \label{asn:regularity-star}
  \hfill
  \begin{enumerate}[nosep,label=(\roman*)]
  \item With probability approaching \(1\), \(\{x:\hat\zeta_k(d\mid x)>0\}\subset\{x:\zeta_0(d\mid x)>0\}\) for every \(d\in[m]\);
  \item There exist \(L_2(P_0)\)-functions \(\bar\pi\), \(\bar\zeta\), \(\bar{w}^*\), \(\bar{e}^*\), \(\bar\mu^*\) and \(\bar\tau\) such that
    \begin{align*}
      &\|(\hat{\pi}_k-\bar{\pi})(X)\|_{P_0}=o_{P_0}(1),\\
      &\|(1-G)(\hat{\zeta}_k-\bar{\zeta})(d\mid X)\|_{P_0}=o_{P_0}(1),\\
      &\|I(D=d)(\hat{w}^*_k-\bar{w}^*)(X,d)\|_{P_0}=o_{P_0}(1),\\
      &\|I(D=d)(\hat{e}^*_k-\bar{e}^*)(a\mid X_d,X,d)\|_{P_0}=o_{P_0}(1),\\
      &\|I(D=d)(\hat{\mu}^*_k-\bar{\mu}^*)(a,X_d,X,d)\|_{P_0}=o_{P_0}(1),\\
      &\|I(D=d)(\hat\tau_k-\bar\tau)(X,d)\|_{P_0}=o_{P_0}(1);
    \end{align*}
  \item There exists a universal constant \(C>1\) such that
    \begin{align*}
      &\alpha_0\geq C^{-1},\hat\alpha_k\geq C^{-1},\hat\pi_k(x)\leq 1-C^{-1},e^*_0(a\mid x_d,x,d)\geq C^{-1},\hat{e}^*_k(a\mid x_d,x,d)\geq C^{-1},\\
      &|\hat\mu^*_k(a,x_d,x,d)|\leq C, V^*_0(a,x_d,x,d)\leq C,|\hat{w}^*_k(x,d)|\leq C,|\bar{w}^*(x,d)|\leq C,\\
      &\big|\textstyle\sum_{d'\in[m]}\hat\zeta_k(d'\mid x)\hat{w}^*_k(x,d)\big|\geq C^{-1},\big|\textstyle\sum_{d'\in[m]}\zeta_0(d'\mid x)\bar{w}^*(x,d)\big|\geq C^{-1}.
    \end{align*}
  \end{enumerate}
\end{assumption}

\stepcounter{assumption}
\begin{subassumption}[Correct specifications]
  \label{asn:model-consistency-star}
  \hfill
  \begin{enumerate}[label=(\roman*)]
  \item \(\bar{\mu}^*=\mu^*_0\) and \(\bar{\tau}=\delta^*\); or
  \item \(\bar{e}^*=e^*_0\), \(\bar{\zeta}=\zeta_0\), and \(\bar{\pi}=\pi_0\).
  \end{enumerate}
\end{subassumption}

\begin{subassumption}[Rate conditions]
  \label{asn:model-linearity-star}
  \hfill
  \begin{enumerate}[label=(\roman*)]
  \item\label{asn:plim-star} \(\bar{\mu}^*=\mu^*_0\), \(\bar\tau=\delta^*\), \(\bar{e}^*=e^*_0\), \(\bar{\zeta}=\zeta_0\), and \(\bar{\pi}=\pi_0\);
  \item \(\|I(D=d)(\hat{\mu}^*_k-\bar{\mu}^*)(a,X_d,X,d)\|_{P_0}\|I(D=d)(\hat{e}^*_k-\bar{e}^*)(a\mid X_d,X,d)\|_{P_0}=o_{P_0}(n^{-1/2})\) and \(\|I(D=d)(\hat\tau_k-\bar\tau)(X,d)\|_{P_0}\big\{\|(\hat{\zeta}_k-\bar{\zeta})(d\mid X)\|_{P_0}+\|(\hat{\pi}_k-\bar{\pi})(X)\|_{P_0}\big\}=o_{P_0}(n^{-1/2})\).
  \end{enumerate}
\end{subassumption}

\begin{proposition}[Asymptotic behaviour]
  \label{ppn:asymptotic-star}
  Suppose assumption~\ref{asn:regularity-star} holds.
  Then:
  \begin{enumerate}[label=(\roman*)]
  \item \(\hat{\theta}^*-\theta_0=o_{P_0}(1)\) under assumption~\ref{asn:model-consistency-star};
  \item \(\hat{\theta}^*-\theta_0=n^{-1}\sum_{i=1}^{n}\varphi^*(O^*_i,\bar{w}^*)+o_{P_0}(n^{-1/2})\) under assumption~\ref{asn:model-linearity-star}.
  \end{enumerate}
\end{proposition}

\subsection{Relative efficiency and choice of weight function}

While we are not able to ascertain the semiparametric efficiency bound of \(\theta_0\) at the distribution \(P_0\) under the model \(\mathcal{P}^*\), we can characterize the estimator that has relative efficiency no smaller than \(1\) against all other estimators in the class of estimators of \(\theta_0\) with different weight functions.

By observing the structrue of the class of functions \(\ell^*(o,\eta^*)\) and comparing it to the results obtained for \(\ell(o,\eta)\) in the main text, we consider the ``residual''
\begin{multline*}
  U_0^*=\frac{2A-1}{e^*_0(A\mid X_D,X,D)}\{Y-\mu^*_0(A,X_D,X,D)\} \\
  + \{\mu^*_0(1,X_D,X,D)-\mu^*_0(0,X_D,X,D)-\delta^*_0(X)\}.
\end{multline*}
Intuitively, the optimal weight function in the class of estimators associated with \(\ell^*\) should be the inverse conditional variance
\[
  w^*_0(x,d)=\{\mathrm{var}_0(U^*_0\mid X=x,D=d)\}^{-1}.
\]
This function can be expanded as
\begin{multline*}
  w^*_{0}(x,d) = \bigg[E_0\bigg\{\frac{V_0^*(1,X_D,X,D)}{e_0^*(1\mid X_D,X,D)}+\frac{V_0^*(0,X_D,X,D)}{e_0^*(0\mid X_D,X,D)}\biggm\vert X=x,D=d\bigg\}\\
  +\mathrm{var}_0[\mu^*_0(1,X_D,X,D)-\mu^*_0(0,X_D,X,D)\mid X=x,D=d]\bigg]^{-1}.
\end{multline*}

Analogously, we may obtain an estimator of the optimal weight function as the minimizer of the empirical risk in some function class \(\mathcal{F}^*\):
\[
  \hat{w}^*_{k0}(x,d) = \arg\min_{f^*\in \mathcal{F}^*} \sum_{i:i\in\mathcal{I}_k,G_i=0}[-2f^*(X_i,D_i)+(\hat{U}^*_{ki})^2\{f^{*}(X_i,D_i)\}^2],
\]
where
\begin{multline*}
\hat{U}^*_{ki}=\frac{2A_i-1}{\hat{e}^*_k(A_i\mid (X_D)_i,X_i,D_i)}[Y_i-\hat{\mu}^*_k\{A_i,(X_D)_i,X_i,D_i\}] \\
  + [\hat{\mu}^*_k\{1,(X_D)_i,X_i,D_i\}-\hat\mu^*_k\{0,(X_D)_i,X_i,D_i\}-\hat\delta^*_k(X_i)].  
\end{multline*}
Now consider the estimator
\[
  \hat\theta^*_{0} = \frac{1}{n}\sum_{k\in[K]}\sum_{i\in\mathcal{I}_k}\ell^*(O_i,\hat\eta^*_{k0}),
\]
where \(\hat\eta^*_{k0}\) differs from \(\hat\eta^*_{k}\) only in that \(\hat{w}^*_k\) is replaced by \(\hat{w}^*_{k0}\).

\begin{corollary}[Relative efficiency]
  \label{cor:relative-eff}
  Suppose assumptions~\ref{asn:regularity-star} and \ref{asn:model-linearity-star} hold.
  Assume further that \(\|I(D=d)\{\hat{w}^*_{k0}(X,d)-c(X)w^*_0(X,d)\}\|_{P_0}=o_{P_0}(1)\) for some nonrandom function \(c(x)\neq 0\).
  Then the asymptotic variance of \(\hat\theta^*_{0}\) is no larger than that of \(\hat\theta^*\).
\end{corollary}

\section{Asymptotic variance reduction with study-specific weights}
\label{sec:diff-asymp-var-app}

We present details for the difference of the asymptotic variances when \(\tilde{w}(x,d)=\tilde{w}(x)\neq 0\) that corresponds to the discussion in \S\ref{sec:estimation} of the main text.

By direct calculation, the difference is
\begin{align*}
  \MoveEqLeft E_0\{\varphi^{2}(O,w_0)-\varphi^{2}(O,\tilde{w})\} \\
  &= \begin{multlined}[t][.9\textwidth]
    E_0\bigg(\frac{(1-G)\{\pi_0(X)\}^{2}}{\alpha_0^{2} \{1-\pi_0(X)\}^{2}}\frac{1}{\{e_0(A\mid X,D)\}^{2}}\bigg[\bigg\{\frac{w_0(X,D)}{\sum_{d\in[m]}\zeta_0(d\mid X)w_0(X,d)}\bigg\}^{2}-1\bigg]\\
    \{Y-\mu_0(A,X,D)\}^{2}\bigg)
  \end{multlined}\\
  &= E_0\bigg(\frac{(1-G)\{\pi_0(X)\}^{2}}{\alpha_0^{2} \{1-\pi_0(X)\}^{2}}\frac{V_0(A,X,D)}{\{e_0(A\mid X,D)\}^{2}}\bigg[\bigg\{\frac{w_0(X,D)}{\sum_{d\in[m]}\zeta_0(d\mid X)w_0(X,d)}\bigg\}^{2}-1\bigg]\bigg) \\
  &= \begin{multlined}[t][.9\textwidth]
    E_0\bigg(\frac{(1-G)\{\pi_0(X)\}^{2}}{\alpha_0^{2} \{1-\pi_0(X)\}^{2}}\underbrace{\bigg\{\frac{V_0(1,X,D)}{e_0(1\mid X,D)}+\frac{V_0(0,X,D)}{e_0(0\mid X,D)}\bigg\}}_{\{w_0(X,D)\}^{-1}}\\
    \bigg[\bigg\{\frac{w_0(X,D)}{\sum_{d\in[m]}\zeta_0(d\mid X)w_0(X,d)}\bigg\}^{2}-1\bigg]\bigg)
  \end{multlined}\\
  &= \begin{multlined}[t][.9\textwidth]
    E_0\bigg(\frac{(1-G)\{\pi_0(X)\}^{2}}{\alpha_0^{2} \{1-\pi_0(X)\}^{2}}\sum_{d\in[m]}\zeta_0(d\mid X)\{w_0(X,d)\}^{-1}\\
    \bigg[\bigg\{\frac{w_0(X,d)}{\sum_{d'\in[m]}\zeta_0(d'\mid X)w_0(X,d')}\bigg\}^{2}-1\bigg]\bigg)
  \end{multlined}\\
  &= \begin{multlined}[t][.9\textwidth]
    E_0\bigg(\frac{\{\pi_0(X)\}^{2}}{\alpha_0^{2}\{1-\pi_0(X)\}}\bigg[\bigg\{\sum_{d'\in[m]}w_0(X,d')\zeta_0(d'\mid X)\bigg\}^{-1}\\
    -\bigg\{\sum_{d'\in[m]}\zeta_0(d'\mid X)\{w_0(X,d')\}^{-1}\bigg\}\bigg]\bigg).
  \end{multlined}
\end{align*}
Alternatively, using corollary~\ref{cor:tangent-space-complement}, we have that \(\varphi(o,\tilde{w})-\varphi(o,w_0)\) is orthogonal to the tangent space at \(P_0\) but \(\varphi(o,{w_0})\) lies in the tangent space, so that \(E_0\{\varphi^{2}(O,w_0)-\varphi^{2}(O,\tilde{w})\} = -E_0\{\varphi(O,w_0)-\varphi(O,\tilde{w})\}^{2}\leq 0\) directly by the Pythagorean theorem.

The difference of the asymptotic variances for \(w_0(x,d)\) and \(\tilde{w}(x,d)\) such that \(\tilde{w}(x,d)\neq 0\) for some \(d\) is
\begin{multline*}
  E_0\{\varphi^{2}(O,w_0)-\varphi^{2}(O,\tilde{w})\} = E_0\bigg(\frac{\{\pi_0(X)\}^{2}}{\alpha_0^{2}\{1-\pi_0(X)\}}\bigg[\frac{1}{\sum_{d\in[m]}\zeta_0(d\mid X)w_0(X,d)}\\
  -\frac{\sum_{d\in[m]}\zeta_0(d\mid X)\{w_0(X,d)\}^{-1}\{\tilde{w}(X,d)\}^{2}}{\{\sum_{d'\in[m]}\zeta_0(d'\mid X)\tilde{w}(X,d')\}^{2}}\bigg]\bigg).
\end{multline*}

\section{Equivalence of the orthocomplements of the tangent space}
\label{sec:li-app}

In this section, we consider the special case where the number of source studies \(m\) is \(2\); that is, the model restriction is
\[
  \mu(1,x,1)-\mu(0,x,1)=\mu(1,x,2)-\mu(0,x,2),
\]
for all \(x\) such that \(\zeta(1\mid x)\zeta(2\mid x)>0\).
We show that the orthocomplement of the tangent space in corollary~\ref{cor:tangent-space-complement} of the main text and that in corollary~1 of \citetsuppmat{li2024efficient} are equivalent in this case.
We ignore the differences in the local regularity conditions imposed at \(P_0\).
The orthocomplement of the tangent space obtained by \citetsuppmat{li2024efficient} with the identity link function is
\[
  \Omega_0= \{(1-g)f(x)f_0(a,x,d)\{y-\mu_0(a,x,d)\}:h(x)\text{ arbitrary}\}\subset L_2^0(P_0),
\]
where
\begin{multline*}
  f_0(a,x,d)=I(d=1)a-I(d=1)(1-a)\frac{e_0(1\mid x,1)}{e_0(0\mid x,1)}\\
  -I(d=2)a\frac{\zeta_0(1\mid x)}{\zeta_0(2\mid x)}\frac{e_0(1\mid x,1)}{e_0(1\mid x,2)}+I(d=2)(1-a)\frac{\zeta_0(1\mid x)}{\zeta_0(2\mid x)}\frac{e_0(1\mid x,1)}{e_0(0\mid x,2)}
\end{multline*}
from Equation (31) in \citetsuppmat{li2024efficient}.
Clearly,
\[
  f_0(a,x,d)=e_0(1\mid x,1)\zeta_0(1\mid x)\frac{I(d=1)-I(d=2)}{\zeta_0(d\mid x)}\frac{2a-1}{e_0(a\mid x,d)}.
\]
Now we show that
\[
  \Lambda_0=\bigg\{(1-g)h(x,d)\frac{2a-1}{e_0(a\mid x,d)}\{y-\mu_0(a,x,d)\}:E_0\{h(X,D)\mid X=x,G=0\}=0\bigg\}
\]
equals \(\Omega_0\) if \(\zeta_0(1\mid x)\zeta_0(2\mid x)>0\) and \(e_0(1\mid x,1)>0\) for all \(x\).
For any \(f(x)\),
\[
  E_0\bigg\{f(X)e_0(1\mid X,1)\zeta_0(1\mid X)\frac{I(D=1)-I(D=2)}{\zeta_0(D\mid X)}\biggm\vert X=x,G=0\bigg\}=0,
\]
so that \(\Omega_0\subset \Lambda_0\).
Conversely, any function \(h(x,d)\) such that \(E_0\{h(X,D)\mid X=x,G=0\}=0\) can be written as \(\{I(d=1)-\zeta_0(1\mid x)\}h(x)\) for some \(h(x)\).
This can be further rewritten as
\begin{multline*}
  e_0(1\mid x,1)\frac{I(d=1)-\zeta_0(1\mid x)}{\zeta_0(2\mid x)}\frac{h(x)\zeta_0(2\mid x)}{e_0(1\mid x,1)}\\
  =e_0(1\mid x,1)\zeta_0(1\mid x)\frac{I(d=1)-I(d=2)}{\zeta_0(d\mid x)}\frac{h(x)\zeta_0(2\mid x)}{e_0(1\mid x,1)},
\end{multline*}
so that \(\Lambda_0\subset\Omega_0\).

\section{Parametric conditional average treatment effect}
\label{sec:parametric-app}
Consider the class of functions of the baseline covariates \(\{\delta(x,\gamma):\gamma\in\mathbb{R}^{q}\}\), where \(\delta(x,\gamma)\) is a known, smooth function of \(\gamma\), and the true CATE is \(\delta_0(x)= \delta(x,\gamma_0)\) for some unique \(\gamma_0\).
Under the transportability assumption, the conditional mean of the outcome can be expressed as \(\mu_0(a,x,d)=\mu_0(0,x,d)+a\delta(x,\gamma_0)\).
Define the semiparametric model
\[
  \mathcal{P}_{\mathrm{sp}}=\{P\in\mathcal{P}:\mu(1,x,d)-\mu(0,x,d)=\delta(x,\gamma)\}.
\]
Let \(\dot\delta(x,\gamma_0)=(\partial/\partial\gamma)\delta(x,\gamma)\vert_{\gamma=\gamma_0}\).
\begin{lemma}
  \label{lem:eif-sp}
  Suppose there exists a universal constant \(C>0\) such that \(|Y-\mu_0(0,X,D)-A\delta(X,\gamma_0)|\leq C\) and \(\dot{\delta}(X,\gamma_0)\neq 0\) almost surely under \(P_0\).
  The efficient score of \(\gamma_0\) at the distribution \(P_0\) under the semiparametric model \(\mathcal{P}_{\mathrm{sp}}\) is
  \begin{multline}
    \label{eqn:eff-score}
    (1-g)\dot{\delta}(x,\gamma_0)\bigg\{a-\frac{\{V_0(1,x,d)\}^{-1}e_0(1\mid x,d)}{\{V_0(1,x,d)\}^{-1}e_0(1\mid x,d)+\{V_0(0,x,d)\}^{-1}e_0(0\mid x,d)}\bigg\}\\
    \{V_0(a,x,d)\}^{-1}\big[y-\{\mu_0(0,x,d)+a\delta_{\gamma_0}(x)\}\big].
  \end{multline}
\end{lemma}

\begin{proof}
  Define the transformation \(R= Y-\{\mu(0,X,D)+A\delta(X,\gamma)\}\) if \(G=0\) and denote its density by \(p(r\mid a,x,d)\).
  The likelihood of the data can be factorized as
  \begin{align*}
    p_0(o) &= p_0(x)\pi_0(x)^{g}\{1-\pi_0(x)\}^{1-g}\{\zeta_0(d\mid x)e_0(a\mid x,d)p_0(y\mid a,x,d)\}^{1-g} \\
           &= p_0(x)\pi_0(x)^{g}\{1-\pi_0(x)\}^{1-g}\{\zeta_0(d\mid x)e_0(a\mid x,d)p_{0}(r\mid a,x,d)\}^{1-g}.
  \end{align*}
  Let \(h_r(r_0,a,x,d)= (\d/\d r) p_{0}(r_0\mid a,x,d) /p_{0}(r_0\mid a,x,d)\).
  The nuisance tangent space at \(P_0\in\mathcal{P}_{\mathrm{sp}}\) is
  \[
    \dot{\mathcal{P}}_{\mathrm{sp}} = \dot{\mathcal{P}}_{\mu}+(\dot{\mathcal{P}}_{r}\oplus\dot{\mathcal{P}}_{a}\oplus\dot{\mathcal{P}}_{d}\oplus\dot{\mathcal{P}}_{g}\oplus\dot{\mathcal{P}}_{x}),
  \]
  where \(\dot{\mathcal{P}}_{a}\), \(\dot{\mathcal{P}}_{d}\), \(\dot{\mathcal{P}}_{g}\), and \(\dot{\mathcal{P}}_{x}\) are as defined in the proof of lemma~\ref{lem:eif}, and
  \begin{align*}
    \dot{\mathcal{P}}_{r} &= \begin{multlined}[t][.8\textwidth]
      \big\{(1-g)h(r_0,a,x,d): E_0\{h(R_0,A,X,D)\mid A,X,D\}=0,\\
      E_0\{R_0 h(R_0,A,X,D)\mid A,X,D\}=0\big\},
    \end{multlined}\\
    \dot{\mathcal{P}}_{\mu} &= \big\{(1-g)h_r(r_0,a,x,d)h(x,d):h(x,d)\text{ arbitrary}\big\}.
  \end{align*}
  A standard result of semiparametric regression gives that
  \[
    \tilde{\dot{\mathcal{P}}}_{r}=\{(1-g)r_0 h(a,x,d):h(a,x,d)\text{ arbitrary}\}
  \]
  is the orthogonal complement of the subspace \(\dot{\mathcal{P}}_{r}\oplus\dot{\mathcal{P}}_{a}\oplus\dot{\mathcal{P}}_{d}\oplus\dot{\mathcal{P}}_{g}\oplus\dot{\mathcal{P}}_{x}\).

  Define \(\nu_0(a,x,d)= \{E_0(r_0^{2}\mid A=a,X=x,D=d)\}^{-1}=V_0^{-1}(a,x,d)\).
  The orthogonal projection of the subspace \(\dot{\mathcal{P}}_{\mu}\) onto \(\tilde{\dot{\mathcal{P}}}_{r}\) is
  \[
    \Pi\{\dot{\mathcal{P}}_{\mu}\mid \tilde{\dot{\mathcal{P}}}_{r}\}=\{(1-g)h(x,d)\nu_0(a,x,d)r_0:h(x,d)\text{ arbitrary}\}.
  \]
  Now we show that this is true.
  Take an arbitrary element \(J(o)= (1-g)h_{r}(r_0,a,x,d)h(x,d)\in\dot{\mathcal{P}}_{\mu}\).
  We can verify that \(J^*(o)=-(1-g)h(x,d)\nu_0(a,x,d)r_0\) is indeed the projection \(\Pi\{J\mid \tilde{\dot{\mathcal{P}}}_{r}\}=J^{*}(o)\).
  This is because the difference
  \[
    J(o)-J^{*}(o)=(1-g)h(x,d)\{h_{u}(r_0,a,x,d)+\nu_0(a,x,d)r_0\}
  \]
  is orthogonal to \(\tilde{\dot{\mathcal{P}}}_{r}\), since for arbitrary \(h(a,x,d)\),
  \begin{align*}
    \MoveEqLeft E_0[(1-G)R_0 h(A,X,D)h(X,D)\{h_{r}(R_0,A,X,D)+\nu_0(A,X,D)R_0\}] \\
    &= E_0[(1-G)h(A,X,D)h(X,D)E_0\{R_0 h_{r}(R_0,A,X,D)+\nu_0(A,X,D)R_0^{2}\mid A,X,D\}] \\
    &= E_0[(1-G)h(A,X,D)h(X,D)\{(-1) + 1\}] = 0,
  \end{align*}
  where we have used the equality that \(E_0\{R_0 h_{r}(R_0,A,X,D)\mid A,X,D\}=-1\), as a result of differentiating the moment restriction \(\int r_0 p_{0}(r_0\mid a,x,d)\d r_0=0\) with respect to \(\gamma\) and evaluating at \(\gamma_0\).

  The score function of \(\gamma_0\) is
  \[
    S(o)=-(1-g)h_{r}(r_0,a,x,d)a\dot{\delta}(x,\gamma_0).
  \]
  The projection is
  \begin{multline*}
    \Pi\{S\mid \tilde{\dot{\mathcal{P}}}_{r}\}=(1-g)E_0\{S(O)R_0\mid A=a,X=x,D=d\}\nu_0(a,x,d)r_0\\
    =(1-g)a\dot{\delta}(x,\gamma_0)\nu_0(a,x,d)r_0.
  \end{multline*}
  We further project \(\Pi\{S\mid \tilde{\dot{\mathcal{P}}}_r\}\) onto the space \(\Pi\{\dot{\mathcal{P}}_{\mu}\mid \tilde{\dot{\mathcal{P}}}_{r}\}\).
  Assume the projection takes the form
  \[
    \Pi\big[\Pi\{S\mid \tilde{\dot{\mathcal{P}}}_{r}\}\,\big\vert\, \Pi(\dot{\mathcal{P}}_{\mu}\mid \tilde{\dot{\mathcal{P}}}_{r})\big]=(1-g)h^{*}(x,d)\nu_0(a,x,d)r_0.
  \]
  The function \(h^{*}(x,d)\) is the solution to the equation
  \[
    E_0\{\nu_0(A,X,D)R_0\{A\dot{\delta}(X,\gamma_0)\nu_0(A,X,D)R_0-h^{*}(X,D)\nu_0(A,X,D)R_0\}\mid X,D\}=0,
  \]
  which yields
  \begin{align*}
    h^{*}(x,d) &= \frac{E_0\{A\dot{\delta}(X,\gamma_0)\nu_0(A,X,D)\mid X=x,D=d\}}{E_0\{\nu_0(A,X,D)\mid X=x,D=d\}} \\
               &=\frac{\dot{\delta}(x,\gamma_0)\nu_0(1,x,d)e_0(1\mid x,d)}{\nu_0(1,x,d)e_0(1\mid x,d)+\nu_0(0,x,d)e_0(0\mid x,d)}.
  \end{align*}
  The efficient score of \(\gamma_0\) is the projection
  \begin{align*}
    S_{\mathrm{eff}}(o) &= \Pi\{S\mid \dot{\mathcal{P}}_{\mathrm{sp}}^{\perp}\} \\
                         &=\Pi\{S\mid \tilde{\dot{\mathcal{P}}}_{r}\} - \Pi\big[\Pi\{S\mid \tilde{\dot{\mathcal{P}}}_{r}\}\,\big\vert\, \Pi(\dot{\mathcal{P}}_{\mu}\mid \tilde{\dot{\mathcal{P}}}_{r})\big] \\
                        &= (1-g)\{a\dot{\delta}(x,\gamma_0)-h^{*}(x,d)\}\nu_0(a,x,d)r_0\\
                        &=(1-g)\dot{\delta}(x,\gamma_0)\bigg\{a-\frac{\nu_0(1,x,d)e_0(1\mid x,d)}{\nu_0(1,x,d)e_0(1\mid x,d)+\nu_0(0,x,d)e_0(0\mid x,d)}\bigg\}\nu_0(a,x,d)r_0.
  \end{align*}.
\end{proof}

\begin{remark}
The efficient score function of \(\gamma_0\) can be represented as \(S_{\mathrm{eff}}(o,\bar\mu,\bar{e},\bar{V})\}\) with the set of nuisance parameters \(\{\bar\mu(0,x,d),\bar{e}(a,x,d),\bar{V}(a,x,d)\}\).
Since the semiparametric model \(\mathcal{P}_{\mathrm{sp}}\) is induced by a conditional mean restriction, it follows that any score \(S_{\mathrm{eff}}(o,\cdot,\cdot,\bar{V})\) obtained by replacing \(V_0\) with an arbitrary \(\bar{V}\neq 0\) is Neyman orthogonal \citepsuppmat{chernozhukov2018double}.
The efficient score suggests a doubly robust estimating equation in the sense that \(E_0\{S_{\mathrm{eff}}(O,\bar{\mu},\bar{e},\bar{V})\}=0\) if either \(\bar\mu(0,x,d)=\mu_0(0,x,d)\) or \(\bar{e}(a\mid x,d)=e_0(a\mid x,d)\).
The double robustness is attractive especially in randomized trials, because the propensity scores are generally known.
\end{remark}

\section{Details in the simulation and the real data example}
\label{sec:details}

\subsection{Data generating mechanism}

The baseline covariates are generated by
\[
  X \sim 2\Phi[\mathrm{Normal}\{(0,0,0)^{\T},0.5\mathrm{Id} + 0.5\cdot 11^\T\}]-1
\]
so that the joint distribution function is a rescaled copula onto \([-1,1]\).
The indicators \(G\) and \((1-G)D\) are generated by
\begin{align*}
  G\mid X &\sim \mathrm{Bernoulli}[\mathrm{expit}\{-\log 3+\log(1.5)X^\T 1\}], \\
  D\mid \{X,G=0\} &\sim \begin{multlined}[t][.7\textwidth]
    \mathrm{Multinomial}[\mathrm{softmax}\{0, \log(1.5)+\log(1.5)X^\T 1,\\ -\log(0.75)+\log(0.75)X^\T 1\}],
  \end{multlined}
\end{align*}
where \(\mathrm{expit}(x)=\{1+\exp(-x)\}^{-1}\) and \(\mathrm{softmax}(x_1,x_2,x_3)=\{\exp(x_1)+\exp(x_2)+\exp(x_3)\}^{-1}\{\exp(x_1),\exp(x_2), \exp(x_3)\}^\T\).

\subsection{Super learning}
We used the implementation of super learning from the R-package SuperLearner \citepsuppmat{polley2024superlearner}.
The base learners and their corresponding aliases (in parentheses) in SuperLearner were the null model (SL.mean), lasso (SL.glmnet), a generalized additive model (SL.gam), random forest (SL.ranger), and gradient boosting trees (SL.xgboost).

Since SuperLearner does not support categorical outcome with more than two levels, to estimate \(\zeta(d\mid x)\), we fitted ensemble models \(\check\zeta_k(d\mid x)\) treating the binary indicator \(I(D=d)\) as outcome for each \(d\in[m]\).
To ensure the estimated probabilities summed up to \(1\), we used the normalized average
\[
  \hat\zeta_k(d\mid x)=\frac{\check\zeta_k(d\mid x)}{\sum_{d'\in[m]}\check\zeta_k(d'\mid x)}.
\]
The conditional outcome means \(\hat\mu_k(a,x,d)\) were fitted separately for each combination of \(a\in\{0,1\}\) and \(d\in[m]\).

\subsection{Estimators of the TATE in smaller models}
We considered alternative estimators of the TATE
\begin{align*}
  \hat\theta^\dagger &= \begin{multlined}[t][.8\textwidth] \frac{1}{n}\sum_{k=1}^{5}\sum_{i\in\mathcal{I}_k}\bigg[\frac{1-G_i}{\hat\alpha_k}\frac{\hat\pi_k(X_i)}{1-\hat\pi_k(X_i)}\frac{(2A_i-1)\hat{w}^\dagger_k(A_i,X_i,D_i)}{\sum_{d\in[m]}\hat{w}^\dagger_k(A_i,X_i,d)\hat\zeta_k(d\mid X_i)e_0(A_i\mid X_i,d)}\\
    \{Y_i-\hat\mu^\dagger_k(A_i,X_i)\}+\frac{G_i}{\hat\alpha_k}\hat\delta^\dagger_k(X_i)\bigg],
  \end{multlined}\\
  \hat\theta^\ddagger &= \frac{1}{n}\sum_{k=1}^{5}\sum_{i\in\mathcal{I}_k}\bigg[\frac{1-G_i}{\hat\alpha_k}\frac{\hat\pi_k(X_i)}{1-\hat\pi_k(X_i)}\frac{2A_i-1}{\hat{e}^\dagger_k(A_i\mid X_i)}\{Y_i-\hat\mu^\dagger_k(A_i,X_i)\}+\frac{G_i}{\hat\alpha_k}\hat\delta^\dagger_k(X_i)\bigg],
\end{align*}
which follow from the efficient influence functions of the parameter under the models \(\mathcal{P}^\dagger\) and \(\mathcal{P}^\ddagger\).
The outcome regression model \(\hat\mu^\dagger_k\) was fitted with super learning separately for each intervention \(a\in\{0,1\}\) in the joint source population.
For both estimators here, the difference function was a simple difference between the estimates
\[
  \hat\delta^\dagger_k(x)=\hat\mu^\dagger_k(1,x)-\hat\mu^\dagger_k(0,x).
\]
Specifically for \(\hat\theta^\dagger\), the oracle conditional outcome variance was substituted for the weight function; that is, \(\hat{w}^\dagger_k(a,x,d)=V_0^{-1}(a,x,d)\).
The propensity score \(e^\dagger_0\) appearing in \(\hat{\theta}^\ddagger\) was estimated by
\[
  \hat{e}^\dagger_k(a\mid x)=\sum_{d\in[m]}e_0(a\mid x,d)\hat{\zeta}_k(d\mid x).
\]

\subsection{Bootstrap for uniform confidence intervals}
For the uniform confidence interval, we used \(t\)-bootstrap \citepsuppmat{belloni2015new} to approximate the distribution of the Gaussian process supremum \(T=\sup_{z\in\mathcal{Z}}|\mathbb{T}(z)|\).
We calculated the plug-in estimate \(\hat\Psi\) of the covariance matrix \(\bar\Psi\).
In the simulation, the support of \(Z=\{X_1\}\) was the compact set \(\mathcal{Z}=[-1, 1]\).
We used an equidistant grid \(\tilde{\mathcal{Z}}\) of size \(1000\) to capture \(\mathcal{Z}\).
Let \(\{\xi_1,\dots,\xi_{1000}\}\) be \(1000\) i.i.d. copies of a \(4\)-dimensional normal random variable with mean zero and identity covariance matrix.
The dimension of random noise should match the dimension of the cubic basis \(b(z)=(1,x_1,x_1,x_1^3)^\T\).
Then we simulated the Gaussian supremum \(T\) by
\[
  T_j= \max_{z\in\tilde{\mathcal{Z}}}\bigg|\frac{b^\T(z)\hat\Psi^{1/2}}{\{b^\T(z)\hat\Psi b(z)\}^{1/2}}\xi_j\bigg|
\]
for \(j=1,\dots,1000\), and the theoretical quantile \(c_T(0.95)\) was approximated by the \(95\%\)-empirical quantile of the sample \(\{T_1,\dots,T_{1000}\}\).

\subsection{Variable scaling}
For the estimation of the projected CATE in the real data example, we formed a cubic basis of baseline body weight (kg), age (year), and BMI (\(\text{kg}\cdot\text{m}^{-2}\)) after scaling these variables with \(1/100\), \(1/50\), and \(1/50\), respectively.
The inverse of the sample covariance matrix using the scaled variables was more stable numerically compared to the inverse using raw variables.
The projected CATE and associated confidence intervals were calculated and presented on the original scale.

\subsection{Cross-fitting and coverage}
In the simulation study, we saw that the plug-in standard error for the non-cross-fitted estimators underestimated their variability.
We hypothesized that using nonparametric base learners in super learning for certain nuisance parameters would undermine consistency of the standard error estimator.

To investigate the impact of cross-fitting on the plug-in standard error estimates of the transported TATE in the real data example, we changed the base learners in super learning for different combinations of nuisance parameters.
Thus, we re-calculated the transported TATE estimators replacing the full set of base learners above with only the null model and a generalized linear model (SL.glm) for \(\{\hat\mu_k\}\), \(\{\hat\zeta_k\}\), \(\{\hat\pi_k\}\), \(\{\hat\mu_k,\hat\zeta_k\}\), \(\{\hat\mu_k,\hat\pi_k\}\), \(\{\hat\zeta_k,\hat\pi_k\}\), and \(\{\hat\mu_k,\hat\zeta_k,\hat\pi_k\}\) in turn, keeping the rest of the others unchanged.
However, if the true nuisance parameters are not generalized linear models, we should not expect the estimators to be asymptotically normal.
Nevertheless, the difference between the standard error estimates obtained with and without cross-fitting may still inform which nuisance parameters are responsible for underestimation of the standard error.
The results are displayed in table~\ref{tab:tate-sensitivity}.
We are not able to single out any nuisance parameters that can explain the underestimation of the standard error.

We notice a recurring pattern that the estimator using the learned optimal weights has just slightly higher plug-in standard errors than the other two estimators.
This observation suggests that the learned optimal weights might have been constant across the source populations.

\section{Proofs}
\label{sec:proof-app}
\subsection{Proof of lemma~\ref{lem:identification}}
We start with the g-formula representation.
We see that for all \(x\in\mathcal{X}\),
\begin{align*}
  \MoveEqLeft E\{Y(1)-Y(0)\mid X=x,G=1\} \\
  &= E\{Y(1)-Y(0)\mid X=x,D=d\} \\
  &= E\{Y(1)\mid A=1,X=x,D=d\}-E\{Y(0)\mid A=0,X=x,D=d\} \\
  &= E(Y\mid A=1,X=x,D=d)-E(Y\mid A=0,X=x,D=d) \\
  &= \mu(1,x,d)-\mu(0,x,d) \\
  &= \delta(x).
\end{align*}

The target parameter
\begin{align*}
  \theta=E\{Y(1)-Y(0)\mid G=1\} &= E[E\{Y(1)-Y(0)\mid G=1,X\}\mid G=1]\\
                                 &= E\{\delta(X)\mid G=1\}.
\end{align*}
Below we show an inverse probability weighting representation via the identification formula above.
For any \(h(x,d)\) such that \(E\{h(X,D)\mid X,G=0\}=1\), we have
\begin{align*}
  \MoveEqLeft \frac{1}{\alpha}E\bigg\{\frac{(1-G)\pi(X)}{1-\pi(X)}h(X,D)\frac{2A-1}{e(A,X,D)}Y\bigg\} \\
  &= \frac{1}{\alpha} E\bigg\{\frac{(1-G)\pi(X)}{1-\pi(X)}h(X,D)\frac{2A-1}{e(A,X,D)}\mu(A,X,D)\bigg\} \\
  &= \frac{1}{\alpha} E\bigg [\frac{(1-G)\pi(X)}{1-\pi(X)}h(X,D)\{\mu(1,X,D)-\mu(0,X,D)\}\bigg] \\
  &= \frac{1}{\alpha} E\bigg \{\frac{(1-G)\pi(X)}{1-\pi(X)}h(X,D)\delta(X)\bigg\} \\
  &= \frac{1}{\alpha} E\bigg \{\frac{(1-G)\pi(X)}{1-\pi(X)}\delta(X)\bigg\} \\
  &= \frac{1}{\alpha} E\{\pi(X)\delta(X)\} \\
  &= E\{\delta(X)\mid G=1\}.
\end{align*}
The alternative representation in the lemma is immediate.

\subsection{Proof of lemma~\ref{lem:eif}}

Consider the linear subspace of \(L_2^0(P_0)\)
\[
  \dot{\mathcal{P}} = \dot{\mathcal{P}}_{y}\oplus \dot{\mathcal{P}}_{a}\oplus \dot{\mathcal{P}}_{d} \oplus \dot{\mathcal{P}}_{g} \oplus \dot{\mathcal{P}}_{x},
\]
where \(\Lambda_1\oplus\Lambda_2\) denotes the direct sum of the spaces \(\Lambda_1\) and \(\Lambda_2\), and
\begin{align*}
  \dot{\mathcal{P}}_{y} &= \big\{(1-g)h(y,a,x,d):E_0\{h(Y,A,X,D)\mid A,X,D\}=0, \\
                        &\qquad E_0\{Yh(Y,A,X,D)\mid A=1,X,D=d\}-E_0\{Yh(Y,A,X,D)\mid A=0,X,D=d\}= \\
                        &\qquad E_0\{Yh(Y,A,X,D)\mid A=1,X,D=d'\}-E_0\{Yh(Y,A,X,D)\mid A=0,X,D=d'\}\big\}, \\
  \dot{\mathcal{P}}_{a} &= \big\{(1-g)h(a,x,d):E_0\{h(A,X,D)\mid X,D\}=0\big\}, \\
  \dot{\mathcal{P}}_{d} &= \big\{(1-g)h(d,x):E_0\{h(D,X)\mid X,G=0\}=0\big\}, \\
  \dot{\mathcal{P}}_{g} &= \big\{h(g,x):E_0\{h(G,X)\mid X\}=0\big\}, \\
  \dot{\mathcal{P}}_{x} &= \big\{h(x):E_0\{h(X)\}=0\big\}.
\end{align*}

The proof of lemma~\ref{lem:eif} consists of two parts.
In the first part (lemma~\ref{lem:orthogonal}), we show the orthocomplement of the space \(\dot{\mathcal{P}}\) and how to calculate the projection of functions in \(\dot{\mathcal{P}}_y\) onto this space.
In the second part (lemma~\ref{lem:eif}), we show that \(\dot{\mathcal{P}}\) is exactly the tangent space of the model \(\mathcal{P}\) at \(P_0\) under the local regularity conditions (assumption~\ref{asn:bounded}), and the construction of the tangent space depends on the projection results from lemma~\ref{lem:orthogonal}.

\begin{lemma}
  \label{lem:orthogonal}
  The linear space
  \[
    \Lambda_0= \bigg\{(1-g)q(x,d)\frac{2a-1}{e_0(a\mid x,d)}\{y-\mu_0(a,x,d)\}:E_0\{q(X,D)\mid X,G=0\}=0\bigg\}
  \]
  is the orthocomplement of \(\dot{\mathcal{P}}\) in \(L_2^0(P_0)\).
  Moreover, it is also the orthocomplement of \(\dot{\mathcal{P}}_y\) in \(\tilde{\dot{\mathcal{P}}}_{y}= \{(1-g)h(y,a,x,d):E_0\{h(Y,A,X,D)\mid A,X,D\}=0\}\).
  The projection of \((1-g)h\in\tilde{\dot{\mathcal{P}}}_y\) onto \(\Lambda\) is
  \[
    \Pi\{(1-g)h\mid\Lambda_0\} = (1-g)q_{h}(x,d)\frac{2a-1}{e_0(a\mid x,d)}\{y-\mu_0(a,x,d)\},
  \]
  where
  \begin{align*}
    r_{h}(x,d) &= w_0(x,d)\{\mathrm{cov}_0(Y,h\mid A=1,X=x,D=d)-\mathrm{cov}_0(Y,h\mid A=0,X=x,D=d)\},\\
    q_{h}(x,d) &= r_{h}(x,d)-w_0(x,d)\frac{E_0\{r_{h}(X,D)\mid X=x,G=0\}}{E_0\{w_0(X,D)\mid X=x,G=0\}}.
  \end{align*}
\end{lemma}

\begin{proof}
  It is trivial to see that \(\Lambda_0\subset\tilde{\dot{\mathcal{P}}}_{y}\). We will show that \(\Lambda_0\) is the orthocomplement of \(\dot{\mathcal{P}}_{y}\) in \(\tilde{\dot{\mathcal{P}}}_{y}\). Then, since the space \(L_2^0(P_0)\) decomposes as the direct sum of orthogonal subspaces
  \[
    \tilde{\dot{\mathcal{P}}}_y\oplus \dot{\mathcal{P}}_a \oplus \dot{\mathcal{P}}_d\oplus \dot{\mathcal{P}}_g\oplus \dot{\mathcal{P}}_x,
  \]
it follows that \(\Lambda_0\) is also the orthocomplement of \(\dot{\mathcal{P}}\) in \(L_2^0(P_0)\).
  
  We start by showing that any element in \(\Lambda_0\) is orthogonal to all elements in \(\dot{\mathcal{P}}_{y}\). For any
  \begin{align*}
    \lambda_{q}(y,a,x,d,g)&=(1-g)\frac{2a-1}{e_0(a\mid x,d)}q(x,d)\{y-\mu_0(a,x,d)\}\in\Lambda_0, \\ \ell_{h}(y,a,x,d,g)&=(1-g)h(y,a,x,d)\in\dot{\mathcal{P}}_y,
  \end{align*}
  their inner product is
  \begin{align*}
    \MoveEqLeft\langle\lambda_{q}(Y,A,X,D,G),\ell_h(Y,A,X,D,G)\rangle_{L_{2}(P_0)} \\
    &= E_0\bigg[(1-G)\frac{2A-1}{e_0(A\mid X,D)}q(X,D)\{Y-\mu_0(A,X,D)\}h(Y,A,X,D)\bigg] \\
    &= E_0\bigg[(1-G)\frac{2A-1}{e_0(A\mid X,D)}q(X,D)Yh(Y,A,X,D)\bigg] \\
    &= E_0\bigg[(1-G)\frac{2A-1}{e_0(A\mid X,D)}q(X,D)E_0\{Yh(Y,A,X,D)\mid A,X,D\}\bigg], \\
    \intertext{and if we integrate over \(A\) and \(D\) conditionally on \(G=0\) and \(X\) inside the outermost expectation, the inner product is}
    &= E_0\bigg[(1-G)\sum_{d\in[m]}\zeta_0(d\mid X)q(X,d)\kappa_h(X)\bigg], \\
    \intertext{where \(\kappa_h(x)=E_0\{Yh(Y,A,X,D)\mid A=1,X=x,D=d\}-E_0\{Yh(Y,A,X,D)\mid A=0,X=x,D=d\}\) does not depend on the value of \(d\), so}
    &= E_0\big[(1-G)\kappa_h(X)E_0\{q(X,D)\mid X,G=0\}\big] =0.
  \end{align*}
  It follows that \(\Lambda_0\perp\dot{\mathcal{P}}_y\), or equivalently stated, \(\Lambda_0\subset\dot{\mathcal{P}}_y^{\perp}\). It remains to show the opposite direction, \(\dot{\mathcal{P}}_y^{\perp}\subset\Lambda_0\), or equivalently, \(\Lambda_0^{\perp}\subset\dot{\mathcal{P}}_y\). Every element \(\tilde\ell_{h}=(1-g)h(y,a,x,d)\in\tilde{\dot{\mathcal{P}}}_y\) can be written as
  \[
    \tilde\ell_h = \Pi\{\tilde\ell_h\mid \Lambda_0\} + \Pi\{\tilde\ell_h\mid \Lambda_0^\perp\}.
  \]
We will verify that \(\Pi\{\tilde\ell_h\mid \Lambda_0^\perp\}\in\dot{\mathcal{P}}_y\) in a sequence of steps. First we find \(\Pi\{\tilde\ell_h\mid \Lambda_0\}\), establishing the form presented in the lemma, by an intermediate step where \(\tilde\ell_h\) is projected onto a space larger than \(\Lambda_0\) and then the projection is further projected onto \(\Lambda_0\). Finally, we verify by direct calculations that \(\Pi\{\tilde\ell_h\mid \Lambda_0^\perp\}=\tilde\ell_h-\Pi\{\tilde\ell_h\mid \Lambda_0\}\) is an element of \(\dot{\mathcal{P}}_y\). 

Consider the linear subspace, larger than \(\Lambda_0\),
  \[
    \tilde{\Lambda}= \bigg\{(1-g)\frac{2a-1}{e_0(a\mid x,d)}r(x,d)\{y-\mu_0(a,x,d)\}:r(x,d)\text{ arbitrary}\bigg\} \subset L_2^0(P_0).
  \]
For any function \(r(x,d)\), the projection of
  \[
    \tilde{\lambda}_{r}(y,a,x,d,g)= (1-g)\frac{2a-1}{e_0(a\mid x,d)}r(x,d)\{y-\mu_0(a,x,d)\}\in\tilde{\Lambda}
  \]
  onto the subspace \(\Lambda_0\) is
  \begin{multline*}
    \Pi\{\tilde{\lambda}_r\mid \Lambda\}=(1-g)\frac{2a-1}{e_0(a\mid x,d)}\bigg[r(x,d)-w_0(x,d)\frac{E_0\{r(X,D)\mid X=x,G=0\}}{E_0\{w_0(X,D)\mid X=x,G=0\}}\bigg]\\
    \{y-\mu_0(a,x,d)\}.
  \end{multline*}
  The legitimacy of the projection can be established by checking for every \(\lambda_q\in\Lambda_0\), the inner product
  \begin{align*}
    \MoveEqLeft\langle\tilde{\lambda}_r-\Pi\{\tilde{\lambda}_r\mid \Lambda\},\lambda_q(Y,A,X,D,G)\rangle_{L_{2}(P_0)} \\
    &= E_0\bigg[(1-G)\frac{\{Y-\mu_0(A,X,D)\}^{2}}{\{e_0(A\mid X,D)\}^{2}}w_0(X,D)\frac{E_0\{r(X,D)\mid X,G=0\}}{E_0\{w_0(X,D)\mid X,G=0\}}q(X,D)\bigg] \\
    &= E_0\bigg[(1-G)\frac{V_0(A, X,D)}{\{e_0(A\mid X,D)\}^{2}}w_0(X,D)\frac{E_0\{r(X,D)\mid X,G=0\}}{E_0\{w_0(X,D)\mid X,G=0\}}q(X,D)\bigg] \\
    &= E_0\bigg[(1-G)\frac{E_0\{r(X,D)\mid X,G=0\}}{E_0\{w_0(X,D)\mid X,G=0\}}q(X,D)\bigg] \\
    &= E_0\bigg[(1-G)\frac{E_0\{r(X,D)\mid X,G=0\}}{E_0\{w_0(X,D)\mid X,G=0\}}E_0\{q(X,D)\mid X,G=0\}\bigg] \\
    &= 0,
  \end{align*}
  and that indeed
  \[
    E_0\bigg[r(X,D)-w_0(X,D)\frac{E_0\{r(X,D)\mid X,G=0\}}{E_0\{w_0(X,D)\mid X,G=0\}}\biggm\vert X,G=0\bigg]=0.
  \]

  Take an arbitrary element \(\tilde\ell_h\in\tilde{\dot{\mathcal{P}}}_{y}\).
  Then suppose the projection onto \(\tilde\Lambda\) is
  \[
    \Pi\{\tilde\ell_h\mid \tilde{\Lambda}\}=(1-g)\frac{2a-1}{e_0(a\mid x,d)}r_{h}(x,d)\{y-\mu_0(a,x,d)\},
  \]
  so that \(r_{h}(x,d)\) fulfills the equation
  \begin{multline*}
    E_0\bigg(\frac{2A-1}{e_0(A\mid X,D)}\{Y-\mu_0(A,X,D)\}\\
    \bigg[h(Y,A,X,D)-\frac{2A-1}{e_0(A\mid X,D)}r_{h}(X,D)\{Y-\mu_0(A,X,D)\}\bigg]\,\bigg\vert\, X,D\bigg)=0.
  \end{multline*}
  Direct calculation yields the solution
  \begin{multline*}
    r_{h}(x,d) = w_0(x,d)\big[E_0\{Yh(Y,A,X,D)\mid A=1,X=x,D=d\}\\
    -E_0\{Yh(Y,A,X,D)\mid A=0,X=x,D=d\}\big]\\
    =w_0(x,d)\{\mathrm{cov}_0(Y,h\mid A=1,X=x,D=d)-\mathrm{cov}_0(Y,h\mid A=0,X=x,D=d)\}.
  \end{multline*}
  Then further projecting \(\Pi\{\tilde\ell_h\mid\tilde{\Lambda}\}\) onto \(\Lambda_0\), we determine that
  \[
    \Pi\{\tilde\ell_h\mid\Lambda_0\} = (1-g)\frac{2a-1}{e_0(a\mid x,d)}q_{h}(x,d)\{y-\mu_0(a,x,d)\},
  \]
  where
  \[
    q_{h}(x,d)= r_{h}(x,d)-w_0(x,d)\frac{E_0\{r_{h}(X,D)\mid X=x,G=0\}}{E_0\{w_0(X,D)\mid X=x,G=0\}}.
  \]
  In the following, we verify that
  \[
    \Pi\{\tilde\ell_h\mid \Lambda_0^{\perp}\} =(1-g)s_{h}(y,a,x,d),
  \]
  where
  \[
    s_{h}(y,a,x,d)= h(y,a,x,d)-\frac{2a-1}{e_0(a\mid x,d)}q_{h}(x,d)\{y-\mu_0(a,x,d)\},
  \]
  is indeed an element of \(\dot{\mathcal{P}}_{y}\).
  It is trivial that \(E_0\{s_{h}(Y,A,X,D)\mid A,X,D\}=0\).
  Furthermore,
  \begin{align*}
    \MoveEqLeft E_0\{Ys_{h}\mid A=1,X=x,D=d\} - E_0\{Ys_{h}\mid A=0,X=x,D=d\} \\
    &= E_0\{Yh(Y,A,X,D)\mid A=1,X=x,D=d\}-E_0\{Yh(Y,A,X,D)\mid A=0,X=x,D=d\} \\
    &\hphantom{=}\quad- q_{h}(x,d)\sum_{a\in\{0,1\}}\frac{E_0[Y\{Y-\mu_0(a,X,D)\}\mid A=a,X=x,D=d]}{e_0(a\mid x,d)} \\
    &= E_0\{Yh(Y,A,X,D)\mid A=1,X=x,D=d\}-E_0\{Yh(Y,A,X,D)\mid A=0,X=x,D=d\} \\
    &\hphantom{=}\quad- \frac{q_{h}(x,d)}{w_0(x,d)} \\
    &= \frac{E_0\{r_{h}(X,D)\mid X=x,G=0\}}{E_0\{w_0(X,D)\mid X=x,G=0\}}
  \end{align*}
  is constant in the level of \(d\).
  This ascertains that \(\Pi\{\tilde\ell_h\mid \Lambda_0^{\perp}\}\in\dot{\mathcal{P}}_{y}\), and the proof is complete
\end{proof}

\begin{proof}[Proof of lemma~\ref{lem:eif}]
  The observed data distribution is
  \begin{align*}
    p_0(y,a,x,d,g) &= \big[p_0(y\mid a,x,d)e_0(a\mid x,d)\zeta_0(d\mid x)\{1-\pi_0(x)\}\big]^{(1-g)}\{\pi_0(x)\}^{g}p_0(x).
  \end{align*}

  \paragraph{Describing the structure of the maximal tangent space}
  We claim that the tangent space of the model \(\mathcal{P}\) at \(P_0\) is \(\dot{\mathcal{P}}\).
  To see that the tangent space must have this structure, we consider an arbitrary, smooth one-dimensional submodel \(\{P_{\varepsilon}\}\subset\mathcal{P}\) such that \(P_\varepsilon\vert_{\varepsilon=0}=P_0\), whose score function at \(P_0\) is
  \[
    \frac{\dfrac{\d}{\d\varepsilon}p_{\varepsilon}(y,a,x,d,g)\bigg\vert_{\varepsilon=0}}{p_0(y,a,x,d,g)}=h(y,a,x,d,g)\in L_{2}^0(P_0).
  \]
  By the structure of the observed data density, the score function must be decomposable as the sum
  \[
    h(y,a,x,d,g)=(1-g)\{h(y,a,x,d)+h(a,x,d)+h(d,x)\}+h(g,x)+h(x)
  \]
  such that
  \begin{align*}
    E_0\{h(Y,A,X,D)\mid A=a,X=x,D=d\}&=0,\\
    E_0\{h(A,X,D)\mid X=x,D=d\}&=0,\\
    E_0\{h(D,X)\mid X=x,G=0\}&=0,\\
    E_0\{h(G,X)\mid X=x\}&=0,\\
    E_0\{h(X)\}&=0.
  \end{align*}
  The restriction on the tangent space \(\dot{\mathcal{P}}_{y}\) comes from the conditional moment restriction on the observed data distribution that \(E_0(Y\mid A=1,X=x,D=d)-E_0(Y\mid A=0,X=x,D=d)\) does not depend on \(d\); that is,
  \[
    \int y \big\{p_0(y\mid 1,x,d)-p_0(y\mid 0,x,d)\big\}\d y
  \]
  is constant in \(d\).
  Differentiating the term
  \[
    \int y \big\{p_\varepsilon(y\mid 1,x,d)-p_\varepsilon(y\mid 0,x,d)\big\}\d y,
  \]
  with respect to \(\varepsilon\), we have that
  \[
    \int y \big\{p_0(y\mid 1,x,d)h(y,A=1,x,D=d)-p_0(y\mid 0,x,d)h(y,A=0,x,D=d)\big\}\d y
  \]
  is constant in \(d\).
  Equivalently, defining \(\upsilon_{h}(a,x,d)=E_0\{Yh(Y,A,X,D)\mid A=a,X=x,D=d\}\),
  \begin{equation}
    \label{eqn:product-score-y}
    \upsilon_h(1,x,d)-\upsilon_h(0,x,d)
  \end{equation}
  is constant in \(d\).
  We also have \(\kappa_h(x)= \upsilon_h(1,x,d)-\upsilon_h(0,x,d)\) for any \(x\).

  \paragraph{Constructing the maximal tangent space}
  We now show that we can construct parametric submodels, so that the closed linear span of the scores of these submodels is exactly \(\dot{\mathcal{P}}\).
  The construction of \(\dot{\mathcal{P}}_{\dott}\) is standard for \(\dott\in\{a,x,d,g\}\) and omitted here.
  To construct the tangent subspace \(\dot{\mathcal{P}}_{y}\), we only need to find parametric submodels whose score functions form a dense subset of \(\dot{\mathcal{P}}_y\).
  For any \(\ell_{h}=(1-g)h(y,a,x,d)\in\dot{\mathcal{P}}_y\), consider the bounded version
  \[
    \ell_{h,M}(y,a,x,d,g)=(1-g)\big[h_M-E_0\{h_M\mid A=a,X=x,D=d\}\big]\in\tilde{\dot{\mathcal{P}}}_y,
  \]
  where \(h_{M}=hI(|h|\leq M)\), for some finite \(M\).
  Then because \(\ell_{h,M}\in\tilde{\dot{\mathcal{P}}}_y\), its projection onto \(\dot{\mathcal{P}}_y\) is its projection onto \(\Lambda^\perp\).
  By the second part of lemma~\ref{lem:orthogonal}, this is
  \begin{multline}
    \ell^{\perp}_{h,M}(y,a,x,d,g)=\Pi\{\ell_{h,M}\mid \dot{\mathcal{P}}_y\}=\Pi\{\ell_{h,M}\mid \Lambda^\perp\}\\
    =(1-g)\bigg[\{h_M-E_0(h_M\mid A=a,X=x,D=d)\}\\
    -\frac{2a-1}{e_0(a\mid x,d)}q_{h,M}(x,d)\{y-\mu_0(a,x,d)\}\bigg],
  \end{multline}
  where
  \begin{align*}
    r_{h,M}(x,d) &= \begin{multlined}[t][.8\textwidth]
      w_0(x,d)\big[\mathrm{cov}_{P_0}(Y,h_M\mid A=1,X=x,D=d)\\
      -\mathrm{cov}_{P_0}(Y,h_M\mid A=0,X=x,D=d)\big],
    \end{multlined}\\
    q_{h,M}(x,d) &= r_{h,M}(x,d)-w_0(x,d)\frac{E_0\{r_{h,M}(X,D)\mid X=x,G=0\}}{E_0\{w_0(X,D)\mid X=x,G=0\}}.
  \end{align*}

  Assumption~\ref{asn:bounded} implies that \(V_0(a,x,d)\lesssim 1\), \(P_0\)-almost surely.
  We bound the following quantities:
  \begin{align}
    |r_{h,M}(x,d)| &\leq w_0(x,d)\sum_{a\in\{0,1\}}V_0^{1/2}(a,x,d)\{\mathrm{var}_{P_0}(h_M\mid A=a,X=x,D=d)\}^{1/2} \nonumber\\
                   &\leq w_0(x,d)\sum_{a\in\{0,1\}}V_0^{1/2}(a,x,d)\{E_0(h_M^2\mid A=a,X=x,D=d)\}^{1/2} \nonumber\\
                   &\lesssim_{M} 1, \label{eqn:bound-r}
  \end{align}
  so that \(|q_{h,M}(x,d)| \lesssim_{M} 1\) and thus \(|\ell^{\perp}_{h,M}| \lesssim_{M} 1\) is bounded by some constant dependent on \(M\).
  Then consider the parametric submodel \(\{P_{\varepsilon}(h,M):\varepsilon\in\Gamma\}\) with density
  \[
    p_\varepsilon(y,a,x,d,g) = \big\{p_\varepsilon(y\mid a,x,d)p_\varepsilon(a\mid x,d)p_\varepsilon(d\mid x)\big\}^{(1-g)}p_\varepsilon(g\mid x)p_\varepsilon(x),
  \]
  with
  \begin{align*}
    p_\varepsilon(y\mid a,x,d) &= p_0(y\mid a,x,d)\{1+\varepsilon\ell^{\perp}_{h,M}(y,x,a,d,0)\},\\
    p_\varepsilon(a\mid x,d) &= \frac{\chi\{\varepsilon h(a,x,d)\}e_0(a\mid x,d)}{\sum_{a'}\chi\{\varepsilon h(a',x,d)\}e_0(a'\mid x,d)},\\
    p_\varepsilon(d\mid x) &= \frac{\chi\{\varepsilon h(d,x)\}\zeta_0(d\mid x)}{\sum_{d'}\chi\{\varepsilon h(d',x)\}\zeta_0(d'\mid x)},\\
    p_\varepsilon(g\mid x) &=\frac{\chi\{\varepsilon h(g,x)\}\{\pi_0(x)\}^{g}\{1-\pi_0(x)\}^{(1-g)}}{\sum_{g'}\chi\{\varepsilon h(g',x)\}\{\pi_0(x)\}^{g'}\{1-\pi_0(x)\}^{(1-g')}},\\
    p_\varepsilon(x) &=\frac{\chi\{\varepsilon h(x)\}p_0(x)}{\int\chi\{\varepsilon h(x')\}p_0(x')\d x'},
  \end{align*}
  and
  \begin{align*}
    E_0\{h(Y,A,X,D)\mid A=a,X=x,D=d\}&=0,\\
    E_0\{h(A,X,D)\mid X=x,D=d\}&=0,\\
    E_0\{h(D,X)\mid X=x,G=0\}&=0,\\
    E_0\{h(G,X)\mid X=x\}&=0,\\
    E_0\{h(X)\}&=0,
  \end{align*}
  where \(\chi(x)=2\{1+\exp(-2x)\}^{-1}\) (see, for example, \citealpsuppmat[p. 53]{bickel1993efficient}), \(\Gamma\) is an open neighborhood around \(0\) such that \(p_\varepsilon(o)\geq 0\).
  Such a set \(\Gamma\) exists because \(\ell^{\perp}_{h,M}\) is bounded.
  We verify that \(\{P_{\varepsilon}(h,M):\varepsilon\in\Gamma\}\subset\mathcal{P}\).
  It is obvious that \(P_{0}(h,M)=P_0\).
  Additionally, for any \(x\),
  \[
    E_{P_\varepsilon}(Y\mid A=1,X=x,D=d)-E_{P_\varepsilon}(Y\mid A=0,X=x,D=d)
  \]
  does not vary with \(d\), because
  \begin{align*}
    \MoveEqLeft E_{P_\varepsilon}(Y\mid A=1,X=x,D=d)-E_{P_\varepsilon}(Y\mid A=0,X=x,D=d) \\
    &= \{E_0(Y\mid A=1,X=x,D=d)-E_0(Y\mid A=0,X=x,D=d)\}\\
    &\hphantom{=}\quad +\varepsilon \{E_0(Y\ell^{\perp}_{h,M}\mid A=1,X=x,D=d)-E_0(Y\ell^{\perp}_{h,M}\mid A=0,X=x,D=d)\}
  \end{align*}
  is a quantity not depending on \(d\).

  Furthermore, \(\{\ell^{\perp}_{h,M}:\ell_{h}\in\dot{\mathcal{P}}_y, M<\infty\}\) is dense in \(\dot{\mathcal{P}}_y\) in the \(L_2(P_0)\)-sense, which we show below.
  We bound the \(L_2(P_0)\)-distance between \(\ell_h\) and \(\ell^\perp_{h,M}\) by
  \begin{align}
    \|\ell_{h}-\ell^{\perp}_{h,M}\|_{P_0} &\leq \|(1-G)(h-h_M)\|_{P_0} \label{eqn:dense-1}\\
                                          &\hphantom{\leq}\quad + \|(1-G)E_0(h_M\mid A,X,D)\|_{P_0}\label{eqn:dense-2}\\
                                          &\hphantom{\leq}\quad +\bigg\|(1-G)\frac{2A-1}{e_0(A\mid X,D)}q_{h,M}(X,D)\{Y-\mu_0(A,X,D)\}\bigg\|_{P_0}.\label{eqn:dense-3}
  \end{align}
  We argue that every term in the display above tends to zero as \(M\) tends to infinity.
  The limit of the norm \eqref{eqn:dense-1} is zero because bounded functions are dense in \(L_2(P_0)\).
  We have \(|h_M|\leq |h|\), so by dominated convergence, \(\lim_{M\uparrow\infty}E_0(h_{M}\mid A=a,X=x,D=d)=0\), because \(\lim_{M\uparrow\infty}h_M=h\) and \(E_0(h\mid A=a,X=x,D=d)=0\) by definition.
  Since
  \begin{multline*}
    \{E_0(h_M\mid A=a,X=x,D=d)\}^2\leq E_0(h_M^2\mid A=a,X=x,D=d)\\
    \leq E_0(h^2\mid A=a,X=x,D=d),
  \end{multline*}
  \(P_0\)-almost surely and \((1-g)h\in L_2(P_0)\), dominated convergence now shows that \(\lim_{M\uparrow\infty}\|(1-G)E_0(h_M\mid A,X,D)\|_{P_0}=0\), so the limit of \eqref{eqn:dense-2} is zero.

  Since
  \begin{multline*}
    |\{y-\mu_0(a,x,d)\}\{h_M-E_0(h_M\mid A=a,X=x,D=d)\}|\\
    \lesssim |h_M|+E_0(|h_M|\mid A=a,X=x,D=d)|\lesssim |h|+E_0(|h|\mid A=a,X=x,D=d)
  \end{multline*}
  and the rightmost function is integrable with respect to \(P_0(Y\mid A=a,X=x,D=d)\), \(P_0(A,X,D\mid G=0)\)-almost surely, we have
  \[
    \lim_{M\uparrow\infty}\mathrm{cov}_{P_0}(Y,h_M\mid A=a,X=x,D=d)=E_0(Yh\mid A=a,X=x,D=d)
  \]
  from \(\lim_{M\uparrow\infty}\{h_M-E_0(h_M\mid A=a,X=x,D=d)\}=h\).
  Then we have
  \begin{equation}
    \lim_{M\uparrow\infty}E_0\{r_{h,M}(X,D)\mid X=x,G=0\}=\kappa_h(x)E_0\{w_0(X,D)\mid X=x,G=0\}.
    \label{eq:r-bound}
  \end{equation}
  Moreover,
  \begin{align*}
    r_{h,M}^2(x,d) &= \begin{multlined}[t][.8\textwidth]
      w_0^2(x,d)[\mathrm{cov}_{P_0}(Y,h_{M}\mid A=1,X=x,D=d)\\
      -\mathrm{cov}_{P_0}(Y,h_{M}\mid A=0,X=x,D=d)]^2
    \end{multlined}
    \\
                   &\lesssim \sum_{a\in\{0,1\}}\mathrm{cov}_{P_0}^2(Y,h_{M}\mid A=a,X=x,D=d)\\
                   &\leq \sum_{a\in\{0,1\}}V_0(a,x,d)E_0(h_M^2\mid A=a,X=x,D=d)\\
                   &\lesssim \sum_{a\in\{0,1\}}E_0(h^2\mid A=a,X=x,D=d) \\
                   &\lesssim \sum_{a\in\{0,1\}}e_0(a\mid x,d)E_0(h^2\mid A=a,X=x,D=d) \\
                   &=E_0(h^2\mid X=x,D=d),
  \end{align*}
  so that
  \begin{align*}
    q_{h,M}^2(x,d) &= \bigg[r_{h,M}(x,d)-w_0(x,d)\frac{E_0\{r_{h,M}(X,D)\mid X=x,G=0\}}{E_0\{w_0(X,D)\mid X=x,G=0\}}\bigg]^2 \\
                   &\leq 2r_{h,M}^2(x,d)+2w_0^2(x,d)\bigg[\frac{E_0\{r_{h,M}(X,D)\mid X=x,G=0\}}{E_0\{w_0(X,D)\mid X=x,G=0\}}\bigg]^2 \\
                   &\lesssim E_0(h^2\mid X=x,D=d)+E_0\{r_{h,M}^2(X,D)\mid X=x,G=0\} \\
                   &\lesssim E_0(h^2\mid X=x,D=d)+E_0(h^2\mid X=x,G=0) \in L_1(P_0).
  \end{align*}
  Expression \eqref{eqn:dense-3} can be bounded by \(O(\|(1-G)q_{h,M}(X,D)\|_{P_0})\).
  Another application of dominated convergence yields \(\lim_{M\uparrow\infty}\|(1-G)q_{h,M}(X,D)\|_{P_0}=0\), because by \eqref{eq:r-bound},
  \[
    \lim_{M\uparrow\infty}q_{h,M}^2(x,d)=\Big\{\lim_{M\uparrow\infty}q_{h,M}(x,d)\Big\}^2=0,
  \]
  so the limit of \eqref{eqn:dense-3} is zero.
  The denseness follows from \(\lim_{M\uparrow\infty}\|\ell_{h}-\ell^{\perp}_{h,M}\|_{P_0}=0\) for any \(\ell_{h}\in\dot{\mathcal{P}}_y\).

  Therefore, the closed linear span of \(\{\ell^{\perp}_{h,M}:\ell_{h}\in\dot{\mathcal{P}}_y, M<\infty\}\) is exactly \(\dot{\mathcal{P}}_{y}\).

  \paragraph{Calculating the pathwise derivative}
  We next compute the pathwise derivative of \(\theta_\varepsilon\), the target parameter on \(P_{\varepsilon}\), along the submodel \(\{P_{\varepsilon}:\varepsilon\in\Gamma\}\) at \(P_0\).
  To express the target parameter as a function of only the observed data and its density, we notice that
  \begin{align*}
    \theta_\varepsilon &= E_{P_{\varepsilon}}\{\delta_{\varepsilon}(X)\mid G=1\} \\
                       &= \frac{1}{\alpha_{\varepsilon}}\int \delta_{\varepsilon}(x) \pi_{\varepsilon}(x)p_{\varepsilon}(x)\d x \\
                       &= \frac{1}{\alpha_{\varepsilon}}\int\sum_{d\in[m]}\zeta_{\varepsilon}(d\mid x)\delta_{\varepsilon}(x) \pi_{\varepsilon}(x)p_{\varepsilon}(x)\d x \\
                       &= \frac{1}{\alpha_{\varepsilon}}\int\sum_{d\in[m]}\zeta_{\varepsilon}(d\mid x)\{\mu_{\varepsilon}(1,x,d)-\mu_{\varepsilon}(0,x,d)\} \pi_{\varepsilon}(x)p_{\varepsilon}(x)\d x \\
                       &= \frac{1}{\alpha_{\varepsilon}}\int\sum_{d\in[m]}\zeta_{\varepsilon}(d\mid x)\int y\{p_{\varepsilon}(y\mid 1,x,d)-p_{\varepsilon}(y\mid 0,x,d)\}\d y \pi_{\varepsilon}(x)p_{\varepsilon}(x)\d x.
  \end{align*}

  The pathwise derivative of \(\theta_{\varepsilon}\) evaluated at the true model is
  \begin{align}
    \MoveEqLeft\frac{\d}{\d\varepsilon}\theta_{\varepsilon}\bigg\vert_{\varepsilon=0} \nonumber\\
    &= \frac{\d}{\d\varepsilon}\frac{\int\sum_{d\in[m]}\zeta_\varepsilon(d\mid x)\int y\{p_\varepsilon(y\mid 1,x,d)-p_\varepsilon(y\mid 0,x,d)\}\d y \pi_\varepsilon(x)p_\varepsilon(x)\d x}{\int\pi_\varepsilon(x)p_\varepsilon(x)\d x}\bigg\vert_{\varepsilon=0}, \nonumber\\
    \intertext{which by the product rule is}
    &= \frac{1}{\alpha_0}\frac{\d}{\d\varepsilon}\int\sum_{d\in[m]}\zeta_\varepsilon(d\mid x)\int y\{p_\varepsilon(y\mid 1,x,d)-p_\varepsilon(y\mid 0,x,d)\}\d y \pi_\varepsilon(x)p_\varepsilon(x)\d x\bigg\vert_{\varepsilon=0} \nonumber\\
    &\qquad - \frac{\theta_0}{\alpha_0}\frac{\d}{\d\varepsilon}\int\pi_\varepsilon(x)p_\varepsilon(x)\d x\bigg\vert_{\varepsilon=0} \nonumber\\
    &=\begin{multlined}[t][.8\textwidth]
      \frac{1}{\alpha_0}\int\sum_{d\in[m]}\zeta_0(d\mid x)\int y\bigg(\frac{\d}{\d\varepsilon}\{p_0(y\mid 1,x,d;\varepsilon)-p_0(y\mid 0,x,d;\varepsilon)\}\bigg\vert_{\varepsilon=0}\bigg)\\
      \d y \pi_0(x)p_0(x)\d x
    \end{multlined}\label{eqn:pathwise-1}\\
    &\hphantom{=}\quad+ \frac{1}{\alpha_0}\int\frac{\d}{\d\varepsilon}\sum_{d\in[m]}\zeta_\varepsilon(d\mid x)\bigg\vert_{\varepsilon=0}\delta_0(x)\pi_0(x)p_0(x)\d x \label{eqn:pathwise-2}\\
    &\hphantom{=}\quad+ \frac{1}{\alpha_0}\int \delta_0(x) \bigg(\frac{\d}{\d\varepsilon}\pi_\varepsilon(x)p_\varepsilon(x)\bigg\vert_{\varepsilon=0}\bigg)\d x \label{eqn:pathwise-3}\\
    &\hphantom{=}\quad - \frac{\theta_0}{\alpha_0}\int\bigg(\frac{\d}{\d\varepsilon}\pi_\varepsilon(x)p_\varepsilon(x)\bigg\vert_{\varepsilon=0}\bigg)\d x. \label{eqn:pathwise-4}
  \end{align}

  We study the expressions separately.
  We have
  \begin{align*}
    \eqref{eqn:pathwise-1} &= \frac{1}{\alpha_0}\int\sum_{d\in[m]}\zeta_0(d\mid x) \int y \big\{p_0(y\mid 1,x,d)h(y,A=1,x,D=d) \\
                           &\hphantom{=\frac{1}{\alpha_0}\int\sum_{d\in[m]}\zeta_0(d\mid x) \int y \big\{}\quad -p_0(y\mid 0,x,d)h(y,A=0,x,D=d)\big\}\d y \pi_0(x)p_0(x)\d x \\
                           &= \frac{1}{\alpha_0}\int\sum_{d\in[m]}\zeta_0(d\mid x) \kappa_h(x) \pi_0(x)p_0(x)\d x \\
                           &= \frac{1}{\alpha_0}\int \kappa_h(x)\pi_0(x)p_0(x)\d x, \\
    \eqref{eqn:pathwise-2} &= \frac{1}{\alpha}\int\bigg(\frac{\d}{\d\varepsilon}1\bigg\vert_{\varepsilon=0}\bigg)\delta_0(x)\pi_0(x)p_0(x)\d x =0.
  \end{align*}
  We observe that
  \begin{align*}
    \frac{\d}{\d\varepsilon}\pi_\varepsilon(x)p_\varepsilon(x)\bigg\vert_{\varepsilon=0} &= \pi_0(x)p_0(x)\{h(G=1,x)+h(x)\},
  \end{align*}
  and therefore
  \begin{align*}
    \eqref{eqn:pathwise-3}+\eqref{eqn:pathwise-4} &= \frac{1}{\alpha_0}\int \{\delta_0(x)-\theta_0\}\pi_0(x)p_0(x)\{h(G=1,x)+h(x)\}\d x.
  \end{align*}
  Collecting the terms, the pathwise derivative is
  \[
    \frac{\d}{\d\varepsilon}\theta_{\varepsilon}\bigg\vert_{\varepsilon=0} =  \frac{1}{\alpha_0}\int \kappa_h(x)\pi_0(x)p_0(x)\d x + \frac{1}{\alpha_0}\int \{\delta_0(x)-\theta_0\}\pi_0(x)p_0(x)\{h(G=1,x)+h(x)\}\d x
  \]

  \paragraph{Finding the efficient influence function}
  We claim that the efficient influence function of \(\theta_0\) in the model \(\mathcal{P}\) is as displayed in lemma~\ref{lem:eif}.

  The inner product of \(\varphi(o,w_0)\) and any score \(h(o)\in\dot{\mathcal{P}}\) of the model \(\mathcal{P}\) at \(P_0\) is
  \begin{align}
    \MoveEqLeft E_0\{\varphi(O,w_0)h(O)\} \nonumber\\
    &= \begin{multlined}[t][.8\textwidth]E_0\bigg[\frac{(1-G)(2A-1)\pi_0(X)}{\alpha_0 e_0(A\mid X,D)\{1-\pi_0(X)\}}\frac{w_0(X,D)}{\sum_{d\in[m]}\zeta_0(d\mid X)w_0(X,d)}\\
      \{Y-\mu_0(A,X,D)\}h(Y,A,X,D)\bigg]
    \end{multlined}\label{eqn:innerproduct-1}\\
    &\hphantom{=}\quad + E_0\bigg[\frac{G}{\alpha_0}\{\delta_0(X)-\theta_0\}\{h(G,X)+h(X)\}\bigg]. \label{eqn:innerproduct-2}
  \end{align}

  Since the score \(h(y,a,x,d)\) must satisfy \(E_0\{h(Y,A,X,D)\mid A,X,D\}=0\), it follows that
  \begin{align*}
    \eqref{eqn:innerproduct-1} &= E_0\bigg\{\frac{(1-G)(2A-1)\pi_0(X)}{\alpha_0 e_0(A\mid X,D)\{1-\pi_0(X)\}}\frac{w_0(X,D)}{\sum_{d\in[m]}\zeta_0(d\mid X)w_0(X,d)}Yh(Y,A,X,D)\bigg\} & \\
                               &= E_0\bigg [\frac{(1-G)(2A-1)\pi_0(X)}{\alpha_0 e_0(A\mid X,D)\{1-\pi_0(X)\}}\frac{w_0(X,D)}{\sum_{d\in[m]}\zeta_0(d\mid X)w_0(X,d)}\upsilon_h(A,X,D)\bigg ] \\
                               &= E_0\bigg [\frac{(1-G)\pi_0(X)}{\alpha_0\{1-\pi_0(X)\}}\frac{w_0(X,D)}{\sum_{d\in[m]}\zeta_0(d\mid X)w_0(X,d)} \{\upsilon_h(1,X,D)-\upsilon_h(0,X,D)\}\bigg]\\
                               &= E_0\bigg \{\frac{(1-G)\pi_0(X)}{\alpha_0\{1-\pi_0(X)\}}\kappa_h(X)E_0\bigg(\frac{w_0(X,D)}{\sum_{d\in[m]}\zeta_0(d\mid X)w_0(X,d)}\biggm |X,G=0\bigg)\bigg\}\\
                               &= E_0\bigg \{\frac{(1-G)\pi_0(X)}{\alpha_0\{1-\pi_0(X)\}} \kappa_h(X)\bigg\} \\
                               &= E_0\bigg \{\frac{\pi_0(X)}{\alpha_0} \kappa_h(X)\bigg\} \\
                               &= \frac{1}{\alpha_0}\int \kappa_h(x) \pi_0(x)p_0(x)\d x. 
  \end{align*}

  On the other hand, we have
  \begin{align*}
    \eqref{eqn:innerproduct-2} &= E_0\bigg[\frac{G}{\alpha_0}\{\delta_0(X)-\theta_0\}\{h(G,X) + h(X)\}\bigg] \\
                               &= \frac{1}{\alpha_0}\int \{\delta_0(x)-\theta_0\}\{h(G=1,x)+h(x)\}\pi_0(x)p_0(x)\d x.
  \end{align*}
  Therefore, it is clear that
  \[
    E_0\{\varphi(O,w_0)h(O)\} = \eqref{eqn:innerproduct-1} + \eqref{eqn:innerproduct-2} = \frac{\d}{\d\varepsilon}\theta_{\varepsilon}\bigg\vert_{\varepsilon=0},
  \]
  so \(\varphi(o,w_0)\) is an influence function.
  To show that \(\varphi(o,w_0)\) is the efficient influence function, it remains to verify that \(\varphi(o,w_0)\) lies in the tangent space \(\dot{\mathcal{P}}\).

  The function \(\varphi(o,w_0)\) can be decomposed into the sum
  \[
    \varphi(o,w_0) = (1-g)h^*(y,a,x,d) + h^*(g,x) + h^*(x),
  \]
  where 
  \begin{align*}
    h^*(y,a,x,d) &= \frac{(2a-1)\pi_0(x)}{\alpha_0 e_0(a\mid x,d)\{1-\pi_0(x)\}}\frac{w_0(x,d)}{\sum_{d'\in[m]}\zeta_0(d'\mid x)w_0(x,d')}\{y-\mu_0(a,x,d)\}, \\
    h^*(g,x) &= \frac{g-\pi_0(x)}{\alpha_0}\{\delta_0(x)-\theta_0\}, \\
    h^*(x) &= \frac{\pi_0(x)}{\alpha_0}\{\delta_0(x)-\theta_0\}.
  \end{align*}
  It is trivial to verify that \(E_0\{h^*(Y,A,X,D)\mid A,X,D\}=0\), \(E_0\{h^*(G,X)\mid X\}=0\), as well as \(E_0\{h^*(X)\}=0\), so \(h^*(G,X)\in\dot{\mathcal{P}}_{g}\) and \(h^*(X)\in\dot{\mathcal{P}}_{x}\).
  Now,
  \begin{align*}
    \MoveEqLeft E_0\{Yh^*(Y,A,X,D)\mid A=1,X=x,D=d\} - E_0\{Yh^*(Y,A,X,D)\mid A=0,X=x,D=d\} \\
    &= E_0[\{Y-\mu_0(A,X,D)\}h^*(Y,A,X,D)\mid A=1,X=x,D=d] \\
    &\hphantom{=}\quad - E_0[\{Y-\mu_0(A,X,D)\}h^*(Y,A,X,D)\mid A=0,X=x,D=d] \\
    &= \frac{\pi_0(x)}{\alpha_0 \{1-\pi_0(x)\}}\frac{w_0(x,d)}{\sum_{d'\in[m]}\zeta_0(d'\mid x)w_0(x,d')}\bigg\{\frac{V_0(1,x,d)}{e_0(1\mid x,d)}+\frac{V_0(0,x,d)}{e_0(0\mid x,d)}\bigg\} \\
    &= \frac{\pi_0(x)}{\alpha_0 \{1-\pi_0(x)\}}\frac{1}{\sum_{d'\in[m]}\zeta_0(d'\mid x)w_0(x,d')},
  \end{align*}
  which does not depend on the value of \(d\).
  Thus, \((1-g)h^*(y,a,x,d)\in\dot{\mathcal{P}}_{y}\).
\end{proof}

\subsection{Proof of corollary~\ref{cor:tangent-space-complement}}
The first part of corollary~\ref{cor:tangent-space-complement} is a direct result from the proof of lemma~\ref{lem:eif}.
To prove the second part of corollary~\ref{cor:tangent-space-complement}, note that the space of the influence functions of the parameter \(\theta_0\) can be characterized by the translation \(\varphi(o,w_0)+\dot{\mathcal{P}}^{\perp}\), where \(\varphi(o,w_0)\) is the EIF in lemma~\ref{lem:eif}.
Therefore, if we choose
\begin{multline*}
  u(o,\tilde{w}) = \frac{(1-g)\pi_0(x)}{\alpha_0\{1-\pi_0(x)\}}\frac{2a-1}{e_0(a\mid x,d)}\\
  \bigg\{\frac{\tilde{w}(x,d)}{\sum_{d'\in[m]}\zeta_0(d'\mid x)\tilde{w}(x,d')}
  -\frac{w_0(x,d)}{\sum_{d'\in[m]}\zeta_0(d'\mid x)w_0(x,d')}\bigg\}\{y-\mu_0(a,x,d)\},
\end{multline*}
for \(\tilde{w}\) as stated in corollary~\ref{cor:tangent-space-complement}, then \(u(o,\tilde{w})\in\dot{\mathcal{P}}^{\perp}\), and \(\varphi(o,\tilde{w})=\varphi(o,w_0)+u(o,\tilde{w})\) is an influence function of \(\theta_0\).

\subsection{Proof of theorem~\ref{thm:asymptotic}}

We quote a lemma from \citetsuppmat{kennedy2024semiparametric}.
\begin{lemma}
  \label{lem:cross-fitting}
  Let \(f_{k}\) be a random function which only depends on the sample \(\mathcal{O}_{-k}=\{O_i:i\notin\mathcal{I}_{k}\}\).
  Then \((\mathbb{P}_{n,k}-P_{0})f_{k}=O_{P_0}(n^{-1/2}\|f_k\|_{P_0})\).
\end{lemma}

\begin{proof}
  By the Markov inequality conditional on \(\mathcal{O}_{-k}\), for \(t> 0\),
  \[
    \Pr\bigg\{\frac{|(n/K)^{1/2}(\mathbb{P}_{n,k}-P_{0})f_{k}|^2}{\|f_k\|_{P_0}^2}\geq t\biggm\vert \mathcal{O}_{-k}\bigg\}\leq \frac{P_0(f_{k}-P_0f_k)^2}{t\|f_k\|_{P_0}^2}\leq \frac{1}{t}.
  \]
  Marginally, we have \(\Pr\{|(n/K)^{1/2}(\mathbb{P}_{n,k}-P_{0})f_{k}|/\|f_k\|_{P_0}\geq t^{1/2}\}\leq t^{-1}\).
  Therefore, we also have \(|(n/K)^{1/2}(\mathbb{P}_{n,k}-P_{0})f_{k}|/\|f_k\|_{P_0}=O_{P_0}(1)\), and the lemma follows, since \(K\) does not depend on \(n\).
\end{proof}


\begin{lemma}
  \label{lem:boundedness-ell}
  If assumption~\ref{asn:regularity} is satisfied, then \(\|\ell(O,\hat\eta_k)\|_{P_0}=O_{P_0}(1)\) for every \(k\in[K]\).
\end{lemma}

\begin{proof}
  Let
  \[
    H_k(a,x,d)=\frac{\hat{\pi}_k(x)}{1-\hat\pi_k(x)}\frac{\hat{w}_k(x,d)}{\sum_{d'\in[m]}\hat\zeta_k(d'\mid x)\hat{w}_k(x,d')}\frac{1}{\hat{e}_k(a\mid x,d)}\lesssim 1.
  \]
  We also have
  \[
    |\hat\delta_k(x)| \leq \sum_{d\in[m]}\sum_{a\in\{0,1\}}\frac{\big|\hat{w}_k(x,d)\hat{\zeta}_k(d\mid x)\hat{\mu}_k(a,x,d)\big|}{\big|\sum_{d'\in[m]}\hat{w}_k(x,d')\hat{\zeta}_k(d'\mid x)\big|} \lesssim 1.
  \]
  Then
  \begin{align*}
    \MoveEqLeft\|\ell(O,\hat\eta_k)\|_{P_0}^2 \\
    &\lesssim P_0[(1-G)H_k^2(A,X,D)\{Y-\hat{\mu}_k(A,X,D)\}^2] + P_0\{\pi_0(X)\hat\delta_k^2(X)\}\\
    &\lesssim P_0[(1-G)\{Y-\mu_0(A,X,D)\}^2] + P_0[(1-G)(\hat{\mu}_k-\bar\mu)^2(A,X,D)] \\
    &\hphantom{\lesssim}\quad + P_0[(1-G)(\bar\mu^2+\mu_0^2)(A,X,D)] + 1\\
    &\lesssim P_0[(1-G)V_0(A,X,D)] + \max_{d\in[m]}\max_{a\in\{0,1\}}\|I(D=d)(\hat{\mu}_k-\bar\mu)(a,X,d)\|_{P_0}^2 + 1\\
    &\lesssim o_{P_0}(1) + 1 = O_{P_0}(1).
  \end{align*}
\end{proof}

Let
\[
  \bar\eta=\{\alpha_0,\pi_0,\zeta_0,\bar{w},e_0,\mu_0\}
\]
be the set of nuisance parameters at the truth, where the weight function \(\bar{w}\) may not be optimal.

\begin{lemma}
  \label{lem:convergence-ell}
  If assumptions~\ref{asn:regularity} and \ref{asn:model-linearity}\ref{asn:plim} are satisfied, then \(\|\ell(O,\hat\eta_k)-\ell(O,\bar\eta)\|_{P_0}=o_{P_0}(1)\) for every \(k\in[K]\).
\end{lemma}

\begin{proof}
  We first show consistency of \(\hat\delta_k\).
  By the triangular inequality,
  \begin{align*}
    \MoveEqLeft\|(\hat\delta_k-\delta_0)(X)\|_{P_0} \\
    &= \begin{multlined}[t][.8\textwidth]
      \bigg\|\frac{\sum_{d\in[m]}\hat{w}_k(X,d)\hat{\zeta}_k(d\mid X)\{\hat{\mu}_k(1,X,d)-\hat{\mu}_k(0,X,d)\}}{\sum_{d'\in[m]}\hat{w}_k(X,d')\hat{\zeta}_k(d'\mid X)}\\
      -\frac{\sum_{d\in[m]}\bar{w}(X,d)\zeta_0(d\mid X)\{\mu_0(1,X,d)-\mu_0(0,X,d)\}}{\sum_{d'\in[m]}\bar{w}(X,d')\zeta_0(d'\mid X)}\bigg\|_{P_0}
    \end{multlined}
    \\
    &\leq \sum_{a\in\{0,1\}}\sum_{d\in[m]}\bigg\{\bigg(P_0\bigg[\frac{\big\{\hat{w}_k(X,d)(\hat{\zeta}_k-\zeta_0)(d\mid X)\hat{\mu}_k(a,X,d)\big\}^2}{\big\{\sum_{d'\in[m]}\hat{w}_k(X,d')\hat{\zeta}_k(d'\mid X)\big\}^2}\bigg]\bigg)^{1/2}\\
    &\hphantom{\leq}\quad +\sum_{a\in\{0,1\}}\sum_{d\in[m]}\bigg(P_0\bigg[\frac{\big\{(\hat{w}_k-\bar{w})(X,d)\zeta_0(d\mid X)\hat{\mu}_k(a,X,d)\big\}^2}{\big\{\sum_{d'\in[m]}\hat{w}_k(X,d')\hat{\zeta}_k(d'\mid X)\big\}^2}\bigg]\bigg)^{1/2}\\
    &\hphantom{\leq}\quad+\begin{multlined}[t][.85\textwidth]
      \sum_{a\in\{0,1\}}\sum_{d\in[m]}\bigg(P_0\bigg[\bigg\{\frac{1}{\sum_{d'\in[m]}\hat{w}_k(X,d')\hat{\zeta}_k(d'\mid X)}\\
      -\frac{1}{\sum_{d'\in[m]}\bar{w}(X,d')\zeta_0(d'\mid X)}\bigg\}^2\\
      \big\{\bar{w}(X,d)\zeta_0(d\mid X)\hat{\mu}_k(a,X,d)\big\}^2\bigg]\bigg)^{1/2}
    \end{multlined}\\
    &\hphantom{\leq}\quad +\sum_{a\in\{0,1\}}\sum_{d\in[m]}\bigg(P_0\bigg[\frac{\big\{\bar{w}(X,d)\zeta_0(d\mid X)(\hat{\mu}_k-\mu_0)(a,X,d)\big\}^2}{\big\{\sum_{d'\in[m]}\bar{w}(X,d')\zeta_0(d'\mid X)\big\}^2}\bigg]\bigg)^{1/2}\\
    &\lesssim \max_{d\in[m]}\|(\hat{\zeta}_k-\zeta_0)(d\mid X)\|_{P_0} + \max_{d\in[m]}\|I(D=d)(\hat{w}_k-\bar{w})(X,d)\|_{P_0} \\
    &\hphantom{\lesssim}\quad + \max_{d\in[m]}\max_{a\in\{0,1\}}\|I(D=d)(\hat{\mu}_k-\mu_0)(a,X,d)\|_{P_0} = o_{P_0}(1).
  \end{align*}

  We write \(\ell(o,\hat\eta_k)-\ell(o,\bar\eta)\) as a sum such that
  \begin{align*}
    \MoveEqLeft \ell(o,\hat\eta_k)-\ell(o,\bar\eta) \\
    &= \frac{1-g}{\hat\alpha_k}\frac{\hat\pi_k(x)}{1-\hat\pi_k(x)}\frac{\hat{w}_k(x,d)}{\sum_{d'\in[m]}\hat\zeta_k(d'\mid x)\hat{w}_k(x,d')}\frac{2a-1}{\hat{e}_k(a\mid x,d)}(\mu_0-\hat\mu)(a,x,d)\\
                                      &\hphantom{=}\quad- \frac{1-g}{\hat\alpha_k}\frac{\hat\pi_k(x)}{1-\hat\pi_k(x)}\frac{\hat{w}_k(x,d)}{\sum_{d'\in[m]}\hat\zeta_k(d'\mid x)\hat{w}_k(x,d')}\frac{2a-1}{\hat{e}_k(a\mid x,d)e_0(a\mid x,d)}\\
                                      &\hphantom{=}\quad\hphantom{+}\qquad(\hat{e}_k-e_0)(a\mid x,d)\{y-\mu_0(a,x,d)\} \\
                                      &\hphantom{=}\quad+\frac{1-g}{\hat\alpha_k}\frac{\hat\pi_k(x)}{1-\hat\pi_k(x)}\frac{(\hat{w}_k-\bar{w})(x,d)}{\sum_{d'\in[m]}\hat\zeta_k(d'\mid x)\hat{w}_k(x,d')}\frac{2a-1}{e_0(a\mid x,d)}\{y-\mu_0(a,x,d)\}\\
                                      &\hphantom{=}\quad+\frac{1-g}{\hat\alpha_k}\frac{\hat\pi_k(x)}{1-\hat\pi_k(x)}\frac{\bar{w}(x,d)\sum_{d'\in[m]}(\zeta_0-\hat\zeta_k)(d'\mid x)\hat{w}_k(x,d')}{\sum_{d'\in[m]}\hat\zeta_k(d'\mid x)\hat{w}_k(x,d')\sum_{d'\in[m]}\zeta_0(d'\mid x)\bar{w}(x,d')}\\
                                      &\hphantom{=}\quad\hphantom{+}\qquad\frac{2a-1}{e_0(a\mid x,d)}\{y-\mu_0(a,x,d)\}\\
                                      &\hphantom{=}\quad+\frac{1-g}{\hat\alpha_k}\frac{\hat\pi_k(x)}{1-\hat\pi_k(x)}\frac{\bar{w}(x,d)\sum_{d'\in[m]}\zeta_0(d'\mid x)(\bar{w}-\hat{w}_k)(x,d')}{\sum_{d'\in[m]}\hat\zeta_k(d'\mid x)\hat{w}_k(x,d')\sum_{d'\in[m]}\zeta_0(d'\mid x)\bar{w}(x,d')}\\
                                      &\hphantom{=}\quad\hphantom{+}\qquad\frac{2a-1}{e_0(a\mid x,d)}\{y-\mu_0(a,x,d)\}\\
                                      &\hphantom{=}\quad+\frac{1-g}{\hat\alpha_k}\frac{(\hat\pi_k-\pi_0)(x)}{\{1-\pi_0(x)\}\{1-\hat\pi_k(x)\}}\frac{\bar{w}(x,d)}{\sum_{d'\in[m]}\zeta_0(d'\mid x)\bar{w}(x,d')}\\
                                      &\hphantom{=}\quad\hphantom{+}\qquad\frac{2a-1}{e_0(a\mid x,d)}\{y-\mu_0(a,x,d)\}\\
                                      &\hphantom{=}\quad+\begin{multlined}[t][.8\textwidth]
                                        \frac{(1-g)(\alpha_0-\hat\alpha_k)}{\alpha_0\hat\alpha_k}\frac{\pi_0(x)}{1-\pi_0(x)}\frac{\bar{w}(x,d)}{\sum_{d'\in[m]}\zeta_0(d'\mid x)\bar{w}(x,d')}\\
                                        \frac{2a-1}{e_0(a\mid x,d)}\{y-\mu_0(a,x,d)\}
                                      \end{multlined}\\
                                      &\hphantom{=}\quad+\frac{g}{\hat\alpha_k}(\hat\delta-\delta_0)(x)+\frac{g(\alpha_0-\hat\alpha_k)}{\alpha_0\hat\alpha_k}\delta_0(x)
  \end{align*}

  By the triangular inequality,
  \begin{align*}
    \MoveEqLeft\|\ell(O,\hat\eta_k)-\ell(O,\bar\eta)\|_{P_0}\\
    &\leq \bigg(P_0\bigg[\frac{1-\pi_0(X)}{\hat\alpha_k^2}\frac{\hat\pi_k^2(X)}{\{1-\hat\pi_k(X)\}^2}\sum_{d\in[m]}\frac{\hat{w}_k^2(X,d)\zeta_0(d\mid X)}{\{\sum_{d'\in[m]}\hat\zeta_k(d'\mid X)\hat{w}_k(X,d')\}^2}\\
    &\hphantom{\leq\bigg(P_0\bigg[}\quad\sum_{a\in\{0,1\}}\frac{e_0(a\mid X,d)}{\hat{e}_k^2(a\mid X,d)}(\mu_0-\hat\mu)^2(a,X,d)\bigg]\bigg)^{1/2}\\
    &\hphantom{\leq}\quad+ \bigg(P_0\bigg[\frac{1-\pi_0(X)}{\hat\alpha_k^2}\frac{\hat\pi_k^2(X)}{\{1-\hat\pi_k(X)\}^2}\sum_{d\in[m]}\frac{\hat{w}_k^2(X,d)\zeta_0(d\mid X)}{\big\{\sum_{d'\in[m]}\hat\zeta_k(d'\mid X)\hat{w}_k(X,d')\big\}^2}\\
    &\hphantom{\leq}\quad\hphantom{+\bigg(P_0\bigg[}\qquad\sum_{a\in\{0,1\}}\frac{(\hat{e}_k-e_0)^2(a\mid X,d)}{\hat{e}_k^2(a\mid X,d)e_0(a\mid X,d)}V_0(a,X,d)\bigg]\bigg)^{1/2}\\
    &\hphantom{\leq}\quad+ \bigg(P_0\bigg[\frac{1-\pi_0(X)}{\hat\alpha_k^2}\frac{\hat\pi_k^2(X)}{\{1-\hat\pi_k(X)\}^2}\sum_{d\in[m]}\frac{(\hat{w}_k-\bar{w})^2(X,d)\zeta_0(d\mid X)w_0^{-1}(X,d)}{\big\{\sum_{d'\in[m]}\hat\zeta_k(d'\mid X)\hat{w}_k(X,d')\big\}^2}\bigg]\bigg)^{1/2}\\
    &\hphantom{\leq}\quad+ \bigg(P_0\bigg[\frac{1-\pi_0(X)}{\hat\alpha_k^2}\frac{\hat\pi_k^2(X)}{\{1-\hat\pi_k(X)\}^2}\bigg\{\sum_{d'\in[m]}(\zeta_0-\hat\zeta_k)(d'\mid X)\hat{w}_k(X,d')\bigg\}^2\\
    &\hphantom{\leq}\quad\hphantom{+\bigg(P_0\bigg[}\qquad\frac{\sum_{d\in[m]}\bar{w}^2(X,d)\zeta_0(d\mid X)w_0^{-1}(X,d)}{\big\{\sum_{d'\in[m]}\hat\zeta_k(d'\mid X)\hat{w}_k(X,d')\big\}^2\big\{\sum_{d'\in[m]}\zeta_0(d'\mid X)\bar{w}(X,d')\big\}^2}\bigg]\bigg)^{1/2}\\
    &\hphantom{\leq}\quad+ \bigg(P_0\bigg[\frac{1-\pi_0(X)}{\hat\alpha_k^2}\frac{\hat\pi_k^2(X)}{\{1-\hat\pi_k(X)\}^2}\bigg\{\sum_{d'\in[m]}\zeta_0(d'\mid X)(\hat{w}_k-\bar{w})(X,d')\bigg\}^2\\
    &\hphantom{\leq}\quad\hphantom{+\bigg(P_0\bigg[}\qquad\frac{\sum_{d\in[m]}\bar{w}^2(X,d)\zeta_0(d\mid X)w_0^{-1}(X,d)}{\big\{\sum_{d'\in[m]}\hat\zeta_k(d'\mid X)\hat{w}_k(X,d')\big\}^2\big\{\sum_{d'\in[m]}\zeta_0(d'\mid X)\bar{w}(X,d')\big\}^2}\bigg]\bigg)^{1/2}\\
    &\hphantom{\leq}\quad+ \bigg(P_0\bigg[\frac{1}{\hat\alpha_k^2}\frac{(\hat\pi_k-\pi_0)^2(X)}{\{1-\pi_0(X)\}\{1-\hat\pi_k(X)\}^2}\sum_{d\in[m]}\frac{\bar{w}^2(X,d)\zeta_0(d\mid X)w_0^{-1}(X,d)}{\big\{\sum_{d'\in[m]}\zeta_0(d'\mid X)\bar{w}(X,d')\big\}^2}\bigg]\bigg)^{1/2}\\
    &\hphantom{\leq}\quad+ \bigg(P_0\bigg[\frac{(\alpha_0-\hat\alpha_k)^2}{\alpha_0^2\hat\alpha_k^2}\frac{\pi_0^2(X)}{1-\pi_0(X)}\sum_{d\in[m]}\frac{\bar{w}^2(X,d)\zeta_0(d\mid X)w_0^{-1}(X,d)}{\big\{\sum_{d'\in[m]}\zeta_0(d'\mid X)\bar{w}(X,d')\big\}^2}\bigg]\bigg)^{1/2}\\
    &\hphantom{\leq}\quad+ \bigg(P_0\bigg[\frac{\pi_0(X)}{\hat\alpha_k^2}(\hat\delta-\delta_0)^2(X)\bigg]\bigg)^{1/2}+ \bigg(P_0\bigg[\frac{\pi_0(X)(\alpha_0-\hat\alpha_k)^2}{\alpha_0^2\hat\alpha_k^2}\delta^2(X)\bigg]\bigg)^{1/2}\\
    &\lesssim \max_{d\in[m]}\max_{a\in\{0,1\}}\|I(D=d)(\hat\mu_k-\mu_0)(a,X,d)\|_{P_0}\\
    &\hphantom{\lesssim}\quad +\max_{d\in[m]}\max_{a\in\{0,1\}}\|I(D=d)(\hat{e}_k-e_0)(a,X,d)\|_{P_0}\\
    &\hphantom{\lesssim}\quad+\max_{d\in[m]}\|I(D=d)(\hat{w}_k-\bar{w})(X,d)\|_{P_0}\\
    &\hphantom{\lesssim}\quad+\max_{d\in[m]}\|(\hat\zeta_k-\zeta_0)(d\mid X)\|_{P_0}+\|(\hat\pi_k-\pi_0)(X)\|_{P_0}+\|(\hat\delta_k-\delta_0)(X)\|_{P_0}+|\hat\alpha_k-\alpha_0|,
  \end{align*}
  which converges in probability to zero by assumption and by \(\|(\hat\delta_k-\delta_0)(X)\|_{P_0}=o_{P_0}(1)\) shown earlier in the proof.
\end{proof}

\begin{proof}[Proof of theorem~\ref{thm:asymptotic}]
  The difference between the estimator and the target parameter decomposes as
  \begin{equation}
    \label{eqn:decomp-cons}
    \hat\theta-\theta_0=\frac{1}{K}\sum_{k\in[K]}\bigg[(\mathbb{P}_{n,k}-P_0)\ell(O,\hat\eta_k)+\bigg\{P_0\ell(O,\hat\eta_k)-\frac{\alpha_0}{\hat{\alpha}_k}\theta_0\bigg\}-\frac{\hat{\alpha}_k-\alpha_0}{\hat{\alpha}_k}\theta_0\bigg].
  \end{equation}

  The second term in \eqref{eqn:decomp-cons} is, using \(\theta_0=E_0\{\pi_0(X)\delta_0(X)\}/\alpha_0\),
  \begin{align}
    P_0\ell(O,\hat\eta_k)-\frac{\alpha_0}{\hat{\alpha}_k}\theta_0 &=\frac{1}{\hat{\alpha}_k}P_0\bigg\{\frac{\{1-\pi_0(X)\}\hat{\pi}_k(X)}{\{1-\hat{\pi}_k(X)\}}\sum_{d\in[m]}\frac{\hat{w}_k(X,d)\zeta_0(d\mid X)}{\sum_{d'\in[m]}\hat{w}_k(X,d')\hat{\zeta}_k(d'\mid X)}\nonumber\\
                                                                 &\hphantom{=}\quad\hphantom{+\frac{1}{\hat{\alpha}_k}P_0\bigg\{}\quad\sum_{a\in\{0,1\}}\frac{(2a-1)e_0(a\mid X,d)}{\hat{e}_k(a\mid X,d)}(\mu_0-\hat{\mu}_k)(a,X,d)\bigg\}\label{eqn:rep}\\
                                                                 &\hphantom{=}\quad+ P_0\bigg[\frac{\pi_0(X)}{\hat{\alpha}_k}(\hat{\delta}_k-\delta_0)(X)\bigg]. \nonumber
  \end{align}

  Developing from \eqref{eqn:rep},
  \begin{align}
    \eqref{eqn:rep} &= \begin{multlined}[t][.8\textwidth] P_0\bigg[\frac{\{1-\pi_0(X)\}\hat{\pi}_k(X)}{\hat{\alpha}_k\{1-\hat{\pi}_k(X)\}}\sum_{d\in[m]}\frac{\hat{w}_k(X,d)\zeta_0(d\mid X)}{\sum_{d'\in[m]}\hat{w}_k(X,d')\hat{\zeta}_k(d'\mid X)} \\
      \hphantom{P_0\bigg[}\quad\sum_{a\in\{0,1\}}\frac{(2a-1)e_0(a\mid X,d)}{\hat{e}_k(a\mid X,d)}(\mu_0-\hat{\mu}_k)(a,X,d)\bigg]
      \end{multlined}\nonumber\\
                    &=\begin{multlined}[t][.8\textwidth]
                      P_0\bigg[\frac{\{1-\pi_0(X)\}\hat{\pi}_k(X)}{\hat{\alpha}_k\{1-\hat{\pi}_k(X)\}}\frac{\sum_{d\in[m]}\hat{w}_k(X,d)\zeta_0(d\mid X)}{\sum_{d'\in[m]}\hat{w}_k(X,d')\hat{\zeta}_k(d'\mid X)}\\
                      \sum_{a\in\{0,1\}}\frac{(2a-1)(e_0-\hat{e}_k)(a\mid X,d)}{\hat{e}_k(a\mid X,d)}(\mu_0-\hat{\mu}_k)(a,X,d)\bigg]
                      \end{multlined}\nonumber\\
                    &\hphantom{=}\quad+\begin{multlined}[t][.77\textwidth]
                      P_0\bigg(\frac{\{1-\pi_0(X)\}\hat{\pi}_k(X)}{\hat{\alpha}_k\{1-\hat{\pi}_k(X)\}}\frac{\sum_{d\in[m]}\hat{w}_k(X,d)\zeta_0(d\mid X)}{\sum_{d'\in[m]}\hat{w}_k(X,d')\hat{\zeta}_k(d'\mid X)}\\
                      [\delta_0(X)-\{\hat{\mu}_k(1,X,d)-\hat{\mu}_k(0,X,d)\}]\bigg).
                      \end{multlined}\label{eqn:rep-2}
  \end{align}

  Continuing from the last expression,
  \begin{align*}
    \eqref{eqn:rep-2} &= P_0\bigg[\frac{(\hat{\pi}_k-\pi_0)(X)}{\hat{\alpha}_k\{1-\hat{\pi}_k(X)\}}\frac{\sum_{d\in[m]}\hat{w}_k(X,d)\zeta_0(d\mid X)}{\sum_{d'\in[m]}\hat{w}_k(X,d')\hat{\zeta}_k(d'\mid X)}\sum_{a\in\{0,1\}}(2a-1)(\mu_0-\hat{\mu}_k)(a,X,d)\bigg] \\
                      &\hphantom{=}\quad+P_0\bigg(\frac{\pi_0(X)}{\hat{\alpha}_k}\frac{\sum_{d\in[m]}\hat{w}_k(X,d)(\zeta_0-\hat{\zeta}_k)(d\mid X)}{\sum_{d'\in[m]}\hat{w}_k(X,d')\hat{\zeta}_k(d'\mid X)}\sum_{a\in\{0,1\}}(2a-1)(\mu_0-\hat{\mu}_k)(a,X,d)\bigg) \\
                      &\hphantom{=}\quad+P_0\bigg(\frac{\pi_0(X)}{\hat{\alpha}_k}\bigg[\delta_0(X)-\underbrace{\frac{\sum_{d\in[m]}\hat{w}_k(X,d)\hat{\zeta}_k(d\mid X)}{\sum_{d'\in[m]}\hat{w}_k(X,d')\hat{\zeta}_k(d'\mid X)}\{\hat{\mu}_k(1,X,d)-\hat{\mu}_k(0,X,d)\}}_{=\hat{\delta}_k(X)}\bigg]\bigg).
  \end{align*}

  Collecting all relevant terms,
  \begin{align*}
    \MoveEqLeft P_0\ell(O,\hat\eta_k)-\frac{\alpha_0}{\hat{\alpha}_k}\theta_0 \\
    &= \begin{multlined}[t][.86\textwidth]
      P_0\bigg[\frac{\{1-\pi_0(X)\}\hat{\pi}_k(X)}{\hat{\alpha}_k\{1-\hat{\pi}_k(X)\}}\frac{\sum_{d\in[m]}\hat{w}_k(X,d)\zeta_0(d\mid X)}{\sum_{d'\in[m]}\hat{w}_k(X,d')\hat{\zeta}_k(d'\mid X)}\\
      \sum_{a\in\{0,1\}}\frac{(2a-1)(e_0-\hat{e}_k)(a\mid X,d)}{\hat{e}_k(a\mid X,d)}(\mu_0-\hat{\mu}_k)(a,X,d)\bigg]
      \end{multlined}\\
    &\hphantom{=}\quad+P_0\bigg(\frac{\pi_0(X)}{\hat{\alpha}_k}\frac{\sum_{d\in[m]}\hat{w}_k(X,d)(\zeta_0-\hat{\zeta}_k)(d\mid X)}{\sum_{d'\in[m]}\hat{w}_k(X,d')\hat{\zeta}_k(d'\mid X)}\sum_{a\in\{0,1\}}(2a-1)(\mu_0-\hat{\mu}_k)(a,X,d)\bigg)\\
    &\hphantom{=}\quad\begin{multlined}[.85\textwidth]
      +P_0\bigg(\frac{(\hat{\pi}_k-\pi_0)(X)}{\hat{\alpha}_k\{1-\hat{\pi}_k(X)\}}\frac{\sum_{d\in[m]}\hat{w}_k(X,d)\zeta_0(d\mid X)}{\sum_{d'\in[m]}\hat{w}_k(X,d')\hat{\zeta}_k(d'\mid X)}\\
      \sum_{a\in\{0,1\}}(2a-1)(\mu_0-\hat{\mu}_k)(a,X,d)\bigg).
    \end{multlined}
  \end{align*}

  The representation of the second-order remainder above can be bounded as
  \begin{align*}
    \MoveEqLeft\bigg|P_0\ell(O,\hat\eta_k)-\frac{\alpha_0}{\hat{\alpha}_k}\theta_0\bigg|\\
    &\lesssim \max_{d\in[m]}\max_{a\in\{0,1\}}P_0\{I(D=d)|(e_0-\hat{e}_k)(a\mid X,d)||(\mu_0-\hat{\mu}_k)(a,X,d)|\}\\
    &\hphantom{\lesssim }\quad+\max_{d\in[m]}\max_{a\in\{0,1\}}P_0\{I(D=d)|(\zeta_0-\hat{\zeta}_k)(d\mid X)||(\mu_0-\hat{\mu}_k)(a,X,d)|\} \\
    &\hphantom{\lesssim }\quad+\max_{d\in[m]}\max_{a\in\{0,1\}}P_0\{I(D=d)|(\hat{\pi}_k-\pi_0)(X)||(\mu_0-\hat{\mu}_k)(a,X,d)|\} \\
    &\leq \begin{multlined}[t][.8\textwidth]
      \max_{d\in[m]}\max_{a\in\{0,1\}}\|I(D=d)(\hat{\mu}_k-\mu_0)(a,X,d)\|_{P_0}\\
    \bigg\{\max_{d\in[m]}\max_{a\in\{0,1\}}\|I(D=d)(\hat{e}_k-e_0)(a\mid X,d)\|_{P_0}\\
    +\|(\hat{\pi}_k-\pi_0)(X)\|_{P_0}+\max_{d\in[m]}\|(\hat{\zeta}_k-\zeta_0)(d\mid X)\|_{P_0}\bigg\}.
    \end{multlined}
  \end{align*}

  We first show consistency of \(\hat\theta\).
  The estimation error is bounded by
  \[
    |\hat\theta-\theta_0|\leq \frac{1}{K}\sum_{k\in[K]}\bigg\{|(\mathbb{P}_{n,k}-P_0)\ell(O,\hat\eta_{k})|+\bigg|P_0\ell(O,\hat\eta_k)-\frac{\alpha_0}{\hat{\alpha}_k}\theta_0\bigg|+\frac{|\hat{\alpha}_k-\alpha_0|}{\hat{\alpha}_k}\theta_0\bigg\}.
  \]
  Since the number of splits does not scale with \(n\), we focus on the terms in the braces for every \(k\in[K]\).
  The first term converges in probability to zero by lemmas~\ref{lem:cross-fitting} and \ref{lem:boundedness-ell}.
  The second term converges in probability to zero by assumption~\ref{asn:model-consistency}.
  The third term converges in probability to zero by an application of Slutsky's theorem because \(\hat\alpha_k-\alpha_0=o_{P_0}(1)\).
  Therefore, we have \(\hat\theta-\theta_0=o_{P_0}(1)\).

  To show asymptotic linearly, we further decompose \eqref{eqn:decomp-cons} as
  \begin{multline*}
    \hat\theta-\theta_0 = \mathbb{P}_n\varphi(O,\bar{w}) + \frac{1}{K}\sum_{k\in[K]}\bigg[(\mathbb{P}_{n,k}-P_0)\{\ell(O,\hat\eta_k)-\ell(O,\bar\eta)\} \\
    + \bigg\{P_0\ell(O,\hat\eta_k)-\frac{\alpha_0}{\hat\alpha_k}\theta_0\bigg\} + \frac{(\hat\alpha_k-\alpha_0)^2}{\hat\alpha_k\alpha_0}\theta_0\bigg].
  \end{multline*}
  The second term is \(o_{P_0}(n^{-1/2})\) by lemmas~\ref{lem:cross-fitting} and \ref{lem:convergence-ell}.
  The third term is \(o_{P_0}(n^{-1/2})\) by assumption~\ref{asn:model-linearity}.
  By the central limit theorem, \(\hat{\alpha}_k-\alpha_0=O_{P_0}(n^{-1/2})\), and the last term is \(O_{P}(n^{-1})=o_{P_0}(n^{-1/2})\) by Slutsky's theorem.
\end{proof}

\begin{remark}
  When \(\hat{\mu}_k(1,x,d)-\hat{\mu}_k(0,x,d)=\hat{\mu}_k(1,x,d')-\hat{\mu}_k(0,x,d')\) and \(\hat{w}_k(x,d)=\hat{w}_k(x,d')\) for all \(d,d'\in[m]\), the error from nuisance model estimation no longer involves the product term \(\max_{d\in[m]}\|(\hat{\zeta}_k-\zeta_0)(d\mid X)\|_{P_0}\max_{a\in\{0,1\}}\|I(D=d)(\hat{\mu}_k-\mu_0)(a,X,d)\|_{P_0}\).
\end{remark}

\subsection{Proof of proposition~\ref{ppn:eif-li}}

The tangent space at \(P_0\in\mathcal{P}^\dagger\) is
\[
  \dot{\mathcal{P}}^\dagger = \dot{\mathcal{P}}_{y}^\dagger \oplus \dot{\mathcal{P}}_{a}\oplus \dot{\mathcal{P}}_{d} \oplus \dot{\mathcal{P}}_{g} \oplus \dot{\mathcal{P}}_{x},
\]
where \(\dot{\mathcal{P}}_a\), \(\dot{\mathcal{P}}_d\), \(\dot{\mathcal{P}}_g\), and \(\dot{\mathcal{P}}_x\) are as in the proof of lemma~\ref{lem:eif}, and
\begin{multline*}
  \dot{\mathcal{P}}_{y}^\dagger = \big\{(1-g)h(y,a,x,d):E_0\{h(Y,A,X,D)\mid A,X,D\}=0, \\
  E_0\{Yh(Y,A,X,D)\mid A,X,D=d\}=E_0\{Yh(Y,A,X,D)\mid A,X,D=d'\}\big\}.
\end{multline*}
Along the parametric submodel \(\{P_\varepsilon\}\) with score function \(h(o)\), the pathwise derivative of
\[
  \theta_\varepsilon = E_{P_\varepsilon}\{\mu_\varepsilon(1,X)-\mu_\varepsilon(0,X)\mid D=1\}
\]
at \(\varepsilon=0\) is
\begin{multline}
  \label{eqn:pathwise-li} \frac{\d}{\d\varepsilon}\theta_{\varepsilon}\bigg\vert_{\varepsilon=0} =  \frac{1}{\alpha_0}\int \{\upsilon_h(1,x)-\upsilon_h(0,x)\}\pi_0(x)p_0(x)\d x \\
  +\frac{1}{\alpha_0}\int \{\mu(1,x)-\mu(0,x)-\theta_0\}\pi_0(x)p_0(x)\{h(G=1,x)+h(x)\}\d x,
\end{multline}
where \(\upsilon_h(a,x)=E_0\{Yh(Y,A,X,D)\mid A=a,X=x,D=d\}\) does not depend on \(d\).
We will verify that \(\varphi^\dagger(o,w^\dagger_0)\) is an influence function by showing that
\[
  \frac{\d}{\d\varepsilon}\theta_{\varepsilon}\bigg\vert_{\varepsilon=0}=E_0\{\varphi^\dagger(O,w^\dagger_0)h(O)\}.
\]
For any \(h(o)\in\dot{\mathcal{P}}^\dagger\), the inner product
\begin{align}
  \MoveEqLeft E_0\{\varphi^\dagger(O,w^\dagger_0)h(O)\} \nonumber\\
  &= \begin{multlined}[t][.8\textwidth]
    E_0\bigg[\frac{(1-G)\pi_0(X)}{\alpha_0 \{1-\pi_0(X)\}}\frac{2A-1}{e_0(A\mid X)}\frac{w_0^\dagger(A,X,D)}{\sum_{d\in[m]}w_0^\dagger(A,X,d)\zeta_0(d\mid A,X)}\\
    \{Y-\mu_0(A,X)\}h(Y,A,X,D)\bigg]
  \end{multlined}\label{eqn:innerproduct-li-1}\\
  &\hphantom{=}\quad + E_0\bigg[\frac{G}{\alpha_0}\{\delta_0(X)-\theta_0\}\{h(G,X)+h(X)\}\bigg].
\end{align}
The second term in the display above is equal to the second term in the derivative \eqref{eqn:pathwise-li}.
The first term is
\begin{align*}
  \eqref{eqn:innerproduct-li-1} &= \begin{multlined}[t][.78\textwidth]
    E_0\bigg[\frac{(1-G)\pi_0(X)}{\alpha_0 \{1-\pi_0(X)\}}\frac{2A-1}{e_0(A\mid X)}\frac{w_0^\dagger(A,X,D)}{\sum_{d\in[m]}w_0^\dagger(A,X,d)\zeta_0(d\mid A,X)}\\
    E_0\{Yh(Y,A,X,D)\mid A,X,D\}\bigg]
  \end{multlined}\\
                                &= E_0\bigg[\frac{(1-G)\pi_0(X)}{\alpha_0 \{1-\pi_0(X)\}}\frac{2A-1}{e_0(A\mid X)}\frac{w_0^\dagger(A,X,D)}{\sum_{d\in[m]}w_0^\dagger(A,X,d)\zeta_0(d\mid A,X)}\upsilon_{h}(A,X)\bigg] \\
                                &= E_0\bigg[\frac{(1-G)\pi_0(X)}{\alpha_0 \{1-\pi_0(X)\}}\frac{2A-1}{e_0(A\mid X)}\upsilon_{h}(A,X)\bigg] \\
                                &= E_0\bigg[\frac{(1-G)\pi_0(X)}{\alpha_0 \{1-\pi_0(X)\}}\{\upsilon_{h}(1,X)-\upsilon_{h}(0,X)\}\bigg]\\
                                &=\frac{1}{\alpha_0}\int \{\upsilon_{h}(1,x)-\upsilon_{h}(0,x)\}\pi_0(x)p_0(x)\d x.
\end{align*}
Therefore, \(\varphi^\dagger(o,w^\dagger_0)\) is a gradient of \(\theta_0\).
To show that it is the efficient influence function, we check that \(\varphi^\dagger(o,w^\dagger_0)\in\dot{\mathcal{P}}^\dagger\).
This amounts to verifying that
\[
  h^{*}(y,a,x,d) = \frac{1}{\alpha_0}\frac{\pi_0(x)}{1-\pi_0(x)}\frac{2a-1}{e_0(a\mid x)}\frac{w_0^\dagger(a,x,d)}{\sum_{d'\in[m]}w_0^\dagger(a,x,d')\zeta(d'\mid a,x)}\{y-\mu_0(a,x)\}
\]
satisfies \(E_0\{Yh^{*}(Y,A,X,D)\mid A,X,D=d\}=E_0\{Yh^{*}(Y,A,X,D)\mid A,X,D=d'\}\).
By direct calculation,
\begin{align*}
  \MoveEqLeft E_0\{Yh^{*}(Y,A,X,D)\mid A=a,X=x,D=d\} \\
  &= \begin{multlined}[t][.9\textwidth]
    \frac{1}{\alpha_0}E_0\bigg[\frac{\pi_0(X)}{1-\pi_0(X)}\frac{2A-1}{e_0(A\mid X)}\frac{w_0^\dagger(A,X,D)}{\sum_{d'\in[m]}w_0^\dagger(A,X,d')\zeta_0(d'\mid A,X)}\\
    \{Y-\mu_0(A,X)\}Y\biggm\vert A=a,X=x,D=d\bigg] 
  \end{multlined}\\
  &= \begin{multlined}[t][.9\textwidth]
    \frac{1}{\alpha_0}E_0\bigg[\frac{\pi_0(X)}{1-\pi_0(X)}\frac{2A-1}{e_0(A\mid X)}\frac{w_0^\dagger(A,X,D)}{\sum_{d'\in[m]}w_0^\dagger(A,X,d')\zeta_0(d'\mid A,X)}\\
    V_0(A,X,D)\biggm\vert A=a,X=x,D=d\bigg]
  \end{multlined}\\
  &= \frac{1}{\alpha_0}\frac{\pi_0(x)}{1-\pi_0(x)}\frac{2a-1}{\sum_{d'\in[m]}w_0^\dagger(a,x,d')\zeta_0(d'\mid a,x)},
\end{align*}
which is constant in \(d\).
This observation concludes the proof of the first part.

For the second part, we follow the arguments in the proof of lemma~\ref{lem:orthogonal}.
Define
\[
  \tilde\Lambda^\dagger=\bigg\{(1-g)r(a,x,d)\{y-\mu_0(a,x)\}:r(a,x,d)\text{ arbitrary}\bigg\}\subset L_2^0(P_0).
\]
The projection of any
\[
  \tilde\lambda_{r}=(1-g)r(a,x,d)\{y-\mu_0(a,x)\}\in\tilde\Lambda^\dagger
\]
onto \(\Lambda^\dagger\) is
\begin{multline*}
  \Pi\{\tilde\lambda_r\mid \Lambda^\dagger\}=(1-g)\bigg[r(a,x,d)-w_0^\dagger(a,x,d)\frac{E_0\{r(A,X,D)\mid A=a,X=x,G=0\}}{E_0\{w_0^\dagger(A,X,D)\mid A=a,X=x,G=0\}}\bigg]\\
  \{y-\mu_0(a,x)\}.
\end{multline*}

Take any \(\ell_{h}(y,a,x,d,g) = (1-g)h(y,a,x,d)\in\tilde{\dot{\mathcal{P}}}_{y}\), its projection onto \(\tilde\Lambda^\dagger\) is
\[
  \Pi\{\ell_h\mid \tilde\Lambda^\dagger\}=(1-g)r_h(a,x,d)\{y-\mu_0(a,x)\},
\]
where
\[
  r_h(a,x,d)=w_0^\dagger(a,x,d)E_0\{Yh(Y,A,X,D)\mid A=a,X=x,D=d\}.
\]
Hence,
\[
  \Pi\{\ell_h\mid \Lambda^\dagger\}=(1-g)q_h(a,x,d)\{y-\mu_0(a,x)\},
\]
where
\[
  q_h(a,x,d)=r_h(a,x,d)-w_0^\dagger(a,x,d)\frac{E_0\{r_h(A,X,D)\mid A=a,X=x,G=0\}}{E_0\{w_0^\dagger(A,X,D)\mid A=a,X=x,G=0\}}.
\]
Now \(\ell_h-\Pi\{\ell_h\mid \Lambda^\dagger\}=(1-g)s_h(y,a,x,d)\),
where
\[
  s_h(y,a,x,d)=h(y,a,x,d)-q_h(a,x,d)\{y-\mu_0(a,x)\}.
\]
We obviously have \(E_0\{s_h(Y,A,X,D)\mid A=a,X=x,D=d\}=0\), and
\begin{align*}
  \MoveEqLeft E_0\{Ys_h(Y,A,X,D)\mid A=a,X=x,D=d\}\\
  &=E_0\{Yh(Y,A,X,D)\mid A=a,X=x,D=d\}-q_h(a,x,d)V_0(a,x,d)\\
  &=\frac{E_0\{r_h(A,X,D)\mid A=a,X=x,G=0\}}{E_0\{w_0^\dagger(A,X,D)\mid A=a,X=x,G=0\}},
\end{align*}
which does not depend on \(d\).
Therefore, \(\Pi\{\ell_h\mid (\Lambda^{\dagger})^\perp\}\in\dot{\mathcal{P}}_y^\dagger\).
Together with the decomposition \(\ell_h=\ell_h-\Pi\{\ell_h\mid \Lambda^\dagger\}+\Pi\{\ell_h\mid \Lambda^\dagger\}\), we conclude that \(\dot{\mathcal{P}}^\dagger=(\Lambda^\dagger)^\perp\).

\subsection{Proof of proposition~\ref{ppn:eif-pcate}}
Consider the parametric submodel \(\{P_{\varepsilon}:\varepsilon\in\Gamma\}\) with score function \(h(o)\).
Suppose
\[
  \gamma_\varepsilon \in \arg\min_{\tilde{\gamma}\in\mathbb{R}^q}E_{P_{\varepsilon}}[\{\delta_{\varepsilon}(X)-\tilde{\gamma}^\T b(Z)\}^2\mid G=1].
\]
Then \(\gamma_\varepsilon\) must fulfill the first-order condition
\[
  E_{P_{\varepsilon}}\big[b(Z)\{\delta_{\varepsilon}(X)-\gamma_\varepsilon^\T b(Z)\}\bigm\vert G=1\big]=0.
\]
By the implicit function theorem, there exists a function \(\gamma_\varepsilon\) of \(\varepsilon\) such that \(\gamma_{\varepsilon}\vert_{\varepsilon=0}=\gamma_0\) and that it is differentiable at \(\varepsilon=0\) with derivative
\[
  \frac{\d}{\d\varepsilon}\gamma_\varepsilon\bigg\vert_{\varepsilon=0} = \big[E_0\big\{b(Z)b^\T(Z)\mid G=1\big\}\big]^{-1}\frac{\d}{\d\varepsilon}E_{P_{\varepsilon}}\big[b(Z)\{\delta_{\varepsilon}(X)-\gamma_0^\T b(Z)\}\bigm\vert G=1\big]\bigg\vert_{\varepsilon=0}.
\]
The uniqueness of \(\gamma_0\) is ensured by assumption~\ref{asn:invertible}.
The Gateaux derivative
\begin{align*}
  \MoveEqLeft\frac{\d}{\d\varepsilon}E_{P_{\varepsilon}}\big[b(Z)\{\delta_{\varepsilon}(X)-\gamma_0^\T b(Z)\}\bigm\vert G=1\big]\big\vert_{\varepsilon=0}\\
  &= \frac{1}{\alpha_0}E_0\bigg[G b(Z)\frac{\d}{\d\varepsilon}\delta_{\varepsilon}(X)\bigg\vert_{\varepsilon=0}\bigg]+\frac{\d}{\d\varepsilon}\frac{E_{P_{\varepsilon}}\big[G b(Z)\{\delta_0(X)-\gamma_0^\T b(Z)\}\big]}{P_\varepsilon(G =1)}\bigg\vert_{\varepsilon=0}\\
  &= \frac{1}{\alpha_0}\int b(z)\kappa_{h}(x)\{1-\pi_0(x)\}p_0(x)\d x \\
  &\hphantom{=}\quad \begin{multlined}[t][.9\textwidth]
    +\frac{1}{\alpha_0}\int
    \big(b(z)\{\delta_0(x)-\gamma_0^\T b(z)\}-\underbrace{E_0[b(Z)\{\delta_0(X)-\gamma_0^\T b(Z)\}
      \mid G=1]}_{=0}\big)\\
    \pi_0(x)p_0(x)\{h(G=1,x)+h(x)\}\d x
  \end{multlined}\\  
  &=\begin{multlined}[t][.9\textwidth]
    E_0\bigg(\bigg[\frac{1-G}{\alpha_0}b(Z)\frac{\pi_0(X)}{1-\pi_0(X)}\frac{w_0(X,D)}{\sum_{d\in[m]}\zeta_0(d\mid X)w_0(X,d)}\frac{2A-1}{e_0(A\mid X,D)}\\\{Y-\mu_0(A,X,D)\}
    +\frac{G}{\alpha_0}\{\delta_0(X)-\gamma_0^\T b(Z)\}\bigg]h(O)\bigg).
  \end{multlined}
\end{align*}
This shows that \(\phi(o,\gamma_0)\) is a gradient of \(\gamma_0\).
Since \(\phi(o,\gamma_0)\in\dot{\mathcal{P}}\), it is the efficient influence function of \(\gamma_0\).

\subsection{Proof of theorem~\ref{thm:asymptotic-tpcate}}
We first show consistency of \(\hat\gamma\).
Define
\[
  \hat{B} = \frac{1}{n_1}\sum_{i:G_i=1}b(Z_i)b^\T(Z_i),\quad B_0 = E_0\{b(Z)b^\T(Z)\mid G=1\}.
\]
Decompose the error by
\begin{align}
  \hat\gamma-\gamma_0 &= \frac{1}{K}\sum_{k\in[K]}\hat{B}^{-1}\mathbb{P}_{n,k}\{b(Z)\ell(O,\hat{\eta}_k)\} - \gamma_0 \nonumber\\
                    &= \frac{1}{K}\sum_{k\in[K]}\hat{B}^{-1}(\mathbb{P}_{n,k}-P_0)\{b(Z)\ell(O,\hat{\eta}_k)\}\nonumber \\
                    &\hphantom{=}\quad +\frac{1}{K}\sum_{k\in[K]}\hat{B}^{-1}\bigg[P_0\{b(Z)\ell(O,\hat{\eta}_k)\}-\frac{\alpha_0}{\hat\alpha_k}B_0\gamma_0\bigg] \nonumber\\
                    &\hphantom{=} \quad +\frac{1}{K}\sum_{k\in[K]}\bigg[\frac{\alpha_0}{\hat\alpha_k}\hat{B}^{-1}B_0 - \mathrm{Id}\bigg]\gamma_0.\label{eqn:decomp-tpcate}
\end{align}
By the law of large numbers, every entry of the matrix \(\hat{B}_{jk}\) converges in probability to the corresponding entry in \((B_0)_{jk}\) for \(j,k\in[q]\).
Since the dimension \(q\) is finite, we have
\[
  \|\hat{B}-B_0\|\leq \bigg\{\sum_{j,k\in[q]}|\hat{B}_{jk}-(B_0)_{jk}|^2\bigg\}^{1/2}=o_{P_0}(1).
\]
By the continuous mapping theorem, \(\|\hat{B}^{-1}-B_0^{-1}\|=o_{P_0}(1)\).
In the following, we combine the matrix-norm convergence with vector-norm convergence using Slutsky's theorem, since \(\|A_{n}v_{n}\|=o_{P_0}(1)\) if \(\|A_{n}\|\|v_n\|=o_{P_0}(1)\).
The first term of \eqref{eqn:decomp-tpcate} converges in probability to \(0\) due to the boundedness of \(|b_j(z)|\) for \(j\in[q]\) and lemmas~\ref{lem:cross-fitting}--\ref{lem:boundedness-ell}.
Consider the third term of \eqref{eqn:decomp-tpcate}.
It converges in probability to \(0\) because \(\hat\alpha_k-\alpha_0=o_{P_0}(1)\), and by the continuous mapping theorem \(\|(\alpha_0/\hat\alpha_k)\hat{B}^{-1}B_0 - \mathrm{Id}\|=o_{P_0}(1)\).
The second term also converges in probability to \(0\) because
\begin{align*}
  \MoveEqLeft\bigg\|P_0\{b(Z)\ell(O,\hat{\eta}_k)\}-\frac{\alpha_0}{\hat\alpha_k}B_0\gamma_0\bigg\| \\
  &= \bigg\|P_0\bigg[b(Z)\bigg\{\ell(O,\hat{\eta}_k)-\frac{G}{\hat\alpha_k}\delta_0(X)+\frac{G}{\hat\alpha_k}\delta_0(X)-\frac{G}{\hat\alpha_k} b^\T(Z)\gamma_0\bigg\}\bigg]\bigg\| \\
  &= \bigg\|P_0\bigg[b(Z)\bigg\{\ell(O,\hat{\eta}_k)-\frac{G}{\hat\alpha_k}\delta_0(X)\bigg\}\bigg]\bigg\|,
\end{align*}
which can be seen to be \(o_{P_0}(1)\) by modifying the steps of bounding \(|P_0\ell(O,\hat\eta_k)-(\alpha_0/\hat\alpha_k)\theta_0|\) in the proof of theorem~\ref{thm:asymptotic} and by using that \(|b_j(z)|\leq C\).
The consistency of \(\hat\gamma\) is established.

To show asymptotic linearity, we further make the following decomposition:
\begin{align}
  \hat\gamma-\gamma_0 &= (\mathbb{P}_{n}-P_0)\phi(O,\bar{w}) \nonumber\\
                    &\hphantom{=}\quad + (\hat{B}^{-1}-B_0^{-1})(\mathbb{P}_{n}-P_0)\bigg[b(Z)\bigg\{\ell(O,\bar\eta)-\frac{G}{\alpha_0}b^\T(Z)\gamma_0\bigg\}\bigg] \nonumber\\
                    &\hphantom{=}\quad + \frac{1}{K}\sum_{k\in[K]}\hat{B}^{-1}(\mathbb{P}_{n,k}-P_0)[b(Z)\{\ell(O,\hat{\eta}_k)-\ell(O,\bar\eta)\}] \nonumber\\
                    &\hphantom{=}\quad +\frac{1}{K}\sum_{k\in[K]}\hat{B}^{-1}\bigg[P_0\{b(Z)\ell(O,\hat{\eta}_k)\}-\frac{\alpha_0}{\hat\alpha_k}B_0\gamma_0\bigg] \nonumber\\
                    &\hphantom{=} \quad +\frac{1}{K}\sum_{k\in[K]}\frac{\alpha_0-\hat\alpha_k}{\hat\alpha_k}(\hat{B}^{-1}B_0 - \mathrm{Id})\gamma_0\nonumber\\
                    &\hphantom{=}\quad + \frac{1}{K}\sum_{k\in[K]}\frac{(\hat\alpha_k-\alpha_0)^2}{\alpha_0\hat\alpha_k}\gamma_0.\label{eqn:decomp-tpcate-2}
\end{align}
Following the arguments in the proof of theorem~\ref{thm:asymptotic}, all terms of \eqref{eqn:decomp-tpcate-2} are \(o_{P_0}(n^{-1/2})\).

\subsection{Proof of corollary~\ref{cor:asymptotic-tpcate-uniform}}



\begin{lemma}
  \label{lem:consistency-var}
  Suppose assumption~\ref{asn:regularity} holds.
  Then \(\hat\psi-\bar\psi=o_{P_0}(1)\) if \(\hat\theta-\theta_0=o_{P_0}(1)\).
\end{lemma}

\begin{proof}
  Since \(\hat\psi=\sum_{k\in[K]}\hat\psi_k/K\) for a finite number of splits, it suffices to show consistency of every \(\hat\psi_{k}\).
  We make the following decomposition:
  \begin{align}
    \hat\psi_{k} &= \mathbb{P}_{n,k}\bigg\{\ell(O,\hat\eta_k)-\frac{G}{\hat\alpha_k}\hat\theta\bigg\}^2 \nonumber\\
                   &= \mathbb{P}_{n,k}\bigg[\bigg\{\ell(O,\hat\eta_k)-\frac{G}{\hat\alpha_k}\hat\theta\bigg\} - \bigg\{\ell(O,\bar\eta)-\frac{G}{\alpha_0}\theta_0\bigg\}\bigg]^2\nonumber\\
                   &\hphantom{=}\quad + 2\mathbb{P}_{n,k}\bigg[\bigg\{\ell(O,\hat\eta_k)-\frac{G}{\hat\alpha_k}\hat\theta\bigg\} - \bigg\{\ell(O,\bar\eta)-\frac{G}{\alpha_0}\theta_0\bigg\}\bigg]\bigg\{\ell(O,\bar\eta)-\frac{G}{\alpha_0}\theta_0\bigg\}\nonumber\\
                   &\hphantom{=}\quad +\mathbb{P}_{n,k}\bigg\{\ell(O,\bar\eta)-\frac{G}{\alpha_0}\theta_0\bigg\}^2.\label{eqn:plugin-var}
  \end{align}
  The third term in \eqref{eqn:plugin-var} converges in probability to \(\bar\psi\) by the law of large numbers.
  By the Cauchy-Schwarz inequality, the second term in \eqref{eqn:plugin-var} is bounded by the product of the square root of the first term and \(\bar\psi^{1/2}\).
  It remains to show that the first term in \eqref{eqn:plugin-var} is \(o_{P_0}(1)\).
  We have
  \begin{multline}
    \mathbb{P}_{n,k}\bigg[\bigg\{\ell(O,\hat\eta_k)-\frac{G}{\hat\alpha_k}\hat\theta\bigg\} - \bigg\{\ell(O,\bar\eta)-\frac{G}{\alpha_0}\theta_0\bigg\}\bigg]^2 \\
    \leq 2\mathbb{P}_{n,k}\{\ell(O,\hat\eta_k)-\ell(O,\bar\eta)\}^2 + 2\hat\alpha_k\bigg(\frac{\hat\theta}{\hat\alpha_k}-\frac{\theta_0}{\alpha_0}\bigg)^2\label{eq:markov-var}
  \end{multline}
  Applying the Markov inequality conditional on the data \(\mathcal{O}_{-k}=\{O_i:i\notin \mathcal{I}_{k}\}\), for any \(t>0\),
  \[
    \Pr\bigg[\frac{\mathbb{P}_{n,k}\{\ell(O,\hat\eta_k)-\ell(O,\bar\eta)\}^2}{\|\ell(O,\hat\eta_k)-\ell(O,\bar\eta)\|_{P_0}^2}\geq t\biggm\vert \mathcal{O}_{-k}\bigg]\leq \frac{1}{t}.
  \]
  Hence, marginally \(\mathbb{P}_{n,k}\{\ell(O,\hat\eta_k)-\ell(O,\bar\eta)\}^2=O_{P_0}\{\|\ell(O,\hat\eta_k)-\ell(O,\bar\eta)\|_{P_0}^2\}\), which is \(o_{P_0}(1)\) as shown in the proof of theorem~\ref{thm:asymptotic}.
  The second term on the right-hand side of \eqref{eq:markov-var} is also \(o_{P_0}(1)\) from the consistency of \(\hat\alpha\) and \(\hat\theta\) followed by an application of Slutsky's theorem.
\end{proof}

\begin{lemma}
  \label{lem:consistency-var-matrix}
  Suppose assumptions~\ref{asn:regularity}, \ref{asn:invertible}, and \ref{asn:bounded-psi} hold.
  Then \(\|\hat\Psi-\bar\Psi\|=o_{P_0}(1)\) if \(\|\hat\gamma-\gamma_0\|=o_{P_0}(1)\).
\end{lemma}

\begin{proof}
  From the proof of theorem~\ref{thm:asymptotic-tpcate}, we have \(\|\hat{B}^{-1}-B_0^{-1}\|=o_{P_0}(1)\).
  Let
  \begin{align*}
    \hat{Q}_{k} &= \mathbb{P}_{n,k}\bigg[b(Z)b^\T(Z)\bigg\{\ell(O,\hat\eta_k)-\frac{G}{\hat\alpha_k}\hat\gamma^\T b(Z)\bigg\}^2\bigg],\\
    Q &= E_{0}\bigg[b(Z)b^\T(Z)\bigg\{\ell(O,\bar\eta)-\frac{G}{\alpha_0}\gamma_0^\T b(Z)\bigg\}^2\bigg].
  \end{align*}
  Then for \(j,j'\in[q]\), every entry \((\hat{Q}_k)_{jj'}\) can be decomposed as
  \begin{align*}
    \MoveEqLeft(\hat{Q}_k)_{jj'}\\
    &= \mathbb{P}_{n,k}\bigg[b_{j}(Z)b_{j'}(Z)\bigg\{\ell(O,\hat\eta_k)-\frac{G}{\hat\alpha_k}\hat\gamma^\T b(Z)\bigg\}^2\bigg] \nonumber\\
    &= \mathbb{P}_{n,k}\bigg(b_{j}(Z)b_{j'}(Z)\bigg[\bigg\{\ell(O,\hat\eta_k)-\frac{G}{\hat\alpha_k}\hat\gamma^\T b(Z)\bigg\} - \bigg\{\ell(O,\bar\eta)-\frac{G}{\alpha_0}\gamma_0^\T b(Z)\bigg\}\bigg]^2\bigg)\nonumber\\
    &\hphantom{=}\quad + \begin{multlined}[t][.8\textwidth]
      2\mathbb{P}_{n,k}\bigg(b_{j}(Z)b_{j'}(Z)\bigg[\bigg\{\ell(O,\hat\eta_k)-\frac{G}{\hat\alpha_k}\hat\gamma^\T b(Z)\bigg\} - \bigg\{\ell(O,\bar\eta)-\frac{G}{\alpha_0}\gamma_0^\T b(Z)\bigg\}\bigg]\\
      \bigg\{\ell(O,\bar\eta)-\frac{G}{\alpha_0}\gamma_0^\T b(Z)\bigg\}\bigg)
    \end{multlined}
    \nonumber\\
    &\hphantom{=}\quad +\mathbb{P}_{n,k}\bigg[b_{j}(Z)b_{j'}(Z)\bigg\{\ell(O,\bar\eta)-\frac{G}{\alpha_0}\gamma_0^\T b(Z)\bigg\}^2\bigg].\nonumber
  \end{align*}
  Following the proof of lemma~\ref{lem:consistency-var}, we can show that \(|\hat{Q}_{jj'}-Q_{jj'}|=o_{P_0}(1)\) if \(\|\hat\gamma-\gamma_0\|=o_{P_0}(1)\), because \(\sup_{z\in\mathcal{Z}}|b_{j}(z)|\leq C\).
  Since the dimension \(q\) does not depend on \(n\), we have \(\|\hat{Q}-Q\|=o_{P_0}(1)\).
  The consistency of \(\hat\Psi_k\) can now be established by the continuous mapping theorem, hence the consistency of \(\hat\Psi\).
\end{proof}

\begin{proof}[Proof of corollary~\ref{cor:asymptotic-tpcate-uniform}]
  In the following, we treat \(z\in\mathcal{Z}\) as an indexing parameter.
  Let \(\bar\omega^2(z)=b^\T(z)\bar\Psi b(z)\) and \(\hat\omega^2(z)=b^\T(z)\hat\Psi b(z)\).
  Define three stochastic processes
  \begin{align*}
    \hat{\mathbb{T}}_n(z) &= \frac{n^{1/2}b^\T(z)(\hat\gamma-\gamma_0)}{\hat\omega(z)},\\
    \tilde{\mathbb{T}}_n(z) &= \frac{n^{1/2}b^\T(z)(\hat\gamma-\gamma_0)}{\bar\omega(z)},\\
    {\mathbb{T}}_n(z) &= \frac{n^{1/2}(\mathbb{P}_n-P_0)\{b^\T(z)\phi(O,\bar{w})\}}{\bar\omega(z)}\\
  \end{align*}
  Define the function class
  \[
    \mathcal{G}=\bigg\{\frac{b^\T(z)\phi(o,\bar{w})}{\bar\omega(z)}:z\in\mathcal{Z}\bigg\}.
  \]
  By assumption~\ref{asn:bounded-psi}, \(\bar\omega^{-1}(z)\leq \lambda_{\mathrm{min}}^{-1/2}(\bar\Psi)\|b(z)\|^{-1}\lesssim 1\) and \(\|\bar\Psi b(z)b^\T(z)\|\leq \lambda_{\mathrm{max}}(\bar\Psi)\|b(z)\|^2\lesssim 1\).
  The partial derivative satisfies
  \[
    \frac{\partial}{\partial\{b(z)\}}\frac{b^\T(z)\phi(o,\bar{w})}{\bar\omega(z)} \lesssim \|\phi(o,\bar{w})\|.
  \]
  Applying theorems 2.7.17 and 2.5.6 from \citetsuppmat{vandervaart2023weak} in this order, we see that \(\mathcal{G}\) is \(P_0\)-Donsker, since
  \[
    \bigg|\frac{b^\T(z)\phi(o,\bar{w})}{\bar\omega(z)}-\frac{b^\T(z')\phi(o,\bar{w})}{\bar\omega(z')}\bigg|\lesssim \|\phi(o,\bar{w})\|\|b(z)-b(z')\|
  \]
  and \(\{b(z):C^{-1}\leq \|b(z)\|\leq C,z\in\mathcal{Z}\}\) represents a compact set in \(\mathbb{R}^q\).
  We have
  \[
    \mathbb{T}_n\leadsto \mathbb{T}, \qquad\text{in }\ell^{\infty}(\mathcal{Z}_1),
  \]
  where \(\mathbb{T}\) is the mean-zero Gaussian process stated in the corollary.

  We now show that \(\|\hat{\mathbb{T}}_n-\mathbb{T}_n\|_{\mathcal{Z}_1}=o_{P_0}(1)\), which by the continuous mapping theorem \citepsuppmat[theorem 1.3.6]{vandervaart2023weak}, proves the corollary combined with the weak convergence above.
  The standard error estimator is uniformly consistent, because
  \begin{align*}
    \bigg\|\frac{\hat{\omega}}{\bar{\omega}}-1\bigg\|_{\mathcal{Z}} &\leq \bigg\|\frac{\hat{\omega}^2}{\bar{\omega}^2}-1\bigg\|_{\mathcal{Z}} \\
                                                                                  &\leq \sup_{z\in\mathcal{Z}}\bigg|\frac{b^\T(z)(\hat\Psi-\bar\Psi)b(z)}{b^\T(z)\bar\Psi b(z)}\bigg| \\
                                                                                  &\leq \lambda_{\mathrm{\min}}^{-1}(\bar\Psi)\|\hat\Psi-\bar\Psi\|=o_{P_0}(1),
  \end{align*}
  which follows from the consistency of \(\hat\Psi\) in lemma~\ref{lem:consistency-var-matrix}.
  We bound the difference of the two stochastic processes by
  \begin{align*}
    \|\hat{\mathbb{T}}_n-\mathbb{T}_n\|_{\mathcal{Z}} &= \bigg\|(\tilde{\mathbb{T}}_n-\mathbb{T}_n)\frac{\bar{\omega}}{\hat{\omega}}+\mathbb{T}_n\bigg(\frac{\bar{\omega}}{\hat{\omega}}-1\bigg)\bigg\|_{\mathcal{Z}} \\
                                                        &\leq \|\tilde{\mathbb{T}}_n-\mathbb{T}_n\|_{\mathcal{Z}}\bigg\|\frac{\bar{\omega}}{\hat{\omega}}\bigg\|_{\mathcal{Z}}+\|\mathbb{T}_n\|_{\mathcal{Z}}\bigg\|\frac{\bar{\omega}}{\hat{\omega}}-1\bigg\|_{\mathcal{Z}}.
  \end{align*}
  Since \(\|\bar{\omega}/\hat{\omega}\|_{\mathcal{Z}}=1+o_{P_0}(1)=O_{P_0}(1)\) and \(\|\mathbb{T}_n\|_{\mathcal{Z}}=O_{P_0}(1)\) from theorem 2.14.2 in \citetsuppmat{vandervaart2023weak}, it remains to show \(\|\tilde{\mathbb{T}}_n-\mathbb{T}_n\|_{\mathcal{Z}}=o_{P_0}(1)\).
  Clearly,
  \begin{align*}
    \tilde{\mathbb{T}}_n(z)-\mathbb{T}_n(z) &= \frac{n^{1/2}b^\T(z)(\hat\gamma-\gamma_0)}{\bar\omega(z)} - \frac{n^{1/2}(\mathbb{P}_n-P_0)\{b^\T(z)\phi(O,\bar{w})\}}{\bar\omega(z)} \\
                                            &= \frac{n^{1/2}b^\T(z)}{\bar\omega(z)}R_n,
  \end{align*}
  where we have used \eqref{eqn:decomp-tpcate-2} from the proof of theorem~\ref{thm:asymptotic-tpcate}, and \(R_n\) denotes the \(o_{P_0}(n^{-1/2})\) terms in \eqref{eqn:decomp-tpcate-2}.
  Since \(\|b(z)\|\lesssim 1\) and \(\bar\omega^{-1}(z)\leq \lambda_{\mathrm{min}}^{-1/2}(\bar\Psi)\|b(z)\|^{-1}\lesssim 1\), we have \(\|\tilde{\mathbb{T}}_n-\mathbb{T}_n\|_{\mathcal{Z}}=o_{P_0}(1)\).
\end{proof}

\subsection{Proof of proposition~\ref{ppn:identifiability-star}}
  For \(x\in\mathcal{X}\),
  \begin{align*}
    \MoveEqLeft E\{Y(1)-Y(0)\mid X=x,G=1\}\\
    &= E\{Y(1)-Y(0)\mid X=x,D=d\} \\
    &= E[E\{Y(1)-Y(0)\mid X_D,X,D\}\mid X=x,D=d] \\
    &= E[E\{Y(1)\mid A=1,X_D,X,D\}-E\{Y(0)\mid A=0,X_D,X,D\}\mid X=x,D=d] \\
    &= E\{\mu^*(1,X_D,X,D)-\mu^*(0,X_D,X,D)\mid X=x,D=d\}\\
    &= \delta^*(x).
  \end{align*}
  Then by iterated expectation,
  \begin{multline*}
    \theta=E\{Y(1)-Y(0)\mid G=1\}\\
    =E[E\{Y(1)-Y(0)\mid X,G=1\}\mid G=1] = E\{\delta^*(X)\mid G=1\}.
  \end{multline*}

\subsection{Proof of proposition~\ref{ppn:asymptotic-star}}
Before proving proposition~\ref{ppn:asymptotic-star}, we show the following lemmas.

\begin{lemma}
  \label{lem:boundedness-ell-star}
  If assumption~\ref{asn:regularity-star} is satisfied, then \(\|\ell^*(O^*,\hat\eta^*_k)\|_{P_0}=O_{P_0}(1)\) for every \(k\in[K]\).
\end{lemma}

\begin{proof}
  Let
  \begin{align*}
    H^*_{1k}(a,x_d,x,d) &= \frac{\hat\pi_k(x)}{1-\hat\pi_k(x)}\frac{\hat{w}^*_k(x,d)}{\sum_{d'\in[m]}\hat\zeta_k(d'\mid x)\hat{w}^*_k(x,d')}\frac{1}{\hat{e}^*_k(a\mid x_d,x,d)}\lesssim 1,\\
    H^*_{2k}(x,d) &= \frac{\hat\pi_k(x)}{1-\hat\pi_k(x)}\frac{\hat{w}^*_k(x,d)}{\sum_{d'\in[m]}\hat\zeta_k(d'\mid x)\hat{w}^*_k(x,d')}\lesssim 1.
  \end{align*}
  Note that
  \[
    |\hat\delta^*_k(x)|\leq \sum_{d\in[m]}\frac{|\hat{w}^*_k(x,d)\hat\zeta_k(d\mid x)\hat\tau_k(x,d)|}{\big|\sum_{d'\in[m]}\hat{w}^*_k(x,d')\hat\zeta_k(d'\mid x)\big|}\lesssim 1.
  \]
  Then we have
  \begin{align*}
    \MoveEqLeft \|\ell^*(O,\hat\eta_k)\|_{P_0}\\
    &\lesssim{} P_0[(1-G)\{H^*_{1k}(A,X_D,X,D)\}^2\{Y-\hat\mu^*_k(A,X_D,X,D)\}^2] \\
    &\hphantom{\lesssim{}}+ P_0[(1-G)\{H^*_{2k}(X,D)\}^2\{\hat\mu^*_k(1,X_D,X,D)-\hat\mu^*_k(0,X_D,X,D)-\hat\delta^*_k(X)\}^2]\\
    &\hphantom{\lesssim{}}+ P_0[G\{\hat\delta^*_k(X)\}^2]\\
    &\lesssim{} P_0[(1-G)\{Y-\mu^*_0(A,X_D,X,D)\}^2] \\
    &\hphantom{\lesssim{}}+ P_0[(1-G)\{(\hat\mu^*_k-\bar\mu^*)^2+(\bar\mu^*-\mu^*_0)^2\}(A,X_D,X,D)] \\
    &\hphantom{\lesssim{}}+ P_0[(1-G)\{(\hat\mu^*_k-\bar\mu^*)^2(1,X_D,X,D)+(\hat\mu^*_k-\bar\mu^*)^2(0,X_D,X,D)\}]\\
    &\hphantom{\lesssim{}}+ P_0((1-G)[\{\bar\mu^*(1,X_D,X,D)\}^2+\{\bar\mu^*(0,X_D,X,D)\}^2]) \\
    &\hphantom{\lesssim{}}+ P_0[(1-G)\{\hat\delta^*_k(X)\}^2]+ 1\\
    &\lesssim{} P_0[(1-G)V^*_0(A,X_D,X,D)]+\max_{d\in[m]}\max_{a\in\{0,1\}}\|(\hat\mu^*_k-\bar\mu^*)(a,X_d,X,d)\|_{P_0} + 1\\
    &\lesssim{} o_{P_0}(1)+1 = O_{P_0}(1).
  \end{align*}
\end{proof}

Consider the nuisance parameter
\[
  \bar\eta^*=\{\alpha_0,\pi_0(x),\zeta_0(d\mid x),\bar{w}^*(x,d),e^*_0(a\mid x_d,x,d),\mu^*_0(a,x_d,x,d),\tau_0(x,d)\}.
\]

\begin{lemma}
  \label{lem:convergence-ell-star}
  If assumptions~\ref{asn:regularity-star} and \ref{asn:model-linearity-star}\ref{asn:plim-star} are satisfied, then \(\|\ell^*(O^*,\hat\eta^*_k)-\ell^*(O^*,\bar\eta^*)\|_{P_0}=o_{P_0}(1)\) for every \(k\in[K]\).
\end{lemma}

\begin{proof}
  We decompose the difference \(\ell^*(o^*,\hat\eta^*_k)-\ell^*(o^*,\bar\eta^*)\) into
  \begin{align*}
    \MoveEqLeft\ell^*(o^*,\hat\eta^*_k)-\ell^*(o^*,\bar\eta^*)\\
    &= \frac{1-g}{\hat\alpha_k}\frac{\hat\pi_k(x)}{1-\hat\pi_k(x)}\frac{\hat{w}^*_k(x,d)}{\sum_{d'\in[m]}\hat\zeta_k(d'\mid x)\hat{w}^*_k(x,d')}\frac{2a-1}{\hat{e}^*_k(a\mid x_d,x,d)}(\mu^*_0-\hat\mu^*_k)(a,x_d,x,d)\\
    &\hphantom{={}}+\begin{multlined}[t][.85\textwidth]
      \frac{1-g}{\hat\alpha_k}\frac{\hat\pi_k(x)}{1-\hat\pi_k(x)}\frac{\hat{w}^*_k(x,d)}{\sum_{d'\in[m]}\hat\zeta_k(d'\mid x)\hat{w}^*_k(x,d')}\\
      (2a-1)\frac{(e^*_0-\hat{e}^*_k)(a\mid x_d,x,d)}{(e^*_0\hat{e}^*_k)(a\mid x_d,x,d)}\{y-\mu^*_0(a,x_d,x,d)\}
    \end{multlined}\\
    &\hphantom{={}}+\frac{1-g}{\hat\alpha_k}\frac{\hat\pi_k(x)}{1-\hat\pi_k(x)}\frac{(\hat{w}^*_k-\bar{w}^*)(x,d)}{\sum_{d'\in[m]}\hat\zeta_k(d'\mid x)\hat{w}^*_k(x,d')}\frac{2a-1}{e^*_0(a\mid x_d,x,d)}\{y-\mu^*_0(a,x_d,x,d)\}\\
    &\hphantom{={}}+\begin{multlined}[t][.85\textwidth]
      \frac{1-g}{\hat\alpha_k}\frac{\hat\pi_k(x)}{1-\hat\pi_k(x)}\frac{\bar{w}^*(x,d)\sum_{d'\in[m]}(\zeta_0-\hat\zeta_k)(d'\mid x)\hat{w}^*_{k}(x,d')}{\sum_{d'\in[m]}\hat\zeta_k(d'\mid x)\hat{w}^*_k(x,d')\sum_{d'\in[m]}\zeta_0(d'\mid x)\bar{w}^*(x,d')}\\
      \frac{2a-1}{e^*_0(a\mid x_d,x,d)}\{y-\mu^*_0(a,x_d,x,d)\}
    \end{multlined}\\
    &\hphantom{={}}+\begin{multlined}[t][.85\textwidth]
      \frac{1-g}{\hat\alpha_k}\frac{\hat\pi_k(x)}{1-\hat\pi_k(x)}\frac{\bar{w}^*(x,d)\sum_{d'\in[m]}\zeta_0(d'\mid x)(\bar{w}^*-\hat{w}^*_{k})(x,d')}{\sum_{d'\in[m]}\hat\zeta_k(d'\mid x)\hat{w}^*_k(x,d')\sum_{d'\in[m]}\zeta_0(d'\mid x)\bar{w}^*(x,d')}\\
      \frac{2a-1}{e^*_0(a\mid x_d,x,d)}\{y-\mu^*_0(a,x_d,x,d)\}
    \end{multlined}\\
    &\hphantom{={}}+\begin{multlined}[t][.85\textwidth]
      \frac{1-g}{\hat\alpha_k}\frac{(\hat\pi_k-\pi_0)(x)}{\{1-\pi_0(x)\}\{1-\hat\pi_k(x)\}}\frac{\bar{w}^*(x,d)}{\sum_{d'\in[m]}\zeta_0(d'\mid x)\bar{w}^*(x,d')}\\
      \frac{2a-1}{e^*_0(a\mid x_d,x,d)}\{y-\mu^*_0(a,x_d,x,d)\}
    \end{multlined}\\
    &\hphantom{={}}+\begin{multlined}[t][.85\textwidth]
      (1-g)\frac{\alpha_0-\hat\alpha_k}{\alpha_0\hat\alpha_k}\frac{\pi_0(x)}{1-\pi_0(x)}\frac{\bar{w}^*(x,d)}{\sum_{d'\in[m]}\zeta_0(d'\mid x)\bar{w}^*(x,d')}\\
      \frac{2a-1}{e^*_0(a\mid x_d,x,d)}\{y-\mu^*_0(a,x_d,x,d)\}
    \end{multlined}\\
    &\hphantom{={}}+ \frac{1-g}{\hat\alpha_k}\frac{\hat\pi_k(x)}{1-\hat\pi_k(x)}\frac{\hat{w}^*_k(x,d)}{\sum_{d'\in[m]}\hat\zeta_k(d'\mid x)\hat{w}^*_k(x,d')}(\delta^*_0-\hat\delta^*_k)(x)\\
    &\hphantom{={}}+ \frac{1-g}{\hat\alpha_k}\frac{\hat\pi_k(x)}{1-\hat\pi_k(x)}\frac{\hat{w}^*_k(x,d)}{\sum_{d'\in[m]}\hat\zeta_k(d'\mid x)\hat{w}^*_k(x,d')}\sum_{a'\in\{0,1\}}(2a'-1)(\hat\mu^*_k-\mu^*_0)(a',x_d,x,d)\\
    &\hphantom{={}}+\begin{multlined}[t][.85\textwidth]
      \frac{1-g}{\hat\alpha_k}\frac{\hat\pi_k(x)}{1-\hat\pi_k(x)}\frac{(\hat{w}^*_k-\bar{w}^*)(x,d)}{\sum_{d'\in[m]}\hat\zeta_k(d'\mid x)\hat{w}^*_k(x,d')}\\
      \{\mu^*_0(1,x_d,x,d)-\mu^*_0(0,x_d,x,d)-\delta^*_0(x)\}
    \end{multlined}\\
    &\hphantom{={}}+\begin{multlined}[t][.85\textwidth]
      \frac{1-g}{\hat\alpha_k}\frac{\hat\pi_k(x)}{1-\hat\pi_k(x)}\frac{\bar{w}^*(x,d)\sum_{d'\in[m]}(\zeta_0-\hat\zeta_k)(d'\mid x)\hat{w}^*_{k}(x,d')}{\sum_{d'\in[m]}\hat\zeta_k(d'\mid x)\hat{w}^*_k(x,d')\sum_{d'\in[m]}\zeta_0(d'\mid x)\bar{w}^*(x,d')}\\
      \{\mu^*_0(1,x_d,x,d)-\mu^*_0(0,x_d,x,d)-\delta^*_0(x)\}
    \end{multlined}\\
    &\hphantom{={}}+\begin{multlined}[t][.85\textwidth]
      \frac{1-g}{\hat\alpha_k}\frac{\hat\pi_k(x)}{1-\hat\pi_k(x)}\frac{\bar{w}^*(x,d)\sum_{d'\in[m]}\zeta_0(d'\mid x)(\bar{w}^*-\hat{w}^*_{k})(x,d')}{\sum_{d'\in[m]}\hat\zeta_k(d'\mid x)\hat{w}^*_k(x,d')\sum_{d'\in[m]}\zeta_0(d'\mid x)\bar{w}^*(x,d')}\\
     \{\mu^*_0(1,x_d,x,d)-\mu^*_0(0,x_d,x,d)-\delta^*_0(x)\}
    \end{multlined}\\
    &\hphantom{={}}+\begin{multlined}[t][.85\textwidth]
      \frac{1-g}{\hat\alpha_k}\frac{(\hat\pi_k-\pi_0)(x)}{\{1-\pi_0(x)\}\{1-\hat\pi_k(x)\}}\frac{\bar{w}^*(x,d)}{\sum_{d'\in[m]}\zeta_0(d'\mid x)\bar{w}^*(x,d')}\\
      \{\mu^*_0(1,x_d,x,d)-\mu^*_0(0,x_d,x,d)-\delta^*_0(x)\}
    \end{multlined}\\
    &\hphantom{={}}+\begin{multlined}[t][.85\textwidth]
      (1-g)\frac{\alpha_0-\hat\alpha_k}{\alpha_0\hat\alpha_k}\frac{\pi_0(x)}{1-\pi_0(x)}\frac{\bar{w}^*(x,d)}{\sum_{d'\in[m]}\zeta_0(d'\mid x)\bar{w}^*(x,d')}\\
      \{\mu^*_0(1,x_d,x,d)-\mu^*_0(0,x_d,x,d)-\delta^*_0(x)\}
    \end{multlined}\\
    &\hphantom{={}}+\frac{g}{\hat\alpha_k}(\hat\delta^*_k-\delta^*_0)(x)+g\frac{\alpha_0-\hat\alpha_k}{\alpha_0\hat\alpha_k}\delta^*_0(x).
  \end{align*}

  Then by the triangular inequality,
  \begin{align*}
    \MoveEqLeft \|\ell^*(O^*,\hat\eta^*_k)-\ell^*(O^*,\bar\eta^*)\|_{P_0}\\
    &\leq \begin{multlined}[t][.85\textwidth]
      \bigg(P_0\bigg[\frac{1-\pi_0(X)}{\hat\alpha_k^2}\bigg\{\frac{\hat\pi_k(X)}{1-\hat\pi_k(X)}\bigg\}^2\sum_{d\in[m]}\frac{\{\hat{w}^*_k(X,d)\}^2\zeta_0(d\mid X)}{\big\{\sum_{d'\in[m]}\hat\zeta_k(d'\mid X)\hat{w}^*_k(X,d')\big\}^2}\\
      E_0\bigg\{\sum_{a\in\{0,1\}}\frac{e^*_0(a\mid X_d,X,d)}{\{\hat{e}^*_k(a\mid X_d,X,d)\}^2}(\mu^*_0-\hat\mu^*_k)^2(a,X_d,X,d)\biggm\vert X,D=d\bigg\}\bigg]\bigg)^{1/2}
    \end{multlined}\\
    &\hphantom{\leq{}}+\begin{multlined}[t][.85\textwidth]
      \bigg(P_0\bigg[\frac{1-\pi_0(X)}{\hat\alpha_k^2}\bigg\{\frac{\hat\pi_k(X)}{1-\hat\pi_k(X)}\bigg\}^2\sum_{d\in[m]}\frac{\{\hat{w}^*_k(X,d)\}^2\zeta_0(d\mid X)}{\big\{\sum_{d'\in[m]}\hat\zeta_k(d'\mid X)\hat{w}^*_k(X,d')\big\}^2}\\
      E_0\bigg\{\sum_{a\in\{0,1\}}\frac{(e^*_0-\hat{e}^*_k)^2(a\mid X_d,X,d)}{(e^*_0\hat{e}^*_k)^2(a\mid X_d,X,d)}V^*_0(a,X_d,X,d)\biggm\vert X,D=d\bigg\}\bigg]\bigg)^{1/2}
    \end{multlined}\\
    &\hphantom{\leq{}}+\begin{multlined}[t][.85\textwidth]
      \bigg(P_0\bigg[\frac{1-\pi_0(X)}{\hat\alpha_k^2}\bigg\{\frac{\hat\pi_k(X)}{1-\hat\pi_k(X)}\bigg\}^2\\
      \sum_{d\in[m]}\frac{(\hat{w}^*_k-\bar{w}^*)^2(X,d)\zeta_0(d\mid X)\{w^*_0(X,d)\}^{-1}}{\big\{\sum_{d'\in[m]}\hat\zeta_k(d'\mid X)\hat{w}^*_k(X,d')\big\}^2}\bigg]\bigg)^{1/2}
    \end{multlined}\\
    &\hphantom{\leq{}}+\begin{multlined}[t][.85\textwidth]
      \bigg(P_0\bigg[\frac{1-\pi_0(X)}{\hat\alpha_k^2}\bigg\{\frac{\hat\pi_k(X)}{1-\hat\pi_k(X)}\bigg\}^2\bigg\{\sum_{d'\in[m]}(\zeta_0-\hat\zeta_k)(d'\mid X)\hat{w}^*_k(X,d')\bigg\}^2\\
      \sum_{d\in[m]}\frac{\{\bar{w}^*(X,d)\}^2\zeta_0(d\mid X)\{w^*_0(X,d)\}^{-1}}{\big\{\sum_{d'\in[m]}\hat\zeta_k(d'\mid X)\hat{w}^*_k(X,d')\big\}^2\big\{\sum_{d'\in[m]}\zeta_0(d'\mid X)\bar{w}^*(X,d')\big\}^2}\bigg]\bigg)^{1/2}
    \end{multlined}\\
    &\hphantom{\leq{}}+\begin{multlined}[t][.85\textwidth]
      \bigg(P_0\bigg[\frac{1-\pi_0(X)}{\hat\alpha_k^2}\bigg\{\frac{\hat\pi_k(X)}{1-\hat\pi_k(X)}\bigg\}^2\bigg\{\sum_{d'\in[m]}\zeta_0(d'\mid X)(\hat{w}^*_k-\bar{w}^*)(X,d')\bigg\}^2\\
      \sum_{d\in[m]}\frac{\{\bar{w}^*(X,d)\}^2\zeta_0(d\mid X)\{w^*_0(X,d)\}^{-1}}{\big\{\sum_{d'\in[m]}\hat\zeta_k(d'\mid X)\hat{w}^*_k(X,d')\big\}^2\big\{\sum_{d'\in[m]}\zeta_0(d'\mid X)\bar{w}^*(X,d')\big\}^2}\bigg]\bigg)^{1/2}
    \end{multlined}\\
    &\hphantom{\leq{}}+\begin{multlined}[t][.85\textwidth]
      \bigg(P_0\bigg[\frac{1}{\hat\alpha_k^2}\frac{(\hat\pi_k-\pi_0)^2(X)}{\{1-\pi_0(X)\}\{1-\hat\pi_k(X)\}^2}\\
      \sum_{d\in[m]}\frac{\{\bar{w}^*(X,d)\}^2\zeta_0(d\mid X)\{w^*_0(X,d)\}^{-1}}{\big\{\sum_{d'\in[m]}\zeta_0(d'\mid X)\bar{w}^*(X,d')\big\}^2}\bigg]\bigg)^{1/2}
    \end{multlined}\\
    &\hphantom{\leq{}}+
      \bigg(P_0\bigg[\frac{(\alpha_0-\hat\alpha_k)^2}{\alpha_0^2\hat\alpha_k^2}\frac{\pi_0^2(X)}{1-\pi_0(X)}
      \sum_{d\in[m]}\frac{\{\bar{w}^*(X,d)\}^2\zeta_0(d\mid X)\{w^*_0(X,d)\}^{-1}}{\big\{\sum_{d'\in[m]}\zeta_0(d'\mid X)\bar{w}^*(X,d')\big\}^2}\bigg]\bigg)^{1/2}\\
    &\hphantom{\leq{}}+\begin{multlined}[t][.85\textwidth]
      \bigg(P_0\bigg[\frac{1-\pi_0(X)}{\hat\alpha_k^2}\bigg\{\frac{\hat\pi_k(X)}{1-\hat\pi_k(X)}\bigg\}^2\sum_{d\in[m]}\frac{\{\hat{w}^*_k(X,d)\}^2\zeta_0(d\mid X)}{\big\{\sum_{d'\in[m]}\hat\zeta_k(d'\mid X)\hat{w}^*_k(X,d')\big\}^2}\\(\delta^*_0-\hat\delta^*_k)^2(X)\bigg]\bigg)^{1/2}
    \end{multlined}\\
    &\hphantom{\leq{}}+\bigg[P_0\bigg\{\frac{\pi_0(X)}{\hat\alpha_k^2}(\hat\delta^*_k-\delta^*_0)^2(X)\bigg\}\bigg]^{1/2}+\bigg(P_0\bigg[\pi_0(X)\frac{(\alpha_0-\hat\alpha_k)^2}{\alpha_0^2\hat\alpha_k^2}\{\delta^*_0(X)\}^2\bigg]\bigg)^{1/2}\\
    &\lesssim \max_{d\in[m]}\max_{a\in\{0,1\}}\|1_{\{D=d\}}(\hat{\mu}^*_k-\mu^*_0)(a,X_d,X,d)\|_{P_0}\\
    &\hphantom{{}\lesssim}+ \max_{d\in[m]}\max_{a\in\{0,1\}}\|1_{\{D=d\}}(\hat{e}^*_k-e^*_0)(a\mid X_d,X,d)\|_{P_0}\\
    &\hphantom{{}\lesssim}+ \max_{d\in[m]}\|1_{\{D=d\}}(\hat{w}^*_k-\bar{w}^*)(X,d)\|_{P_0}+\max_{d\in[m]}\|(\hat\zeta_k-\zeta_0)(d\mid X)\|_{P_0}\\
    &\hphantom{{}\lesssim}+ \|(\hat{\pi}_k-\pi_0)(X)\|_{P_0} + \|(\hat\delta^*_k-\delta^*_0)(X)\|_{P_0} + |\hat\alpha_k-\alpha_0|.
  \end{align*}
  Since \(\|(\hat\delta^*_k-\delta^*_0)(X)\|_{P_0}\lesssim\max_{d\in[m]}\|1_{\{D=d\}}(\hat\tau_k-\tau_0)(X,d)\|_{P_0}\) and \(|\hat\alpha_k-\alpha_0|=o_{P_0}(1)\), we conclude the proof under the stated assumptions.
\end{proof}

\begin{proof}[Proof of proposition~\ref{ppn:asymptotic-star}]
  The bias of the estimator can be decomposed as follows:
  \begin{equation}
    \label{eq:bias-star}
    \hat\theta^*-\theta_0=\frac{1}{K}\sum_{k\in[K]}\bigg[(\mathbb{P}_{n,k}-P_0)\ell^*(O^*,\hat\eta^*_k)+\bigg\{P_0\ell^*(O^*,\hat\eta^*_k)-\frac{\alpha_0}{\hat\alpha_k}\theta_0\bigg\}-\frac{\hat\alpha_k-\alpha_0}{\hat\alpha_k}\theta_0\bigg].
  \end{equation}
  The second term in \eqref{eq:bias-star} is
  \begin{align*}
  \MoveEqLeft P_0\ell^*(O^*,\hat\eta^*_k)-\frac{\alpha_0}{\hat\alpha_k}\theta_0\nonumber\\
  &=\begin{multlined}[t][.9\textwidth]
    P_0\bigg[\frac{1-\pi_0(X)}{\hat\alpha_k}\frac{\hat\pi_k(X)}{1-\hat\pi_k(X)}\frac{\hat{w}^*(X,D)}{\sum_{d'\in[m]}\hat{w}^*(X,d')\hat\zeta_k(d'\mid X)}\\
    \bigg\{\sum_{a\in\{0,1\}}(2a-1)\frac{e^*_0(a\mid X_d,X,D)}{\hat{e}^*_k(a\mid X_d,X,D)}(\mu^*_0-\hat\mu^*_k)(a,X_d,X,D)\\
    +\hat\mu^*_k(1,X_d,X,D)-\hat\mu^*_k(0,X_d,X,D)-\hat\delta^*_k(X)\bigg\}\bigg]
  \end{multlined}\nonumber\\
  &\hphantom{={}}+P_0\bigg\{\frac{\pi_0(X)}{\hat\alpha_k}(\hat\delta^*_k-\delta^*_0)(X)\bigg\}\nonumber\\
  &=\begin{multlined}[t][.9\textwidth]
    P_0\bigg\{\frac{1-\pi_0(X)}{\hat\alpha_k}\frac{\hat\pi_k(X)}{1-\hat\pi_k(X)}\frac{\hat{w}^*(X,D)}{\sum_{d'\in[m]}\hat{w}^*(X,d')\hat\zeta_k(d'\mid X)}\\
    \sum_{a\in\{0,1\}}(2a-1)\frac{(e^*_0-\hat{e}^*_k)(a\mid X_d,X,D)}{\hat{e}^*_k(a\mid X_d,X,D)}(\mu^*_0-\hat\mu^*_k)(a,X_d,X,D)\bigg\}
  \end{multlined}\nonumber\\
  &\hphantom{={}}+P_0\bigg[\frac{1-\pi_0(X)}{\hat\alpha_k}\frac{\hat\pi_k(X)}{1-\hat\pi_k(X)}\sum_{d\in[m]}\frac{\hat{w}^*(X,d)\zeta_0(d\mid X)}{\sum_{d'\in[m]}\hat{w}^*(X,d')\hat\zeta_k(d'\mid X)}\{\delta^*_0(X)-\hat\delta^*_k(X)\}\bigg]\\
  &\hphantom{={}}+P_0\bigg\{\frac{\pi_0(X)}{\hat\alpha_k}(\hat\delta^*_k-\delta^*_0)(X)\bigg\}\nonumber\\
  &=\begin{multlined}[t][.9\textwidth]
    P_0\bigg\{\frac{1-\pi_0(X)}{\hat\alpha_k}\frac{\hat\pi_k(X)}{1-\hat\pi_k(X)}\frac{\hat{w}^*(X,D)}{\sum_{d'\in[m]}\hat{w}^*(X,d')\hat\zeta_k(d'\mid X)}\\
    \sum_{a\in\{0,1\}}(2a-1)\frac{(e^*_0-\hat{e}^*_k)(a\mid X_d,X,D)}{\hat{e}^*_k(a\mid X_d,X,D)}(\mu^*_0-\hat\mu^*_k)(a,X_d,X,D)\bigg\}
  \end{multlined}\nonumber\\
  &\hphantom{={}}+P_0\bigg\{\frac{1}{\hat\alpha_k}\frac{(\hat\pi_k-\pi_0)(X)}{1-\hat\pi_k(X)}\sum_{d\in[m]}\frac{\hat{w}^*(X,d)\zeta_0(d\mid X)}{\sum_{d'\in[m]}\hat{w}^*(X,d')\hat\zeta_k(d'\mid X)}(\delta^*_0-\hat\delta^*_k)(X)\bigg\}\\
  &\hphantom{={}}+P_0\bigg\{\frac{\pi_0(X)}{\hat\alpha_k}\sum_{d\in[m]}\frac{\hat{w}^*(X,d)(\zeta_0-\hat\zeta_k)(d\mid X)}{\sum_{d'\in[m]}\hat{w}^*(X,d')\hat\zeta_k(d'\mid X)}(\delta^*_0-\hat\delta^*_k)(X)\bigg\}.
\end{align*}

With assumption~\ref{asn:regularity-star}, we can bound the \(L_2(P_0)\) distance between \(\hat\delta^*_k\) and \(\delta^*_0\) by
\begin{align*}
  \MoveEqLeft \|(\hat\delta^*_k-\delta^*_0)(X)\|_{P_0}\\
  &= \bigg\|\sum_{d\in[m]}\frac{\hat{w}^*(X,d)\hat\zeta_k(d\mid X)}{\sum_{d'\in[m]}\hat{w}^*(X,d')\hat\zeta_k(d'\mid X)}\{\hat{\tau}_k(X,d)-\delta^*_0(X)\}\bigg\|_{P_0}\\
  &\lesssim \max_{d\in[m]}\|I(D=d)\{\hat{\tau}_k(X,d)-\delta^*_0(X)\}\|_{P_0}.
\end{align*}
So by assumption~\ref{asn:regularity-star} again and the Cauchy-Schwarz inequality, we have
\begin{align*}
  \MoveEqLeft\bigg|P_0\ell^*(O^*,\hat\eta^*_k)-\frac{\alpha_0}{\hat\alpha_k}\theta_0\bigg|\\
  &\lesssim{} \max_{d\in[m]}\max_{a\in\{0,1\}}\|I(D=d)(\hat{\mu}^*_k-\mu^*_0)(a,X_d,X,d)\|_{P_0}\|I(D=d)(\hat{e}^*_k-e^*_0)(a\mid X_d,X,d)\|_{P_0}\\
  &\phantom{{}\lesssim{}}+\|(\hat\pi_k-\pi_0)(X)\|_{P_0}\max_{d\in[m]}\|I(D=d)\{\hat{\tau}_k(X,d)-\delta^*_0(X)\}\|_{P_0}\\
  &\phantom{{}\lesssim{}}+\max_{d\in[m]}\|(\hat\zeta_k-\zeta_0)(d\mid X)\|_{P_0}\|I(D=d)\{\hat{\tau}_k(X,d)-\delta^*_0(X)\}\|_{P_0}.
\end{align*}

The consistency follows immediately from assumption~\ref{asn:model-consistency-star} and lemmas~\ref{lem:cross-fitting} and \ref{lem:boundedness-ell-star}.

For asymptotic linearity, we consider the following bias decomposition:
\begin{multline*}
  \hat\theta^*-\theta_0=\mathbb{P}_n\varphi^*(O^*,\bar{w}^*)+\frac{1}{K}\sum_{k\in[K]}\bigg[(\mathbb{P}_{n,k}-P_0)\{\ell^*(O^*,\hat\eta^*_k)-\ell^*(O^*,\bar\eta^*)\}\\
  +\bigg\{P_0\ell^*(O,\hat\eta^*_k)-\frac{\alpha_0}{\hat\alpha_k}\theta_0\bigg\}-\frac{(\hat\alpha_k-\alpha_0)^2}{\alpha_0\hat\alpha_k}\theta_0\bigg].
\end{multline*}
The second term is \(o_{P_0}(n^{-1/2})\) by lemmas~\ref{lem:cross-fitting} and \ref{lem:convergence-ell-star}.
The third term is \(o_{P_0}(n^{-1/2})\) by assumption~\ref{asn:model-linearity-star}.
The last term is \(o_{P_0}(n^{-1/2})\) since \(\hat\alpha_k-\alpha_0=O_{P_0}(n^{-1/2})\).
\end{proof}

\subsection{Proof of corollary~\ref{cor:relative-eff}}

  By proposition~\ref{ppn:asymptotic-star},
  \begin{align*}
    \hat\theta^*_{0}-\theta_0 &= \mathbb{P}_n\varphi^*(O^*,w^*_0)+o_{P_0}(n^{-1/2}),\\
    \hat\theta^*-\theta_0 &= \mathbb{P}_n\varphi^*(O^*,\bar{w}^*)+o_{P_0}(n^{-1/2}).
  \end{align*}
  Therefore, the difference between the asymptotic variances is
  \begin{multline*}
    E_0[\{\varphi^*(O^*,\bar{w}^*)\}^2]-E_0[\{\varphi^*(O^*,w^*_0)\}^2] \\
    = 2E_0[\{\varphi^*(O^*,\bar{w}^*)-\varphi^*(O^*,w^*_0)\}\varphi^*(O^*,w_0^*)]+E_0[\{\varphi^*(O^*,\bar{w}^*)-\varphi^*(O^*,w^*_0)\}^2].
  \end{multline*}
  We have
  \begin{align*}
    \MoveEqLeft E_0[\{\varphi^*(O^*,\bar{w}^*)-\varphi^*(O^*,w^*_0)\}\varphi^*(O^*,w_0^*)]\\
    &= \begin{multlined}[t][.85\textwidth] E_0\bigg[\frac{1-G}{\alpha_0^2}\bigg\{\frac{\pi_0(X)}{1-\pi_0(X)}\bigg\}^2\\
      \bigg\{\frac{\bar{w}^*(X,D)}{\sum_{d'\in[m]}\zeta_0(d'\mid X)\bar{w}^*(X,d')}-\frac{w^*_0(X,D)}{\sum_{d'\in[m]}\zeta_0(d'\mid X)w^*_0(X,d')}\bigg\}\\
      \frac{w^*_0(X,D)}{\sum_{d'\in[m]}\zeta_0(d'\mid X)w^*_0(X,d')}\bigg\{\frac{Y-\mu^*_0(A,X_D,X,D)}{e^*_0(A\mid X_d,X,D)}\bigg\}^2\bigg]
      \end{multlined}\\
  &\hphantom{={}}+\begin{multlined}[t][.85\textwidth]
    E_0\bigg[\frac{1-G}{\alpha_0^2}\bigg\{\frac{\pi_0(X)}{1-\pi_0(X)}\bigg\}^2\\
      \bigg\{\frac{\bar{w}^*(X,D)}{\sum_{d'\in[m]}\zeta_0(d'\mid X)\bar{w}^*(X,d')}-\frac{w^*_0(X,D)}{\sum_{d'\in[m]}\zeta_0(d'\mid X)w^*_0(X,d')}\bigg\}\\
      \frac{w^*_0(X,D)}{\sum_{d'\in[m]}\zeta_0(d'\mid X)w^*_0(X,d')}\{\mu^*_0(1,X_D,X,D)-\mu^*_0(0,X_D,X,D)-\tau_0(X)\}^2\bigg]
    \end{multlined}\\
  &=\begin{multlined}[t][.85\textwidth]
    E_0\bigg[\frac{1-G}{\alpha_0^2}\bigg\{\frac{\pi_0(X)}{1-\pi_0(X)}\bigg\}^2\\
      \bigg\{\frac{\bar{w}^*(X,D)}{\sum_{d'\in[m]}\zeta_0(d'\mid X)\bar{w}^*(X,d')}-\frac{w^*_0(X,D)}{\sum_{d'\in[m]}\zeta_0(d'\mid X)w^*_0(X,d')}\bigg\}\\
      \frac{1}{\sum_{d'\in[m]}\zeta_0(d'\mid X)w^*_0(X,d')}\bigg]
    \end{multlined}\\
    &=0.
  \end{align*}
  Hence,
  \[
    E_0[\{\varphi^*(O^*,\bar{w}^*)\}^2]-E_0[\{\varphi^*(O^*,w^*_0)\}^2]=E_0[\{\varphi^*(O^*,\bar{w}^*)-\varphi^*(O^*,w^*_0)\}^2]\geq 0,
  \]
  and the corollary follows.

\bibliographysuppmat{./multi-trial-transport.bib}

\newpage

\begin{table}
  \caption{Summary of simulation results for \(\hat\theta\) in setting II with cross-fitting.}
  \label{tab:sim-2}
  \footnotesize
  \centering
  
\begin{tabular}{rllrrrrr}
{\(n\)} & {Experiment} & {Weight} & {Mean} & {Bias} & {RMSE} & {SE} & {Coverage}\\
\(1250\) & 1 & Constant & \(3.11\) & \(5.46\) & \(12.21\) & \(12.13\) & \(95.1\)\\
 &  & Overlap & \(3.11\) & \(5.38\) & \(12.24\) & \(12.18\) & \(95.0\)\\
 &  & Learned & \(3.11\) & \(5.91\) & \(12.35\) & \(11.66\) & \(94.4\)\\
 &  & Oracle & \(3.11\) & \(5.40\) & \(11.91\) & \(11.41\) & \(95.1\)\\
 & 2 & Constant & \(3.09\) & \(-20.79\) & \(15.13\) & \(12.69\) & \(94.2\)\\
 &  & Overlap & \(3.09\) & \(-21.00\) & \(15.30\) & \(12.76\) & \(94.4\)\\
 &  & Learned & \(3.09\) & \(-21.18\) & \(14.22\) & \(12.34\) & \(93.9\)\\
 &  & Oracle & \(3.09\) & \(-18.98\) & \(13.33\) & \(12.05\) & \(93.9\)\\
 & 3 & Constant & \(3.10\) & \(-5.66\) & \(12.93\) & \(12.04\) & \(93.6\)\\
 &  & Overlap & \(3.10\) & \(-5.98\) & \(13.01\) & \(12.09\) & \(93.7\)\\
 &  & Learned & \(3.10\) & \(-2.85\) & \(12.59\) & \(11.55\) & \(93.8\)\\
 &  & Oracle & \(3.11\) & \(-2.53\) & \(12.05\) & \(11.36\) & \(94.2\)\\
 & 4 & Constant & \(2.94\) & \(-169.86\) & \(23.95\) & \(12.55\) & \(68.7\)\\
 &  & Overlap & \(2.94\) & \(-172.23\) & \(24.30\) & \(12.62\) & \(68.1\)\\
 &  & Learned & \(2.96\) & \(-150.93\) & \(21.47\) & \(12.24\) & \(73.4\)\\
 &  & Oracle & \(2.97\) & \(-138.70\) & \(19.50\) & \(11.93\) & \(76.3\)\\
\(2500\) & 1 & Constant & \(3.12\) & \(8.62\) & \(8.40\) & \(8.39\) & \(94.8\)\\
 &  & Overlap & \(3.12\) & \(8.58\) & \(8.42\) & \(8.43\) & \(94.8\)\\
 &  & Learned & \(3.12\) & \(7.85\) & \(8.03\) & \(7.94\) & \(95.0\)\\
 &  & Oracle & \(3.12\) & \(7.70\) & \(7.99\) & \(7.86\) & \(95.1\)\\
 & 2 & Constant & \(3.10\) & \(-5.52\) & \(9.04\) & \(8.71\) & \(94.0\)\\
 &  & Overlap & \(3.10\) & \(-5.75\) & \(9.07\) & \(8.75\) & \(94.0\)\\
 &  & Learned & \(3.10\) & \(-4.73\) & \(8.81\) & \(8.41\) & \(93.8\)\\
 &  & Oracle & \(3.10\) & \(-4.18\) & \(8.77\) & \(8.33\) & \(94.0\)\\
 & 3 & Constant & \(3.11\) & \(1.32\) & \(8.82\) & \(8.33\) & \(93.6\)\\
 &  & Overlap & \(3.11\) & \(1.12\) & \(8.88\) & \(8.36\) & \(93.5\)\\
 &  & Learned & \(3.11\) & \(2.34\) & \(8.10\) & \(7.89\) & \(95.0\)\\
 &  & Oracle & \(3.11\) & \(2.25\) & \(8.02\) & \(7.83\) & \(95.2\)\\
 & 4 & Constant & \(2.94\) & \(-171.97\) & \(19.30\) & \(8.51\) & \(49.0\)\\
 &  & Overlap & \(2.93\) & \(-174.52\) & \(19.54\) & \(8.54\) & \(47.9\)\\
 &  & Learned & \(2.96\) & \(-147.72\) & \(17.06\) & \(8.29\) & \(57.5\)\\
 &  & Oracle & \(2.97\) & \(-137.98\) & \(16.16\) & \(8.20\) & \(60.8\)\\
\end{tabular}

  \medskip
  {Mean: average of estimates; Bias: Monte-Carlo bias, \(10^{-3}\); RMSE: root mean squared error, \(10^{-2}\); \\SE: average of standard error estimates, \(10^{-2}\); Coverage: \(95\%\) confidence interval coverage, \(\%\).}
\end{table}

\begin{table}
  \caption{Summary of simulation results for \(\hat\theta\) in setting III with cross-fitting.}
  \label{tab:sim-3}
  \footnotesize
  \centering
  
\begin{tabular}{rllrrrrr}
{\(n\)} & {Experiment} & {Weight} & {Mean} & {Bias} & {RMSE} & {SE} & {Coverage}\\
\(1250\) & 1 & Constant & \(3.12\) & \(10.38\) & \(12.08\) & \(8.43\) & \(94.5\)\\
 &  & Overlap & \(3.12\) & \(10.39\) & \(12.20\) & \(8.44\) & \(94.7\)\\
 &  & Learned & \(3.12\) & \(9.34\) & \(10.50\) & \(8.47\) & \(94.6\)\\
 &  & Oracle & \(3.12\) & \(10.39\) & \(12.20\) & \(8.44\) & \(94.7\)\\
 & 2 & Constant & \(3.09\) & \(-20.66\) & \(10.59\) & \(8.83\) & \(93.3\)\\
 &  & Overlap & \(3.09\) & \(-20.88\) & \(10.66\) & \(8.85\) & \(93.3\)\\
 &  & Learned & \(3.09\) & \(-21.11\) & \(11.31\) & \(9.07\) & \(92.9\)\\
 &  & Oracle & \(3.09\) & \(-20.88\) & \(10.66\) & \(8.85\) & \(93.3\)\\
 & 3 & Constant & \(3.11\) & \(5.87\) & \(10.70\) & \(8.14\) & \(94.0\)\\
 &  & Overlap & \(3.11\) & \(5.79\) & \(10.80\) & \(8.14\) & \(93.7\)\\
 &  & Learned & \(3.11\) & \(4.95\) & \(9.93\) & \(8.19\) & \(93.1\)\\
 &  & Oracle & \(3.11\) & \(5.79\) & \(10.80\) & \(8.14\) & \(93.7\)\\
 & 4 & Constant & \(2.94\) & \(-170.33\) & \(20.10\) & \(8.38\) & \(44.4\)\\
 &  & Overlap & \(2.94\) & \(-172.72\) & \(20.36\) & \(8.39\) & \(43.5\)\\
 &  & Learned & \(2.94\) & \(-169.23\) & \(20.58\) & \(8.73\) & \(46.5\)\\
 &  & Oracle & \(2.94\) & \(-172.72\) & \(20.36\) & \(8.39\) & \(43.5\)\\
\(2500\) & 1 & Constant & \(3.11\) & \(7.16\) & \(5.65\) & \(5.56\) & \(94.7\)\\
 &  & Overlap & \(3.11\) & \(7.14\) & \(5.65\) & \(5.56\) & \(95.0\)\\
 &  & Learned & \(3.11\) & \(7.15\) & \(5.67\) & \(5.61\) & \(94.9\)\\
 &  & Oracle & \(3.11\) & \(7.14\) & \(5.65\) & \(5.56\) & \(95.0\)\\
 & 2 & Constant & \(3.10\) & \(-6.76\) & \(6.55\) & \(6.08\) & \(92.1\)\\
 &  & Overlap & \(3.10\) & \(-6.97\) & \(6.55\) & \(6.09\) & \(91.9\)\\
 &  & Learned & \(3.10\) & \(-6.03\) & \(6.55\) & \(6.11\) & \(92.0\)\\
 &  & Oracle & \(3.10\) & \(-6.97\) & \(6.55\) & \(6.09\) & \(91.9\)\\
 & 3 & Constant & \(3.11\) & \(3.85\) & \(5.68\) & \(5.37\) & \(94.3\)\\
 &  & Overlap & \(3.11\) & \(3.77\) & \(5.69\) & \(5.37\) & \(94.2\)\\
 &  & Learned & \(3.11\) & \(3.78\) & \(5.72\) & \(5.41\) & \(94.6\)\\
 &  & Oracle & \(3.11\) & \(3.77\) & \(5.69\) & \(5.37\) & \(94.2\)\\
 & 4 & Constant & \(2.93\) & \(-173.06\) & \(18.27\) & \(5.68\) & \(14.1\)\\
 &  & Overlap & \(2.93\) & \(-175.59\) & \(18.51\) & \(5.68\) & \(13.3\)\\
 &  & Learned & \(2.95\) & \(-162.82\) & \(17.35\) & \(5.84\) & \(22.6\)\\
 &  & Oracle & \(2.93\) & \(-175.59\) & \(18.51\) & \(5.68\) & \(13.3\)\\
\end{tabular}

  \medskip
  {Mean: average of estimates; Bias: Monte-Carlo bias, \(10^{-3}\); RMSE: root mean squared error, \(10^{-2}\); \\SE: average of standard error estimates, \(10^{-2}\); Coverage: \(95\%\) confidence interval coverage, \(\%\).
    One simulation run with sample size \(n=1250\) was omitted due to numerical instability.}
\end{table}

\begin{table}
  \centering
  \caption{Comparison of plug-in standard error estimates in simulations.}
  \label{tab:prop-smaller-se}
  \footnotesize
  
\begin{tabular}{lrrrrr}
&& \multicolumn{4}{c}{Comparison} \\
{Setting} & {\(n\)} & {A} & {B} & {C} & {D}\\
I & \(1250\) & \(99.2\) & \(99.2\) & \(90.3\) & \(91.3\)\\
 & \(2500\) & \(100.0\) & \(100.0\) & \(99.3\) & \(99.5\)\\
II & \(1250\) & \(99.7\) & \(99.8\) & \(95.8\) & \(96.8\)\\
 & \(2500\) & \(100.0\) & \(100.0\) & \(99.8\) & \(99.8\)\\
III & \(1250\) & \(61.2\) & \(100.0\) & \(10.2\) & \(9.6\)\\
 & \(2500\) & \(63.2\) & \(100.0\) & \(9.1\) & \(7.8\)\\
\end{tabular}

  \medskip
  {Comparison: A (oracle and constant), B (oracle and overlap), C (learned and constant), D (learned and overlap);
    Values represent the proportions (in percentages) of the simulation runs where the standard error of the estimator using the former weight was smaller than using the latter.}
\end{table}

\begin{table}
  \caption{Summary of simulation results for \(\hat\theta\) in setting I without cross-fitting.}
  \label{tab:sim-1-nocf}
  \footnotesize
  \centering
  
\begin{tabular}{rllrrrrr}
{\(n\)} & {Experiment} & {Weight} & {Mean} & {Bias} & {RMSE} & {SE} & {Coverage}\\
\(1250\) & 1 & Constant & \(3.09\) & \(-13.61\) & \(36.30\) & \(9.19\) & \(92.4\)\\
 &  & Overlap & \(3.09\) & \(-13.50\) & \(35.75\) & \(9.19\) & \(92.3\)\\
 &  & Learned & \(3.10\) & \(-11.04\) & \(27.03\) & \(8.38\) & \(90.0\)\\
 &  & Oracle & \(3.10\) & \(-10.97\) & \(27.51\) & \(8.68\) & \(92.0\)\\
 & 2 & Constant & \(3.06\) & \(-45.48\) & \(11.56\) & \(9.04\) & \(88.3\)\\
 &  & Overlap & \(3.06\) & \(-46.01\) & \(11.60\) & \(9.05\) & \(88.2\)\\
 &  & Learned & \(3.06\) & \(-50.58\) & \(11.81\) & \(8.43\) & \(84.9\)\\
 &  & Oracle & \(3.07\) & \(-42.75\) & \(11.28\) & \(8.84\) & \(87.1\)\\
 & 3 & Constant & \(3.09\) & \(-20.26\) & \(44.58\) & \(9.15\) & \(90.7\)\\
 &  & Overlap & \(3.09\) & \(-20.26\) & \(44.07\) & \(9.15\) & \(90.7\)\\
 &  & Learned & \(3.09\) & \(-16.13\) & \(32.90\) & \(8.31\) & \(87.8\)\\
 &  & Oracle & \(3.09\) & \(-15.89\) & \(32.74\) & \(8.61\) & \(90.2\)\\
 & 4 & Constant & \(2.93\) & \(-181.59\) & \(20.74\) & \(8.64\) & \(44.7\)\\
 &  & Overlap & \(2.92\) & \(-184.06\) & \(20.97\) & \(8.65\) & \(43.9\)\\
 &  & Learned & \(2.94\) & \(-171.11\) & \(19.94\) & \(8.11\) & \(42.9\)\\
 &  & Oracle & \(2.95\) & \(-162.37\) & \(19.01\) & \(8.49\) & \(50.6\)\\
\(2500\) & 1 & Constant & \(3.11\) & \(3.57\) & \(6.71\) & \(6.22\) & \(92.9\)\\
 &  & Overlap & \(3.11\) & \(3.54\) & \(6.71\) & \(6.23\) & \(92.9\)\\
 &  & Learned & \(3.11\) & \(2.84\) & \(6.48\) & \(5.81\) & \(92.2\)\\
 &  & Oracle & \(3.11\) & \(2.49\) & \(6.47\) & \(5.95\) & \(92.5\)\\
 & 2 & Constant & \(3.09\) & \(-22.68\) & \(7.83\) & \(6.53\) & \(89.1\)\\
 &  & Overlap & \(3.08\) & \(-23.09\) & \(7.86\) & \(6.55\) & \(89.2\)\\
 &  & Learned & \(3.08\) & \(-26.12\) & \(7.79\) & \(6.16\) & \(86.5\)\\
 &  & Oracle & \(3.09\) & \(-21.40\) & \(7.71\) & \(6.40\) & \(88.5\)\\
 & 3 & Constant & \(3.11\) & \(-0.24\) & \(6.94\) & \(6.06\) & \(90.4\)\\
 &  & Overlap & \(3.11\) & \(-0.35\) & \(6.97\) & \(6.07\) & \(90.5\)\\
 &  & Learned & \(3.11\) & \(-0.44\) & \(6.59\) & \(5.65\) & \(91.2\)\\
 &  & Oracle & \(3.11\) & \(-0.85\) & \(6.57\) & \(5.80\) & \(92.5\)\\
 & 4 & Constant & \(2.93\) & \(-176.20\) & \(18.93\) & \(6.21\) & \(22.2\)\\
 &  & Overlap & \(2.93\) & \(-178.71\) & \(19.17\) & \(6.22\) & \(21.5\)\\
 &  & Learned & \(2.95\) & \(-160.81\) & \(17.51\) & \(5.93\) & \(28.7\)\\
 &  & Oracle & \(2.95\) & \(-157.11\) & \(17.14\) & \(6.12\) & \(29.7\)\\
\end{tabular}

  \medskip
  {Mean: average of estimates; Bias: Monte-Carlo bias, \(10^{-3}\); RMSE: root mean squared error, \(10^{-2}\); \\SE: average of standard error estimates, \(10^{-2}\); Coverage: \(95\%\) confidence interval coverage, \(\%\).}
\end{table}

\begin{table}
  \caption{Summary of simulation results for \(\hat\theta\) in setting II without cross-fitting.}
  \label{tab:sim-2-nocf}
  \footnotesize
  \centering
  
\begin{tabular}{rllrrrrr}
{\(n\)} & {Experiment} & {Weight} & {Mean} & {Bias} & {RMSE} & {SE} & {Coverage}\\
\(1250\) & 1 & Constant & \(3.09\) & \(-20.07\) & \(49.81\) & \(11.27\) & \(91.4\)\\
 &  & Overlap & \(3.09\) & \(-19.89\) & \(49.04\) & \(11.29\) & \(91.6\)\\
 &  & Learned & \(3.09\) & \(-15.27\) & \(34.75\) & \(10.05\) & \(89.0\)\\
 &  & Oracle & \(3.09\) & \(-16.74\) & \(37.31\) & \(10.44\) & \(91.2\)\\
 & 2 & Constant & \(3.06\) & \(-44.72\) & \(13.11\) & \(10.89\) & \(89.4\)\\
 &  & Overlap & \(3.06\) & \(-45.21\) & \(13.16\) & \(10.92\) & \(89.6\)\\
 &  & Learned & \(3.06\) & \(-46.18\) & \(13.07\) & \(9.97\) & \(86.3\)\\
 &  & Oracle & \(3.07\) & \(-40.56\) & \(12.74\) & \(10.43\) & \(89.3\)\\
 & 3 & Constant & \(3.08\) & \(-31.74\) & \(61.34\) & \(11.44\) & \(89.8\)\\
 &  & Overlap & \(3.08\) & \(-31.77\) & \(60.63\) & \(11.47\) & \(89.9\)\\
 &  & Learned & \(3.08\) & \(-23.93\) & \(44.24\) & \(10.17\) & \(88.3\)\\
 &  & Oracle & \(3.08\) & \(-24.54\) & \(44.40\) & \(10.57\) & \(91.1\)\\
 & 4 & Constant & \(2.93\) & \(-181.77\) & \(21.86\) & \(10.70\) & \(59.0\)\\
 &  & Overlap & \(2.92\) & \(-184.23\) & \(22.09\) & \(10.73\) & \(58.2\)\\
 &  & Learned & \(2.95\) & \(-159.20\) & \(19.99\) & \(9.83\) & \(60.2\)\\
 &  & Oracle & \(2.96\) & \(-149.38\) & \(19.05\) & \(10.33\) & \(65.8\)\\
\(2500\) & 1 & Constant & \(3.11\) & \(4.09\) & \(8.24\) & \(7.62\) & \(92.9\)\\
 &  & Overlap & \(3.11\) & \(4.05\) & \(8.26\) & \(7.65\) & \(92.8\)\\
 &  & Learned & \(3.11\) & \(3.58\) & \(7.85\) & \(6.98\) & \(91.8\)\\
 &  & Oracle & \(3.11\) & \(2.95\) & \(7.83\) & \(7.15\) & \(92.7\)\\
 & 2 & Constant & \(3.09\) & \(-21.31\) & \(9.18\) & \(7.88\) & \(90.8\)\\
 &  & Overlap & \(3.09\) & \(-21.70\) & \(9.22\) & \(7.91\) & \(90.8\)\\
 &  & Learned & \(3.09\) & \(-22.03\) & \(8.87\) & \(7.30\) & \(88.7\)\\
 &  & Oracle & \(3.09\) & \(-18.85\) & \(8.84\) & \(7.54\) & \(90.2\)\\
 & 3 & Constant & \(3.11\) & \(-2.11\) & \(8.75\) & \(7.59\) & \(91.2\)\\
 &  & Overlap & \(3.11\) & \(-2.29\) & \(8.81\) & \(7.62\) & \(91.4\)\\
 &  & Learned & \(3.11\) & \(-0.95\) & \(8.00\) & \(6.95\) & \(91.5\)\\
 &  & Oracle & \(3.11\) & \(-1.58\) & \(7.93\) & \(7.16\) & \(92.4\)\\
 & 4 & Constant & \(2.93\) & \(-175.62\) & \(19.57\) & \(7.71\) & \(39.1\)\\
 &  & Overlap & \(2.93\) & \(-178.14\) & \(19.81\) & \(7.74\) & \(38.3\)\\
 &  & Learned & \(2.96\) & \(-149.93\) & \(17.16\) & \(7.20\) & \(46.8\)\\
 &  & Oracle & \(2.97\) & \(-142.22\) & \(16.47\) & \(7.46\) & \(52.5\)\\
\end{tabular}

  \medskip
  {Mean: average of estimates; Bias: Monte-Carlo bias, \(10^{-3}\); RMSE: root mean squared error, \(10^{-2}\); \\SE: average of standard error estimates, \(10^{-2}\); Coverage: \(95\%\) confidence interval coverage, \(\%\).}
\end{table}

\begin{table}
  \caption{Summary of simulation results for \(\hat\theta\) in setting III without cross-fitting.}
  \label{tab:sim-3-nocf}
  \footnotesize
  \centering
  
\begin{tabular}{rllrrrrr}
{\(n\)} & {Experiment} & {Weight} & {Mean} & {Bias} & {RMSE} & {SE} & {Coverage}\\
\(1250\) & 1 & Constant & \(3.10\) & \(-10.09\) & \(28.05\) & \(7.62\) & \(92.0\)\\
 &  & Overlap & \(3.10\) & \(-10.02\) & \(27.62\) & \(7.61\) & \(92.1\)\\
 &  & Learned & \(3.10\) & \(-8.31\) & \(22.03\) & \(7.26\) & \(91.0\)\\
 &  & Oracle & \(3.10\) & \(-10.02\) & \(27.62\) & \(7.61\) & \(92.1\)\\
 & 2 & Constant & \(3.06\) & \(-46.59\) & \(10.72\) & \(7.64\) & \(85.0\)\\
 &  & Overlap & \(3.06\) & \(-47.15\) & \(10.79\) & \(7.64\) & \(85.1\)\\
 &  & Learned & \(3.05\) & \(-55.80\) & \(11.98\) & \(7.27\) & \(81.7\)\\
 &  & Oracle & \(3.06\) & \(-47.15\) & \(10.79\) & \(7.64\) & \(85.1\)\\
 & 3 & Constant & \(3.09\) & \(-14.09\) & \(34.42\) & \(7.58\) & \(90.7\)\\
 &  & Overlap & \(3.09\) & \(-14.07\) & \(34.02\) & \(7.57\) & \(90.8\)\\
 &  & Learned & \(3.10\) & \(-12.39\) & \(28.50\) & \(7.22\) & \(89.0\)\\
 &  & Oracle & \(3.09\) & \(-14.07\) & \(34.02\) & \(7.57\) & \(90.8\)\\
 & 4 & Constant & \(2.93\) & \(-181.80\) & \(20.21\) & \(7.21\) & \(31.5\)\\
 &  & Overlap & \(2.92\) & \(-184.26\) & \(20.45\) & \(7.21\) & \(30.5\)\\
 &  & Learned & \(2.93\) & \(-180.39\) & \(20.56\) & \(6.96\) & \(30.0\)\\
 &  & Oracle & \(2.92\) & \(-184.26\) & \(20.45\) & \(7.21\) & \(30.5\)\\
\(2500\) & 1 & Constant & \(3.11\) & \(2.59\) & \(5.54\) & \(5.14\) & \(93.1\)\\
 &  & Overlap & \(3.11\) & \(2.58\) & \(5.53\) & \(5.14\) & \(93.0\)\\
 &  & Learned & \(3.11\) & \(2.71\) & \(5.53\) & \(5.01\) & \(92.3\)\\
 &  & Oracle & \(3.11\) & \(2.58\) & \(5.53\) & \(5.14\) & \(93.0\)\\
 & 2 & Constant & \(3.08\) & \(-24.18\) & \(6.90\) & \(5.52\) & \(87.0\)\\
 &  & Overlap & \(3.08\) & \(-24.59\) & \(6.92\) & \(5.53\) & \(87.0\)\\
 &  & Learned & \(3.08\) & \(-29.26\) & \(7.05\) & \(5.29\) & \(85.3\)\\
 &  & Oracle & \(3.08\) & \(-24.59\) & \(6.92\) & \(5.53\) & \(87.0\)\\
 & 3 & Constant & \(3.11\) & \(0.21\) & \(5.63\) & \(4.98\) & \(91.7\)\\
 &  & Overlap & \(3.11\) & \(0.16\) & \(5.64\) & \(4.98\) & \(91.7\)\\
 &  & Learned & \(3.11\) & \(0.43\) & \(5.65\) & \(4.86\) & \(90.9\)\\
 &  & Oracle & \(3.11\) & \(0.16\) & \(5.64\) & \(4.98\) & \(91.7\)\\
 & 4 & Constant & \(2.93\) & \(-177.09\) & \(18.63\) & \(5.18\) & \(9.6\)\\
 &  & Overlap & \(2.93\) & \(-179.60\) & \(18.87\) & \(5.17\) & \(8.4\)\\
 &  & Learned & \(2.94\) & \(-166.13\) & \(17.63\) & \(5.07\) & \(13.3\)\\
 &  & Oracle & \(2.93\) & \(-179.60\) & \(18.87\) & \(5.17\) & \(8.4\)\\
\end{tabular}

  \medskip
  {Mean: average of estimates; Bias: Monte-Carlo bias, \(10^{-3}\); RMSE: root mean squared error, \(10^{-2}\); \\SE: average of standard error estimates, \(10^{-2}\); Coverage: \(95\%\) confidence interval coverage, \(\%\).}
\end{table}

\begin{table}
  \caption{Summary of simulation results for \(\hat\theta^\dagger\) and \(\hat\theta^\ddagger\) with cross-fitting.}
  \label{tab:sim-alt}
  \footnotesize
  \centering
  
\begin{tabular}{rlrrrrrr}
{\(n\)} & {Setting} & {Estimator} & {Mean} & {Bias} & {RMSE} & {SE} & {Coverage}\\
\(1250\) & I & \(\hat\theta^\dagger\) & \(-0.01\) & \(-3113.81\) & \(311.39\) & \(8.43\) & \(0.0\)\\
 &  & \(\hat\theta^\ddagger\) & \(0.00\) & \(-3105.59\) & \(310.56\) & \(8.90\) & \(0.0\)\\
 & II & \(\hat\theta^\dagger\) & \(-0.00\) & \(-3111.88\) & \(311.22\) & \(10.47\) & \(0.0\)\\
 &  & \(\hat\theta^\ddagger\) & \(0.00\) & \(-3105.80\) & \(310.59\) & \(11.25\) & \(0.0\)\\
 & III & \(\hat\theta^\dagger\) & \(0.00\) & \(-3105.60\) & \(310.56\) & \(7.06\) & \(0.0\)\\
 &  & \(\hat\theta^\ddagger\) & \(0.00\) & \(-3105.60\) & \(310.56\) & \(7.06\) & \(0.0\)\\
\(2500\) & I & \(\hat\theta^\dagger\) & \(-0.01\) & \(-3113.98\) & \(311.40\) & \(5.92\) & \(0.0\)\\
 &  & \(\hat\theta^\ddagger\) & \(0.00\) & \(-3105.55\) & \(310.56\) & \(6.26\) & \(0.0\)\\
 & II & \(\hat\theta^\dagger\) & \(-0.00\) & \(-3111.75\) & \(311.19\) & \(7.32\) & \(0.0\)\\
 &  & \(\hat\theta^\ddagger\) & \(0.00\) & \(-3105.38\) & \(310.54\) & \(7.89\) & \(0.0\)\\
 & III & \(\hat\theta^\dagger\) & \(0.00\) & \(-3106.01\) & \(310.60\) & \(4.97\) & \(0.0\)\\
 &  & \(\hat\theta^\ddagger\) & \(0.00\) & \(-3106.01\) & \(310.60\) & \(4.97\) & \(0.0\)\\
\end{tabular}

  \medskip
  {Mean: average of estimates; Bias: Monte-Carlo bias, \(10^{-3}\); RMSE: root mean squared error, \(10^{-2}\); \\SE: average of standard error estimates, \(10^{-2}\); Coverage: \(95\%\) confidence interval coverage, \(\%\).}
\end{table}







\begin{table}
  \caption{Summary of simulation results for \(\hat\gamma_1\) in experiment 1.}
  \label{tab:sim-beta-1}
  \footnotesize
  \centering
  
\begin{tabular}{llrlrrrrr}
{Cross-fitting} & {Setting} & {\(n\)} & {Weight} & {Mean} & {Bias} & {RMSE} & {SE} & {Coverage}\\
\(5\)-fold & I & \(1250\) & Constant & \(3.00\) & \(5.48\) & \(14.51\) & \(13.83\) & \(94.0\)\\
 &  &  & Overlap & \(3.00\) & \(5.39\) & \(14.52\) & \(13.86\) & \(94.0\)\\
 &  &  & Learned & \(3.00\) & \(5.82\) & \(14.74\) & \(13.45\) & \(93.9\)\\
 &  &  & Oracle & \(3.00\) & \(6.70\) & \(14.43\) & \(13.19\) & \(94.6\)\\
 &  & \(2500\) & Constant & \(3.01\) & \(10.93\) & \(9.15\) & \(9.44\) & \(95.9\)\\
 &  &  & Overlap & \(3.01\) & \(10.85\) & \(9.17\) & \(9.46\) & \(95.8\)\\
 &  &  & Learned & \(3.00\) & \(10.03\) & \(8.65\) & \(9.05\) & \(96.2\)\\
 &  &  & Oracle & \(3.01\) & \(10.26\) & \(8.61\) & \(8.96\) & \(95.2\)\\
 & II & \(1250\) & Constant & \(3.00\) & \(7.05\) & \(18.28\) & \(17.61\) & \(94.1\)\\
 &  &  & Overlap & \(3.00\) & \(6.94\) & \(18.32\) & \(17.69\) & \(94.3\)\\
 &  &  & Learned & \(3.00\) & \(8.75\) & \(19.38\) & \(16.91\) & \(94.3\)\\
 &  &  & Oracle & \(3.00\) & \(8.60\) & \(18.40\) & \(16.52\) & \(94.4\)\\
 &  & \(2500\) & Constant & \(3.01\) & \(11.92\) & \(11.68\) & \(12.08\) & \(96.1\)\\
 &  &  & Overlap & \(3.01\) & \(11.82\) & \(11.72\) & \(12.13\) & \(96.3\)\\
 &  &  & Learned & \(3.01\) & \(11.75\) & \(10.85\) & \(11.38\) & \(95.5\)\\
 &  &  & Oracle & \(3.01\) & \(11.76\) & \(10.80\) & \(11.25\) & \(95.3\)\\
 & III & \(1250\) & Constant & \(3.00\) & \(6.70\) & \(12.67\) & \(11.18\) & \(94.0\)\\
 &  &  & Overlap & \(3.00\) & \(6.67\) & \(12.64\) & \(11.18\) & \(94.0\)\\
 &  &  & Learned & \(3.00\) & \(5.74\) & \(12.66\) & \(11.34\) & \(93.6\)\\
 &  &  & Oracle & \(3.00\) & \(6.67\) & \(12.64\) & \(11.18\) & \(94.0\)\\
 &  & \(2500\) & Constant & \(3.00\) & \(9.45\) & \(7.23\) & \(7.47\) & \(95.6\)\\
 &  &  & Overlap & \(3.00\) & \(9.40\) & \(7.22\) & \(7.47\) & \(95.5\)\\
 &  &  & Learned & \(3.00\) & \(9.02\) & \(7.25\) & \(7.54\) & \(95.9\)\\
 &  &  & Oracle & \(3.00\) & \(9.40\) & \(7.22\) & \(7.47\) & \(95.5\)\\
None & I & \(1250\) & Constant & \(2.99\) & \(-1.11\) & \(17.46\) & \(12.45\) & \(91.9\)\\
 &  &  & Overlap & \(2.99\) & \(-1.19\) & \(17.33\) & \(12.46\) & \(92.1\)\\
 &  &  & Learned & \(2.99\) & \(-2.29\) & \(16.32\) & \(11.24\) & \(90.8\)\\
 &  &  & Oracle & \(2.99\) & \(-2.49\) & \(16.65\) & \(11.73\) & \(92.5\)\\
 &  & \(2500\) & Constant & \(3.00\) & \(7.15\) & \(8.90\) & \(8.55\) & \(94.6\)\\
 &  &  & Overlap & \(3.00\) & \(7.09\) & \(8.92\) & \(8.57\) & \(94.5\)\\
 &  &  & Learned & \(3.00\) & \(6.57\) & \(8.34\) & \(7.91\) & \(93.4\)\\
 &  &  & Oracle & \(3.00\) & \(6.38\) & \(8.37\) & \(8.13\) & \(94.3\)\\
 & II & \(1250\) & Constant & \(3.00\) & \(1.40\) & \(22.25\) & \(15.86\) & \(92.4\)\\
 &  &  & Overlap & \(3.00\) & \(1.25\) & \(22.12\) & \(15.90\) & \(92.5\)\\
 &  &  & Learned & \(2.99\) & \(-1.72\) & \(15.98\) & \(13.92\) & \(90.9\)\\
 &  &  & Oracle & \(3.00\) & \(0.65\) & \(19.23\) & \(14.64\) & \(92.3\)\\
 &  & \(2500\) & Constant & \(3.00\) & \(8.34\) & \(11.34\) & \(10.90\) & \(94.5\)\\
 &  &  & Overlap & \(3.00\) & \(8.27\) & \(11.38\) & \(10.94\) & \(94.2\)\\
 &  &  & Learned & \(3.00\) & \(8.46\) & \(10.48\) & \(9.89\) & \(93.2\)\\
 &  &  & Oracle & \(3.00\) & \(7.83\) & \(10.50\) & \(10.17\) & \(94.0\)\\
 & III & \(1250\) & Constant & \(2.99\) & \(-1.32\) & \(14.08\) & \(9.93\) & \(91.8\)\\
 &  &  & Overlap & \(2.99\) & \(-1.38\) & \(13.96\) & \(9.92\) & \(91.7\)\\
 &  &  & Learned & \(2.99\) & \(-2.09\) & \(12.67\) & \(9.39\) & \(91.1\)\\
 &  &  & Oracle & \(2.99\) & \(-1.38\) & \(13.96\) & \(9.92\) & \(91.7\)\\
 &  & \(2500\) & Constant & \(3.00\) & \(5.78\) & \(7.01\) & \(6.79\) & \(94.2\)\\
 &  &  & Overlap & \(3.00\) & \(5.74\) & \(7.01\) & \(6.79\) & \(94.1\)\\
 &  &  & Learned & \(3.00\) & \(5.92\) & \(6.97\) & \(6.59\) & \(93.2\)\\
 &  &  & Oracle & \(3.00\) & \(5.74\) & \(7.01\) & \(6.79\) & \(94.1\)\\
\end{tabular}

  \medskip
  {Mean: average of estimates; Bias: Monte-Carlo bias, \(10^{-3}\); RMSE: root mean squared error, \(10^{-2}\); \\SE: average of standard error estimates, \(10^{-2}\); Coverage: \(95\%\) confidence interval coverage, \(\%\).
    One simulation run with sample size \(n=1250\) using cross-fitting was omitted due to numerical instability.}
\end{table}

\begin{table}
  \caption{Summary of simulation results for \(\hat\gamma_2\) in experiment 1.}
  \label{tab:sim-beta-2}
  \footnotesize
  \centering
  
\begin{tabular}{llrlrrrrr}
{Cross-fitting} & {Setting} & {\(n\)} & {Weight} & {Mean} & {Bias} & {RMSE} & {SE} & {Coverage}\\
\(5\)-fold & I & \(1250\) & Constant & \(0.96\) & \(20.78\) & \(41.86\) & \(40.88\) & \(95.0\)\\
 &  &  & Overlap & \(0.96\) & \(21.22\) & \(41.93\) & \(40.96\) & \(95.0\)\\
 &  &  & Learned & \(0.96\) & \(19.37\) & \(41.83\) & \(39.76\) & \(95.0\)\\
 &  &  & Oracle & \(0.95\) & \(15.96\) & \(40.87\) & \(38.90\) & \(95.0\)\\
 &  & \(2500\) & Constant & \(0.94\) & \(0.35\) & \(27.83\) & \(27.72\) & \(94.6\)\\
 &  &  & Overlap & \(0.94\) & \(0.22\) & \(27.89\) & \(27.77\) & \(94.6\)\\
 &  &  & Learned & \(0.94\) & \(0.38\) & \(26.33\) & \(26.59\) & \(94.8\)\\
 &  &  & Oracle & \(0.94\) & \(-0.27\) & \(26.20\) & \(26.33\) & \(94.2\)\\
 & II & \(1250\) & Constant & \(0.96\) & \(19.50\) & \(51.63\) & \(51.35\) & \(94.8\)\\
 &  &  & Overlap & \(0.96\) & \(20.01\) & \(51.78\) & \(51.57\) & \(94.8\)\\
 &  &  & Learned & \(0.95\) & \(15.94\) & \(52.29\) & \(49.25\) & \(94.9\)\\
 &  &  & Oracle & \(0.95\) & \(12.40\) & \(50.63\) & \(48.12\) & \(95.3\)\\
 &  & \(2500\) & Constant & \(0.94\) & \(-1.54\) & \(35.40\) & \(35.15\) & \(94.7\)\\
 &  &  & Overlap & \(0.94\) & \(-1.71\) & \(35.57\) & \(35.30\) & \(94.7\)\\
 &  &  & Learned & \(0.94\) & \(-0.14\) & \(32.86\) & \(33.14\) & \(94.6\)\\
 &  &  & Oracle & \(0.94\) & \(-1.36\) & \(32.55\) & \(32.77\) & \(94.0\)\\
 & III & \(1250\) & Constant & \(0.95\) & \(7.66\) & \(48.05\) & \(33.87\) & \(94.7\)\\
 &  &  & Overlap & \(0.95\) & \(7.60\) & \(48.37\) & \(33.87\) & \(94.7\)\\
 &  &  & Learned & \(0.95\) & \(14.18\) & \(42.69\) & \(34.15\) & \(95.8\)\\
 &  &  & Oracle & \(0.95\) & \(7.60\) & \(48.37\) & \(33.87\) & \(94.7\)\\
 &  & \(2500\) & Constant & \(0.94\) & \(0.35\) & \(21.97\) & \(21.96\) & \(94.5\)\\
 &  &  & Overlap & \(0.94\) & \(0.27\) & \(21.98\) & \(21.96\) & \(94.4\)\\
 &  &  & Learned & \(0.94\) & \(1.24\) & \(22.08\) & \(22.17\) & \(94.3\)\\
 &  &  & Oracle & \(0.94\) & \(0.27\) & \(21.98\) & \(21.96\) & \(94.4\)\\
None & I & \(1250\) & Constant & \(0.95\) & \(14.37\) & \(48.29\) & \(37.76\) & \(93.6\)\\
 &  &  & Overlap & \(0.95\) & \(14.58\) & \(47.95\) & \(37.76\) & \(93.7\)\\
 &  &  & Learned & \(0.96\) & \(19.67\) & \(48.91\) & \(33.85\) & \(92.7\)\\
 &  &  & Oracle & \(0.96\) & \(17.90\) & \(52.41\) & \(35.42\) & \(93.4\)\\
 &  & \(2500\) & Constant & \(0.94\) & \(-0.02\) & \(26.50\) & \(25.06\) & \(93.2\)\\
 &  &  & Overlap & \(0.94\) & \(-0.15\) & \(26.56\) & \(25.10\) & \(93.3\)\\
 &  &  & Learned & \(0.94\) & \(0.01\) & \(24.63\) & \(23.14\) & \(93.1\)\\
 &  &  & Oracle & \(0.94\) & \(-1.26\) & \(24.92\) & \(23.81\) & \(93.2\)\\
 & II & \(1250\) & Constant & \(0.95\) & \(9.19\) & \(55.27\) & \(47.74\) & \(93.4\)\\
 &  &  & Overlap & \(0.95\) & \(9.68\) & \(55.11\) & \(47.83\) & \(93.5\)\\
 &  &  & Learned & \(0.96\) & \(24.82\) & \(47.47\) & \(41.59\) & \(92.8\)\\
 &  &  & Oracle & \(0.95\) & \(7.47\) & \(48.76\) & \(43.82\) & \(93.5\)\\
 &  & \(2500\) & Constant & \(0.94\) & \(-1.06\) & \(33.70\) & \(31.63\) & \(93.2\)\\
 &  &  & Overlap & \(0.94\) & \(-1.22\) & \(33.85\) & \(31.75\) & \(93.2\)\\
 &  &  & Learned & \(0.94\) & \(-1.17\) & \(30.60\) & \(28.69\) & \(93.1\)\\
 &  &  & Oracle & \(0.94\) & \(-1.64\) & \(30.92\) & \(29.53\) & \(93.3\)\\
 & III & \(1250\) & Constant & \(0.95\) & \(12.31\) & \(40.96\) & \(30.23\) & \(93.7\)\\
 &  &  & Overlap & \(0.95\) & \(12.43\) & \(40.58\) & \(30.17\) & \(93.7\)\\
 &  &  & Learned & \(0.95\) & \(14.33\) & \(36.64\) & \(28.31\) & \(92.8\)\\
 &  &  & Oracle & \(0.95\) & \(12.43\) & \(40.58\) & \(30.17\) & \(93.7\)\\
 &  & \(2500\) & Constant & \(0.94\) & \(-0.49\) & \(20.99\) & \(19.93\) & \(93.4\)\\
 &  &  & Overlap & \(0.94\) & \(-0.57\) & \(21.00\) & \(19.93\) & \(93.4\)\\
 &  &  & Learned & \(0.94\) & \(0.75\) & \(20.68\) & \(19.33\) & \(93.0\)\\
 &  &  & Oracle & \(0.94\) & \(-0.57\) & \(21.00\) & \(19.93\) & \(93.4\)\\
\end{tabular}

  \medskip
  {Mean: average of estimates; Bias: Monte-Carlo bias, \(10^{-3}\); RMSE: root mean squared error, \(10^{-2}\); \\SE: average of standard error estimates, \(10^{-2}\); Coverage: \(95\%\) confidence interval coverage, \(\%\).
    One simulation run with sample size \(n=1250\) using cross-fitting was omitted due to numerical instability.}
\end{table}

\begin{table}
  \caption{Summary of simulation results for \(\hat\gamma_3\) in experiment 1.}
  \label{tab:sim-beta-3}
  \footnotesize
  \centering
  
\begin{tabular}{llrlrrrrr}
{Cross-fitting} & {Setting} & {\(n\)} & {Weight} & {Mean} & {Bias} & {RMSE} & {SE} & {Coverage}\\
\(5\)-fold & I & \(1250\) & Constant & \(0.01\) & \(2.65\) & \(37.88\) & \(32.40\) & \(94.3\)\\
 &  &  & Overlap & \(0.01\) & \(2.80\) & \(38.04\) & \(32.47\) & \(94.3\)\\
 &  &  & Learned & \(0.01\) & \(-0.61\) & \(37.68\) & \(31.57\) & \(94.1\)\\
 &  &  & Oracle & \(0.01\) & \(-2.67\) & \(35.10\) & \(30.72\) & \(94.2\)\\
 &  & \(2500\) & Constant & \(-0.00\) & \(-12.35\) & \(20.75\) & \(21.43\) & \(95.7\)\\
 &  &  & Overlap & \(-0.00\) & \(-12.22\) & \(20.79\) & \(21.48\) & \(95.9\)\\
 &  &  & Learned & \(-0.00\) & \(-12.65\) & \(19.83\) & \(20.55\) & \(95.7\)\\
 &  &  & Oracle & \(-0.00\) & \(-12.90\) & \(19.67\) & \(20.33\) & \(95.4\)\\
 & II & \(1250\) & Constant & \(0.00\) & \(-9.47\) & \(41.48\) & \(40.78\) & \(94.9\)\\
 &  &  & Overlap & \(0.00\) & \(-9.41\) & \(41.61\) & \(40.95\) & \(94.9\)\\
 &  &  & Learned & \(-0.00\) & \(-13.10\) & \(41.79\) & \(39.24\) & \(94.0\)\\
 &  &  & Oracle & \(-0.00\) & \(-13.51\) & \(40.12\) & \(38.22\) & \(94.7\)\\
 &  & \(2500\) & Constant & \(-0.00\) & \(-14.33\) & \(26.51\) & \(27.50\) & \(95.7\)\\
 &  &  & Overlap & \(-0.00\) & \(-14.15\) & \(26.61\) & \(27.61\) & \(95.8\)\\
 &  &  & Learned & \(-0.00\) & \(-16.00\) & \(24.99\) & \(25.98\) & \(96.3\)\\
 &  &  & Oracle & \(-0.00\) & \(-16.36\) & \(24.85\) & \(25.65\) & \(95.5\)\\
 & III & \(1250\) & Constant & \(0.02\) & \(10.96\) & \(44.58\) & \(26.96\) & \(94.2\)\\
 &  &  & Overlap & \(0.02\) & \(11.15\) & \(44.74\) & \(26.97\) & \(94.1\)\\
 &  &  & Learned & \(0.02\) & \(9.70\) & \(40.51\) & \(27.25\) & \(94.1\)\\
 &  &  & Oracle & \(0.02\) & \(11.15\) & \(44.74\) & \(26.97\) & \(94.1\)\\
 &  & \(2500\) & Constant & \(0.00\) & \(-10.89\) & \(16.38\) & \(16.91\) & \(95.7\)\\
 &  &  & Overlap & \(0.00\) & \(-10.78\) & \(16.37\) & \(16.91\) & \(95.6\)\\
 &  &  & Learned & \(0.00\) & \(-9.66\) & \(16.45\) & \(17.09\) & \(95.3\)\\
 &  &  & Oracle & \(0.00\) & \(-10.78\) & \(16.37\) & \(16.91\) & \(95.6\)\\
None & I & \(1250\) & Constant & \(-0.03\) & \(-43.52\) & \(157.16\) & \(30.00\) & \(92.4\)\\
 &  &  & Overlap & \(-0.03\) & \(-42.86\) & \(154.62\) & \(29.99\) & \(92.3\)\\
 &  &  & Learned & \(-0.02\) & \(-30.14\) & \(120.16\) & \(26.78\) & \(91.7\)\\
 &  &  & Oracle & \(-0.02\) & \(-29.09\) & \(122.85\) & \(28.06\) & \(92.5\)\\
 &  & \(2500\) & Constant & \(-0.00\) & \(-13.36\) & \(19.61\) & \(19.27\) & \(94.8\)\\
 &  &  & Overlap & \(-0.00\) & \(-13.27\) & \(19.65\) & \(19.30\) & \(94.8\)\\
 &  &  & Learned & \(-0.00\) & \(-13.41\) & \(18.45\) & \(17.75\) & \(94.3\)\\
 &  &  & Oracle & \(-0.00\) & \(-13.93\) & \(18.64\) & \(18.30\) & \(94.8\)\\
 & II & \(1250\) & Constant & \(-0.06\) & \(-73.22\) & \(214.39\) & \(38.26\) & \(92.6\)\\
 &  &  & Overlap & \(-0.06\) & \(-72.11\) & \(210.91\) & \(38.31\) & \(92.7\)\\
 &  &  & Learned & \(-0.03\) & \(-45.74\) & \(125.73\) & \(32.73\) & \(91.6\)\\
 &  &  & Oracle & \(-0.05\) & \(-58.44\) & \(160.20\) & \(34.99\) & \(92.5\)\\
 &  & \(2500\) & Constant & \(-0.00\) & \(-15.50\) & \(25.04\) & \(24.60\) & \(94.4\)\\
 &  &  & Overlap & \(-0.00\) & \(-15.39\) & \(25.13\) & \(24.70\) & \(94.3\)\\
 &  &  & Learned & \(-0.01\) & \(-17.03\) & \(23.37\) & \(22.28\) & \(93.8\)\\
 &  &  & Oracle & \(-0.00\) & \(-16.95\) & \(23.54\) & \(22.99\) & \(94.7\)\\
 & III & \(1250\) & Constant & \(-0.02\) & \(-30.34\) & \(121.85\) & \(23.93\) & \(92.8\)\\
 &  &  & Overlap & \(-0.02\) & \(-29.88\) & \(119.87\) & \(23.88\) & \(92.8\)\\
 &  &  & Learned & \(-0.01\) & \(-22.01\) & \(93.61\) & \(22.27\) & \(91.2\)\\
 &  &  & Oracle & \(-0.02\) & \(-29.88\) & \(119.87\) & \(23.88\) & \(92.8\)\\
 &  & \(2500\) & Constant & \(-0.00\) & \(-11.97\) & \(15.51\) & \(15.26\) & \(95.0\)\\
 &  &  & Overlap & \(0.00\) & \(-11.90\) & \(15.50\) & \(15.26\) & \(94.8\)\\
 &  &  & Learned & \(0.00\) & \(-11.95\) & \(15.33\) & \(14.78\) & \(94.2\)\\
 &  &  & Oracle & \(0.00\) & \(-11.90\) & \(15.50\) & \(15.26\) & \(94.8\)\\
\end{tabular}

  \medskip
  {Mean: average of estimates; Bias: Monte-Carlo bias, \(10^{-3}\); RMSE: root mean squared error, \(10^{-2}\); \\SE: average of standard error estimates, \(10^{-2}\); Coverage: \(95\%\) confidence interval coverage, \(\%\).
    One simulation run with sample size \(n=1250\) using cross-fitting was omitted due to numerical instability.}
\end{table}

\begin{table}
  \caption{Summary of simulation results for \(\hat\gamma_4\) in experiment 1.}
  \label{tab:sim-beta-4}
  \footnotesize
  \centering
  
\begin{tabular}{llrlrrrrr}
{Cross-fitting} & {Setting} & {\(n\)} & {Weight} & {Mean} & {Bias} & {RMSE} & {SE} & {Coverage}\\
\(5\)-fold & I & \(1250\) & Constant & \(0.10\) & \(-24.83\) & \(70.68\) & \(64.60\) & \(95.8\)\\
 &  &  & Overlap & \(0.10\) & \(-25.60\) & \(70.97\) & \(64.73\) & \(95.9\)\\
 &  &  & Learned & \(0.10\) & \(-23.87\) & \(71.00\) & \(62.91\) & \(95.4\)\\
 &  &  & Oracle & \(0.11\) & \(-18.93\) & \(67.05\) & \(61.31\) & \(95.7\)\\
 &  & \(2500\) & Constant & \(0.13\) & \(5.12\) & \(43.06\) & \(42.95\) & \(95.1\)\\
 &  &  & Overlap & \(0.13\) & \(5.38\) & \(43.16\) & \(43.03\) & \(95.1\)\\
 &  &  & Learned & \(0.13\) & \(4.14\) & \(41.10\) & \(41.19\) & \(94.8\)\\
 &  &  & Oracle & \(0.13\) & \(5.39\) & \(40.92\) & \(40.77\) & \(94.5\)\\
 & II & \(1250\) & Constant & \(0.11\) & \(-12.95\) & \(79.51\) & \(80.25\) & \(95.8\)\\
 &  &  & Overlap & \(0.11\) & \(-13.73\) & \(79.78\) & \(80.58\) & \(95.8\)\\
 &  &  & Learned & \(0.12\) & \(-8.64\) & \(80.28\) & \(77.15\) & \(95.6\)\\
 &  &  & Oracle & \(0.12\) & \(-5.75\) & \(77.51\) & \(75.23\) & \(95.9\)\\
 &  & \(2500\) & Constant & \(0.13\) & \(9.15\) & \(54.70\) & \(54.40\) & \(94.8\)\\
 &  &  & Overlap & \(0.13\) & \(9.50\) & \(54.96\) & \(54.62\) & \(94.8\)\\
 &  &  & Learned & \(0.13\) & \(6.59\) & \(51.40\) & \(51.34\) & \(94.9\)\\
 &  &  & Oracle & \(0.13\) & \(7.59\) & \(50.81\) & \(50.71\) & \(94.1\)\\
 & III & \(1250\) & Constant & \(0.10\) & \(-21.01\) & \(70.26\) & \(52.68\) & \(95.9\)\\
 &  &  & Overlap & \(0.10\) & \(-21.15\) & \(69.70\) & \(52.66\) & \(96.0\)\\
 &  &  & Learned & \(0.10\) & \(-27.59\) & \(69.85\) & \(53.60\) & \(95.7\)\\
 &  &  & Oracle & \(0.10\) & \(-21.15\) & \(69.70\) & \(52.66\) & \(96.0\)\\
 &  & \(2500\) & Constant & \(0.13\) & \(4.56\) & \(34.13\) & \(34.00\) & \(95.1\)\\
 &  &  & Overlap & \(0.13\) & \(4.73\) & \(34.14\) & \(33.99\) & \(95.0\)\\
 &  &  & Learned & \(0.13\) & \(3.47\) & \(34.36\) & \(34.35\) & \(94.6\)\\
 &  &  & Oracle & \(0.13\) & \(4.73\) & \(34.14\) & \(33.99\) & \(95.0\)\\
None & I & \(1250\) & Constant & \(0.14\) & \(19.17\) & \(199.60\) & \(59.87\) & \(94.4\)\\
 &  &  & Overlap & \(0.14\) & \(18.10\) & \(196.45\) & \(59.84\) & \(94.3\)\\
 &  &  & Learned & \(0.12\) & \(-7.30\) & \(161.31\) & \(53.40\) & \(93.4\)\\
 &  &  & Oracle & \(0.12\) & \(-6.78\) & \(169.19\) & \(56.09\) & \(94.5\)\\
 &  & \(2500\) & Constant & \(0.12\) & \(-1.63\) & \(40.45\) & \(38.64\) & \(93.9\)\\
 &  &  & Overlap & \(0.12\) & \(-1.42\) & \(40.54\) & \(38.71\) & \(93.9\)\\
 &  &  & Learned & \(0.12\) & \(-3.39\) & \(37.89\) & \(35.60\) & \(93.1\)\\
 &  &  & Oracle & \(0.12\) & \(-1.27\) & \(38.39\) & \(36.69\) & \(93.7\)\\
 & II & \(1250\) & Constant & \(0.18\) & \(53.98\) & \(263.47\) & \(75.44\) & \(94.9\)\\
 &  &  & Overlap & \(0.18\) & \(52.05\) & \(259.32\) & \(75.54\) & \(94.7\)\\
 &  &  & Learned & \(0.13\) & \(2.73\) & \(120.05\) & \(64.33\) & \(93.5\)\\
 &  &  & Oracle & \(0.16\) & \(35.46\) & \(197.78\) & \(69.03\) & \(94.3\)\\
 &  & \(2500\) & Constant & \(0.12\) & \(0.45\) & \(51.28\) & \(48.72\) & \(93.8\)\\
 &  &  & Overlap & \(0.12\) & \(0.71\) & \(51.51\) & \(48.90\) & \(93.7\)\\
 &  &  & Learned & \(0.12\) & \(-0.58\) & \(46.93\) & \(44.10\) & \(93.1\)\\
 &  &  & Oracle & \(0.12\) & \(-0.63\) & \(47.56\) & \(45.49\) & \(93.8\)\\
 & III & \(1250\) & Constant & \(0.13\) & \(6.19\) & \(157.16\) & \(47.94\) & \(94.1\)\\
 &  &  & Overlap & \(0.13\) & \(5.51\) & \(154.66\) & \(47.83\) & \(94.1\)\\
 &  &  & Learned & \(0.12\) & \(-3.39\) & \(121.97\) & \(44.52\) & \(93.7\)\\
 &  &  & Oracle & \(0.13\) & \(5.51\) & \(154.66\) & \(47.83\) & \(94.1\)\\
 &  & \(2500\) & Constant & \(0.12\) & \(-1.74\) & \(32.21\) & \(30.71\) & \(93.9\)\\
 &  &  & Overlap & \(0.12\) & \(-1.59\) & \(32.22\) & \(30.70\) & \(94.2\)\\
 &  &  & Learned & \(0.12\) & \(-3.91\) & \(31.64\) & \(29.73\) & \(93.4\)\\
 &  &  & Oracle & \(0.12\) & \(-1.59\) & \(32.22\) & \(30.70\) & \(94.2\)\\
\end{tabular}

  \medskip
  {Mean: average of estimates; Bias: Monte-Carlo bias, \(10^{-3}\); RMSE: root mean squared error, \(10^{-2}\); \\SE: average of standard error estimates, \(10^{-2}\); Coverage: \(95\%\) confidence interval coverage, \(\%\).
    One simulation run with sample size \(n=1250\) using cross-fitting was omitted due to numerical instability.}
\end{table}

\begin{table}
  \caption{\(95\%\)-uniform coverage of the target population projected CATE in the simulation study.}
  \label{tab:sim-pcate-uniform-coverage}
  \footnotesize
  \centering
  \begin{tabular}{lrlrr}  & & & \multicolumn{2}{c}{Cross-fitting}\\{Setting} & {\(n\)} & {Weight} & {\(5\)-fold} & {None}\\I & \(1250\) & Constant & \(93.4\) & \(88.9\)\\ &  & Overlap & \(93.1\) & \(88.8\)\\ &  & Learned & \(93.8\) & \(86.3\)\\ &  & Oracle & \(94.3\) & \(88.4\)\\ & \(2500\) & Constant & \(94.8\) & \(91.9\)\\ &  & Overlap & \(94.6\) & \(91.4\)\\ &  & Learned & \(94.8\) & \(90.5\)\\ &  & Oracle & \(94.8\) & \(91.7\)\\II & \(1250\) & Constant & \(93.4\) & \(88.7\)\\ &  & Overlap & \(92.9\) & \(89.4\)\\ &  & Learned & \(94.2\) & \(85.1\)\\ &  & Oracle & \(94.1\) & \(88.7\)\\ & \(2500\) & Constant & \(94.9\) & \(91.3\)\\ &  & Overlap & \(94.7\) & \(91.3\)\\ &  & Learned & \(95.4\) & \(89.5\)\\ &  & Oracle & \(94.7\) & \(91.2\)\\III & \(1250\) & Constant & \(93.8\) & \(88.4\)\\ &  & Overlap & \(93.5\) & \(89.0\)\\ &  & Learned & \(93.1\) & \(86.9\)\\ &  & Oracle & \(93.5\) & \(89.3\)\\ & \(2500\) & Constant & \(94.4\) & \(92.1\)\\ &  & Overlap & \(94.4\) & \(91.9\)\\ &  & Learned & \(95.7\) & \(90.3\)\\ &  & Oracle & \(94.4\) & \(92.1\)\\\end{tabular}

\end{table}

\begin{table}
  \caption{Results for the transported TATE with \(5\)-fold cross-fitting, \(2\)-fold cross-fitting, and without cross-fitting.}
  \label{tab:tate-extra}
  \centering
  \footnotesize
  
\begin{tabular}{llrrr}
{Cross-fitting} & {Weight} & {Estimate} & {SE} & {\(95\%\)-CI}\\
\(5\)-fold & Overlap & \(-13.02\) & \(0.72\) & \((-14.42, -11.62)\)\\
 & Constant & \(-13.02\) & \(0.72\) & \((-14.42, -11.61)\)\\
 & Learned & \(-12.93\) & \(0.76\) & \((-14.41, -11.45)\)\\
\(2\)-fold & Overlap & \(-12.84\) & \(0.80\) & \((-14.41, -11.26)\)\\
 & Constant & \(-12.82\) & \(0.80\) & \((-14.39, -11.24)\)\\
 & Learned & \(-12.69\) & \(0.89\) & \((-14.44, -10.94)\)\\
None & Overlap & \(-13.02\) & \(0.37\) & \((-13.75, -12.29)\)\\
 & Constant & \(-13.00\) & \(0.37\) & \((-13.74, -12.27)\)\\
 & Learned & \(-12.92\) & \(0.35\) & \((-13.61, -12.24)\)\\
\end{tabular}

\end{table}

\begin{table}
  \caption{Sensitivity analysis for the transported TATE with \(10\)-fold cross-fitting and without cross-fitting.}
  \label{tab:tate-sensitivity}
  \centering
  \footnotesize
  
\begin{tabular}{lllrrr}
{Model} & {Cross-fitting} & {Weight} & {Estimate} & {SE} & {\(95\%\)-CI}\\
\(\hat\mu_k\) & \(10\)-fold & Constant & \(-12.80\) & \(0.73\) & \((-14.24, -11.37)\)\\
 &  & Overlap & \(-12.81\) & \(0.73\) & \((-14.25, -11.38)\)\\
 &  & Learned & \(-12.73\) & \(0.74\) & \((-14.18, -11.27)\)\\
 & None & Constant & \(-12.94\) & \(0.47\) & \((-13.86, -12.02)\)\\
 &  & Overlap & \(-12.96\) & \(0.47\) & \((-13.88, -12.04)\)\\
 &  & Learned & \(-12.94\) & \(0.45\) & \((-13.81, -12.06)\)\\
\(\hat\zeta_k\) & \(10\)-fold & Constant & \(-12.76\) & \(0.72\) & \((-14.18, -11.34)\)\\
 &  & Overlap & \(-12.77\) & \(0.72\) & \((-14.19, -11.35)\)\\
 &  & Learned & \(-12.70\) & \(0.75\) & \((-14.17, -11.23)\)\\
 & None & Constant & \(-12.97\) & \(0.35\) & \((-13.66, -12.28)\)\\
 &  & Overlap & \(-12.99\) & \(0.35\) & \((-13.67, -12.30)\)\\
 &  & Learned & \(-12.82\) & \(0.34\) & \((-13.49, -12.15)\)\\
\(\hat\pi_k\) & \(10\)-fold & Constant & \(-12.80\) & \(0.70\) & \((-14.17, -11.42)\)\\
 &  & Overlap & \(-12.80\) & \(0.70\) & \((-14.17, -11.43)\)\\
 &  & Learned & \(-12.72\) & \(0.72\) & \((-14.12, -11.32)\)\\
 & None & Constant & \(-12.97\) & \(0.51\) & \((-13.98, -11.96)\)\\
 &  & Overlap & \(-12.98\) & \(0.51\) & \((-13.99, -11.98)\)\\
 &  & Learned & \(-12.85\) & \(0.49\) & \((-13.82, -11.89)\)\\
\(\hat\mu_k, \hat\zeta_k\) & \(10\)-fold & Constant & \(-12.75\) & \(0.73\) & \((-14.17, -11.33)\)\\
 &  & Overlap & \(-12.76\) & \(0.72\) & \((-14.18, -11.34)\)\\
 &  & Learned & \(-12.68\) & \(0.75\) & \((-14.15, -11.22)\)\\
 & None & Constant & \(-12.94\) & \(0.46\) & \((-13.85, -12.04)\)\\
 &  & Overlap & \(-12.96\) & \(0.46\) & \((-13.87, -12.05)\)\\
 &  & Learned & \(-12.95\) & \(0.45\) & \((-13.83, -12.07)\)\\
\(\hat\mu_k, \hat\pi_k\) & \(10\)-fold & Constant & \(-12.84\) & \(0.70\) & \((-14.21, -11.47)\)\\
 &  & Overlap & \(-12.85\) & \(0.70\) & \((-14.21, -11.48)\)\\
 &  & Learned & \(-12.80\) & \(0.71\) & \((-14.19, -11.42)\)\\
 & None & Constant & \(-12.88\) & \(0.65\) & \((-14.16, -11.60)\)\\
 &  & Overlap & \(-12.90\) & \(0.65\) & \((-14.18, -11.61)\)\\
 &  & Learned & \(-12.94\) & \(0.62\) & \((-14.16, -11.72)\)\\
\(\hat\zeta_k, \hat\pi_k\) & \(10\)-fold & Constant & \(-12.78\) & \(0.70\) & \((-14.15, -11.41)\)\\
 &  & Overlap & \(-12.78\) & \(0.70\) & \((-14.15, -11.42)\)\\
 &  & Learned & \(-12.81\) & \(0.73\) & \((-14.25, -11.38)\)\\
 & None & Constant & \(-12.94\) & \(0.51\) & \((-13.94, -11.94)\)\\
 &  & Overlap & \(-12.96\) & \(0.51\) & \((-13.95, -11.96)\)\\
 &  & Learned & \(-12.88\) & \(0.48\) & \((-13.82, -11.94)\)\\
\(\hat\mu_k, \hat\zeta_k, \hat\pi_k\) & \(10\)-fold & Constant & \(-12.81\) & \(0.70\) & \((-14.18, -11.44)\)\\
 &  & Overlap & \(-12.82\) & \(0.70\) & \((-14.18, -11.46)\)\\
 &  & Learned & \(-12.82\) & \(0.71\) & \((-14.20, -11.44)\)\\
 & None & Constant & \(-12.87\) & \(0.66\) & \((-14.15, -11.58)\)\\
 &  & Overlap & \(-12.88\) & \(0.65\) & \((-14.16, -11.60)\)\\
 &  & Learned & \(-12.89\) & \(0.64\) & \((-14.15, -11.63)\)\\
\end{tabular}

  \medskip
  {Model: the model(s) changed from using all base learners to using only the null model and a generalized linear model.}
\end{table}




\begin{figure}
  \centering
  {
    \scalebox{.8}{\input{./art/tpcate-weight-alt.tex}}
    \scalebox{.8}{\input{./art/tpcate-age-alt.tex}}
    \scalebox{.8}{
\begin{tikzpicture}[x=1pt,y=1pt]
\definecolor{fillColor}{RGB}{255,255,255}
\path[use as bounding box,fill=fillColor,fill opacity=0.00] (0,0) rectangle (216.81,252.94);
\begin{scope}
\path[clip] ( 49.20, 61.20) rectangle (191.61,203.75);
\definecolor{fillColor}{RGB}{0,0,0}

\path[fill=fillColor] ( 98.43,137.76) circle (  2.25);

\path[fill=fillColor] (142.38,154.51) circle (  2.25);
\end{scope}
\begin{scope}
\path[clip] (  0.00,  0.00) rectangle (216.81,252.94);
\definecolor{drawColor}{RGB}{0,0,0}

\path[draw=drawColor,line width= 0.4pt,line join=round,line cap=round] ( 49.20, 61.20) --
	(191.61, 61.20) --
	(191.61,203.75) --
	( 49.20,203.75) --
	cycle;
\end{scope}
\begin{scope}
\path[clip] (  0.00,  0.00) rectangle (216.81,252.94);
\definecolor{drawColor}{RGB}{0,0,0}

\node[text=drawColor,anchor=base,inner sep=0pt, outer sep=0pt, scale=  1.00] at (120.41, 15.60) {Sex};

\node[text=drawColor,rotate= 90.00,anchor=base,inner sep=0pt, outer sep=0pt, scale=  1.00] at ( 10.80,132.47) {Weight change (kg)};
\end{scope}
\begin{scope}
\path[clip] (  0.00,  0.00) rectangle (216.81,252.94);
\definecolor{drawColor}{RGB}{0,0,0}

\path[draw=drawColor,line width= 0.4pt,line join=round,line cap=round] ( 98.43, 61.20) -- (142.38, 61.20);

\path[draw=drawColor,line width= 0.4pt,line join=round,line cap=round] ( 98.43, 61.20) -- ( 98.43, 55.20);

\path[draw=drawColor,line width= 0.4pt,line join=round,line cap=round] (142.38, 61.20) -- (142.38, 55.20);

\node[text=drawColor,anchor=base,inner sep=0pt, outer sep=0pt, scale=  1.00] at ( 98.43, 39.60) {F};

\node[text=drawColor,anchor=base,inner sep=0pt, outer sep=0pt, scale=  1.00] at (142.38, 39.60) {M};

\path[draw=drawColor,line width= 0.4pt,line join=round,line cap=round] ( 49.20, 66.48) -- ( 49.20,198.47);

\path[draw=drawColor,line width= 0.4pt,line join=round,line cap=round] ( 49.20, 66.48) -- ( 43.20, 66.48);

\path[draw=drawColor,line width= 0.4pt,line join=round,line cap=round] ( 49.20, 88.48) -- ( 43.20, 88.48);

\path[draw=drawColor,line width= 0.4pt,line join=round,line cap=round] ( 49.20,110.47) -- ( 43.20,110.47);

\path[draw=drawColor,line width= 0.4pt,line join=round,line cap=round] ( 49.20,132.47) -- ( 43.20,132.47);

\path[draw=drawColor,line width= 0.4pt,line join=round,line cap=round] ( 49.20,154.47) -- ( 43.20,154.47);

\path[draw=drawColor,line width= 0.4pt,line join=round,line cap=round] ( 49.20,176.47) -- ( 43.20,176.47);

\path[draw=drawColor,line width= 0.4pt,line join=round,line cap=round] ( 49.20,198.47) -- ( 43.20,198.47);

\node[text=drawColor,rotate= 90.00,anchor=base,inner sep=0pt, outer sep=0pt, scale=  1.00] at ( 34.80, 66.48) {\(-30\)};

\node[text=drawColor,rotate= 90.00,anchor=base,inner sep=0pt, outer sep=0pt, scale=  1.00] at ( 34.80,110.47) {\(-20\)};

\node[text=drawColor,rotate= 90.00,anchor=base,inner sep=0pt, outer sep=0pt, scale=  1.00] at ( 34.80,154.47) {\(-10\)};

\node[text=drawColor,rotate= 90.00,anchor=base,inner sep=0pt, outer sep=0pt, scale=  1.00] at ( 34.80,198.47) {\(0\)};
\end{scope}
\begin{scope}
\path[clip] ( 49.20, 61.20) rectangle (191.61,203.75);
\definecolor{drawColor}{RGB}{0,0,0}

\path[draw=drawColor,line width= 0.4pt,line join=round,line cap=round] ( 98.43,129.49) -- ( 98.43,146.02);

\path[draw=drawColor,line width= 0.4pt,line join=round,line cap=round] (142.38,139.96) -- (142.38,169.06);
\end{scope}
\end{tikzpicture}}
    \scalebox{.8}{\input{./art/tpcate-bmi-alt.tex}}
  }
  \caption{Transported weight loss effect of semaglutide measured in absolute weight change in the US study population, conditional on baseline body weight, age, sex, and BMI, respectively.
    For body weight, age, and BMI, the solid lines are point estimates, while the pointwise and uniform \(95\%\)-CIs are drawn with dashed and dotted lines.
    For sex, the point estimates and \(95\%\)-CIs are displayed as solid dots and error bars.
    F: female; M: male; p.p.: percentage point.
  }
  \label{fig:tpcate-alt}
\end{figure}



\end{document}